\documentclass[11pt]{article}
\pdfoutput=1

\usepackage{jheppub}
\usepackage{amssymb,amsfonts,amsmath,amsthm,lineno}
\usepackage{enumerate}
\usepackage{mathrsfs}
\usepackage{bbold}
\usepackage{tikz, adjustbox}
\usetikzlibrary{arrows.meta, bending}
\usetikzlibrary{shapes.geometric, arrows}
\usepackage{hyperref}
\usepackage{slashed}

\makeatletter
\newcommand{\Vast}{\bBigg@{4.75}}
\makeatother

\newcommand{\be}{\begin{equation}}
\newcommand{\ee}{\end{equation}}
\newcommand{\bea}{\begin{eqnarray}}
\newcommand{\eea}{\end{eqnarray}}

\newcommand{\CA}{\mathcal{A}}
\newcommand{\CB}{\mathcal{B}}
\newcommand{\CC}{\mathcal{C}}

\newcommand{\CF}{\mathcal{F}}

\newcommand{\CK}{\mathcal{K}}

\newcommand{\CN}{\mathcal{N}}

\newcommand{\CP}{\mathcal{P}}

\newcommand{\CT}{\mathcal{T}}

\newcommand{\lr}{\left (}
\newcommand{\rr}{\right )}
\newcommand{\ls}{\left [}
\newcommand{\rs}{\right ]}

\newcommand\qt\tau

\newcommand{\p}{\partial}

\renewcommand{\tilde}[1]{\widetilde{#1}}
\newcommand{\tr}{\text{tr}}

\makeatletter
\renewcommand{\@seccntformat}[1]{\csname the#1\endcsname.\,\,}
\makeatother
\allowdisplaybreaks

\let \savenumberline \numberline
\def \numberline#1{\savenumberline{#1.}}

\makeatletter
\def\@fpheader{\relax}
\makeatother

\def\bea{\begin{eqnarray}}
\def\eea{\end{eqnarray}}

\usetikzlibrary{shapes,arrows,chains}
\usetikzlibrary{decorations.markings}
\usetikzlibrary{decorations.pathmorphing}
\usetikzlibrary{shapes.multipart}
\tikzset{snake it/.style={decorate, decoration=snake}}

\title{\ \vspace{1.6cm} \\
\scalebox{0.9}{Worldsheet Formalism for Decoupling Limits in String Theory}}
\author[a]{Joaquim Gomis}
\author[b]{and Ziqi Yan\medskip}
\emailAdd{joaquim.gomis@ub.edu}
\emailAdd{ziqi.yan@su.se}
\affiliation[a]{Departament de F\'{i}sica Qu\`{a}ntica i Astrof\'{i}sica and Institut de Ci\`{e}ncies del Cosmos (ICCUB), \\
Universitat de Barcelona, Mart\'{i} i Franqu\`{e}s 1, E-08028 Barcelona, Spain\medskip}
\affiliation[b]{Nordita, KTH Royal Institute of Technology and Stockholm University\\
Hannes Alfv\'{e}ns v\"{a}g 12, SE-106 91 Stockholm, Sweden}
\abstract{We study the bosonic sector of a decoupling limit of type IIA superstring theory, where a background Ramond-Ramond one-form is fined tuned to its critical value, such that it cancels the associated background D0-brane tension.
The light excitations in this critical limit are D0-branes, whose dynamics is described by the Banks-Fischler-Shenker-Susskind (BFSS) Matrix theory that corresponds to M-theory in the Discrete Light-Cone Quantization (DLCQ). We develop the worldsheet formalism for the fundamental string in the same critical limit of type IIA superstring theory. We show that the fundamental string develops singularities on its worldsheet, whose topology is described by nodal Riemann spheres as in ambitwistor string theory. We study the T-duality transformations of this string sigma model 
and provide a worldsheet derivation for the recently revived and expanded duality web that unifies a zoo of decoupling limits in type II superstring theories. 
By matching the string worldsheet actions, we demonstrate how some of these decoupling limits are related to 
tensionless (and ambitwistor) string theory, Carrollian string theory, the Spin Matrix limits of the AdS/CFT correspondence, and more.} 

\begin{document}

\maketitle

\section{Introduction}

Different vacua in string theory are supposed to be unified by a single M-theory in eleven-dimensions. From the ten-dimensional string theoretical perspective, the extra eleventh dimension corresponds to a large string coupling. The nature of M-theory still remains mysterious. In certain decoupling limits of string theory, we zoom in on a self-consistent corner where some states become inaccessible, such that significant simplifications take place. Such simplifications sometimes allow us to probe nonperturbative aspects of string theory. It is therefore highly motivated to classify such decoupling limits and to map out different corners in M-theory, which may eventually be assembled into a bigger picture, while being agnostic to what the fundamental principles are\,\footnote{A complementary approach is to construct potential ultra-violet (UV) completions of the supermembrane sigma model, which is nonrenormalizable. One possible candidate for such a UV completion is known as the quantum critical membrane~\cite{Horava:2008ih, Yan:2022dqk}.}.

The studies of various decoupling limits in string/M-theory have been fruitful during the past decades. For example, the renowned AdS/CFT correspondence~\cite{Maldacena:1997re} relates two different decoupling limits: in the string picture, we decouple the $\CN=4$ super Yang-Mills (SYM) theory from IIB supergravity; in the $p$-brane picture, we decouple the near-horizon AdS geometry from the bulk IIB supergravity modes. Such decoupling limits can be viewed as field-theoretical limits of string/M-theory that lead to supergravities in the target space, gauge theories on D$p$-branes, and \emph{Aharony-Bergman-Jafferis-Maldacena} (ABJM) superconformal field theory on M2-branes~\cite{Aharony:2008ug}. 
Another remarkable decoupling limit that is more directly related to the theme of this paper is the one that leads to the Discrete Light Cone Quantization (DLCQ) of M-theory, \emph{i.e.}~M-theory in spacetime with a lightlike compactification. DLCQ M-theory is usually defined by taking a subtle infinite momentum limit along a spatial circle~\cite{Susskind:1997cw, Seiberg:1997ad, Sen:1997we}. In this limit, all light excitations except the Kaluza-Klein particle states in the lightlike compactificaion are decoupled, and the theory is described by the \emph{Banks-Fischler-Shenker-Susskind} (BFSS) Matrix quantum mechanics~\cite{deWit:1988wri, Banks:1996vh}\,\footnote{Alternatively, the lightlike circle maps to a spatial circle in the T- (U-)dual frame~\cite{Bergshoeff:2018vfn, Gomis:2000bd, Ebert:2023hba}. This relation makes it possible to use the self-consistent theory in the dual frame to define string/M-theory in the DLCQ.}. From the type IIA superstring perspective, the Kaluza-Klein particle states correspond to D0-brane bound states. In the original BFSS paper~\cite{Banks:1996vh}, it is conjectured that the BFSS Matrix theory at the large $N$ limit may describe the full nonperturbative M-theory in asymptotically flat spacetime, with $N$ the size of the matrix. At large $N$, the Matrix quantum mechanics becomes strongly coupled.

\vspace{2mm}

In this paper, we focus on a classification of fundamental strings that arise from various decoupling limits of type II superstring theory, and therefore build a duality web surrounding the BFSS Matrix theory. We will focus on the bosonic sector through the paper. 

\vspace{2mm}

We will start with a simple Galilean limit of the Nambu-Goto string action, where effectively the speed of light in the target space is sent to infinity and the ten-dimensional physics becomes Newtonian. This Galilean limit has been studied in~\cite{Batlle:2016iel, Gomis:2016zur, Batlle:2017cfa}, which leads to the \emph{non-vibrating string} that does \emph{not} admit any worldsheet wave equations. In~\cite{Batlle:2016iel}, the Nambu-Goto and phase-space formulation of the non-vibrating string are studied as part of a formal classification of nonrelativistic limits of the string. Our project grew out from the attempt of deriving the Polyakov formulation of this previously studied non-vibrating string theory, which led us to put forward a new string sigma model with its worldsheet being non-Riemannian\,\footnote{Note that this non-Riemannian feature of the worldsheet is different from the so-called nonrelativistic string theory~\cite{Gomis:2000bd} that arises from a stringy limit, where the worldsheet is Riemannian. See further discussion on nonrelativistic string theory in Section~\ref{sec:sdnrst}.}. We will show that the worldsheet topology of the non-vibrating string is singular: it is described by the nodal Riemann sphere, formed by identifying different pairs of points on a two-sphere. This is reminiscent of the case of ambitwistor string theory with a chiral worldsheet~\cite{Mason:2013sva}, where it is possible to compute stringy amplitudes that have localized moduli space and that describe particle scatterings. In ambitwistor string theory, the pairs of nodes on the Riemann sphere correspond to the loops of Feynman diagrams in a quantum field theory (QFT)~\cite{Geyer:2015bja}. We will show that this resemblance between distinct theories in terms of their worldsheet topology is more than a coincidence, as ambitwistor string theory is naturally connected to the non-vibrating string. We will demonstrate this connection explicitly in Section~\ref{sec:ast} from the worldsheet perspective. The same worldsheet structure also applies to other corners of type II superstring theories that we will later connect to the non-vibrating string via T-duality transformations. 

After this formal study of the non-vibrating string, we will then reveal its surprising relation to the BFSS Matrix theory. This has synergies with the work~\cite{udlstmt} of a different collaboration that also involves one of the current authors, which focuses on the target space perspective. It is explicitly shown in~\cite{udlstmt} that the BFSS Matrix theory arises from a BPS limit of type IIA superstring theory, which zooms in on a background D0-brane. In this decoupling limit, the Ramond-Ramond (RR) one-form coupled to the background D0-brane is fine tuned to cancel the brane tension, and all light excitations except the D0-branes described by the BFSS Matrix theory are decoupled. The resulting corner of the IIA theory is referred to as \emph{Matrix 0-brane Theory} (M0T) in~\cite{udlstmt}. In this current paper, we will show that the fundamental string in M0T is the non-vibrating string in~\cite{Batlle:2016iel}. In retrospective, the fact that the non-vibrating string, \emph{i.e.}~the M0T string, lacks vibrating modes is not surprising, as the dynamics of M0T is supposed to be captured by the light D0-branes described by the BFSS Matrix theory, instead of the fundamental string~\cite{Banks:1996vh}. 

One may thus question why the fundamental string is of any practical use for us to understand the dynamics of M0T. On the contrary, we will show that the M0T string provides an efficient tool for mapping out novel decoupling limits of
type II superstring theory via T-duality transformations. In practice, we will perform a series of T-duality transformations of the Polyakov formulation that we put forward for the non-vibrating string. This study of T-duality using the probe non-vibrating string will lead us to uncover a duality web that connects Matrix theories to tensionless (and ambitwistor) string theory, Carrollian string theory, strings from multicritical field limits, the Spin Matrix limit of the AdS/CFT correspondence, \emph{etc}. See Section~\ref{sec:conclusion} for a summary of the main duality results. 
Our exploration of the string worldsheet theories will recover a major part of the duality web studied in~\cite{udlstmt}, where the relevant decoupling limits are defined using target space techniques\,\footnote{See~\cite{udlstmt} for a road map that schematically depicts the connections between different corners. Some of the dualities there and their related decoupling limits date back to \emph{e.g.}~\cite{Gopakumar:2000ep, Harmark:2000ff, Gomis:2000bd, Danielsson:2000gi}. We also note that this duality web of decoupling limits can be probed by using U-dual invariant BPS mass formulae~\cite{bpslimits}.}. Generically speaking, these are BPS limits that arise from introducing a background brane (or bound branes) with infinite tension, which is canceled by certain $B$-field or RR gauge potentials that are fine-tuned to their critical values, such that all light excitations except the critical brane states are decoupled. The BFSS Matrix theory provides a classic example for such BPS limits. 

This current paper provides an intrinsic worldsheet construction for the duality web centered around the BFSS Matrix theory, which unifies different decoupling limits in string theory. This worldsheet perspective is complementary to the target space approach developed in~\cite{udlstmt, longpaper}: in this paper, we dualize the fundamental string sigma models without necessarily resorting to any limiting procedure, while in~\cite{udlstmt, longpaper} the duality transformations are applied to the reparametrized background fields that are used to define the decoupling limits.  

\vspace{3mm}

The paper is organized as follows. In Section~\ref{sec:nvst}, we derive the Polyakov formulation of the non-vibrating string and show that its worldsheet is non-Riemannian. We then argue that the worldsheet topology is described by nodal Riemann spheres and study the symmetries and gauge fixing of the non-vibrating string. In Section~\ref{sec:mtnlsw}, we reveal the relation between the non-vibrating string and BFSS Matrix theory, and then show how the non-vibrating string arises from a lightlike compactification of M-theory. In Sections~\ref{sec:fdlcq}$\sim$\ref{sec:tnddlcq}, we consider respectively the spacelike, timelike, and lightlike T-duality transformations by starting with the non-vibrating string theory, and uncover the duality web that connects Matrix theories, tensionless (and ambitwistor) string theory, Carrollian string theory, and strings from multicritical field limits.  In Section~\ref{sec:gcb}, we generalize the string actions obtained through this paper to arbitrary (bosonic) background fields. Moreover, we will study a relation between the duality web and the Spin Matrix limit of the AdS/CFT correspondence. In Section~\ref{sec:sdnrst}, we comment on a larger duality web that also involves S-duality, which has been summarized in~\cite{udlstmt} and will be studied in more detail in~\cite{longpaper}. In Section~\ref{sec:conclusion}, we summarize the main results of the paper and provide our outlooks. Some mathematical detail regarding the worldsheet topology is presented in Appendix~\ref{app:nodalcurve}.

\section{Strings in the Galilean Limit} \label{sec:nvst}

We start with reviewing the non-vibrating string in the Nambu-Goto formulation, which effectively arises from a Galilean limit in the target space~\cite{Batlle:2016iel}. We will then build on this previous study of the non-vibrating string and put forward its associated Polyakov formulation. We will find that it is natural to parametrize the worldsheet using nonrelativistic geometry, which corresponds to singular worldsheet topologies that are called nodal Riemann spheres. We will then revisit the symmetries of the non-vibrating string using the Polyakov formulation and recover the result in~\cite{Batlle:2016iel}, which shows that such a string does not vibrate and describes Galilean photons in the target space.  

\subsection{Nambu-Goto Action} \label{sec:nga}

We focus on the bosonic sector of superstring sigma models in flat target space. Define the worldsheet coordinates to be $\sigma^\alpha$\,, $\alpha = 0\,, 1$\,, and define the embedding coordinates that map the worldsheet to the ten-dimensional target space manifold to be $\hat{X}^\mu$, $\mu = 0\,, \, 1\,, \, \cdots, \, 9$\,. In the Nambu-Goto formulation, we write the string sigma model as
\be \label{eq:relngsa0}
    \hat{S} = - \frac{\hat{T}}{c} \int d^2 \sigma \, \sqrt{- \det \Bigl( \p^{}_\alpha \hat{X}^\mu_{\phantom{\dagger}} \, \p^{}_\beta \hat{X}^{}_\mu \Bigr)}\,,
\ee
where
\be \label{eq:xc01}
    \hat{X}^\mu = \bigl( c \, t, \, X^i \bigr)\,, 
        \qquad%
    i = 1\,, \, \cdots, \, 9\,,
\ee
where $t$ is the target space time. Note that we have displayed explicitly the dependence on the speed of light $c$ the target space in this Nambu-Goto action. We use the hatted notation before performing any limit, and unhatted notation after performing the Galilean limit, which essentially sends the speed of light to infinity. However, it does not really sense to directly send a dimensionful parameter, like $c$\,, to infinity. Therefore, in order to facilitate such an infinite speed-of-light limit, we write the action~\eqref{eq:relngsa0} as,
\be \label{eq:relngsa}
	\hat{S} = - \hat{T} \int d^2 \sigma \, \sqrt{- \det \Bigl( \p^{}_\alpha \hat{X}^\mu_{\phantom{\dagger}} \, \p^{}_\beta \hat{X}^{}_\mu \Bigr)}\,,
\ee  
where $c$ is set to 1, but introduce the following reparametrization in terms of a dimensionless control parameter $\omega$\,:
\be \label{eq:txomega}
	\hat{T} = \frac{T}{\omega}\,, 
		\qquad%
	\hat{X}^\mu = \bigl( \omega \, X^0, \, X^i \bigr)\,.
\ee
Note that $\omega$ is introduced at each place where $c$ appears in Eq.~\eqref{eq:relngsa}, and the infinite speed-of-light limit is now captured by sending $\omega$ to infinity. Here, $T$ will be the effective string tension for the theory at infinite $\omega$\,. In this new parametrization, $\omega$ can be viewed as the ratio between the two string tensions. In the $\omega \rightarrow \infty$ limit, we find a finite action that describes the non-vibrating string~\cite{Batlle:2016iel},
\be \label{eq:snvs}
    S = - {T} \int d^2 \sigma \, \sqrt{\epsilon^{\alpha\beta} \, \epsilon^{\gamma\delta} \, \p^{}_\alpha X^0 \, \p^{}_\gamma X^0 \, \p^{}_\beta X^i \, \p^{}_\delta X^i}\,.
\ee
Not surprisingly, under the above infinite speed-of-light limit, the resulting action~\eqref{eq:snvs} is invariant under the Galilean boost transformation,
\be \label{eq:glbt}
    \delta^{}_\text{\scalebox{0.8}{G}} X^0 = 0\,,
        \qquad%
    \delta^{}_\text{\scalebox{0.8}{G}} X^i = \Lambda^i X^0\,,
\ee
where $\Lambda^i$ is the boost velocity. Note that the action~\eqref{eq:snvs} is exactly invariant under the Galilean boost, which suggests that there is \emph{no} central charge. It is in this sense that we are dealing with a Galilean limit (instead of a Bargmann limit, where a central charge is present), which leads to the Galilean algebra in the target space~\cite{levy1969group, marmo1988quasi}. However, later in Section~\ref{sec:mtnlsw}, we will see that it is still possible to extend the Galilean algebra to include a central charge, such that it becomes the Bargmann algebra. This is because the background RR one-form potential also develops a divergence in $\omega$\,, which is essential for defining the infinite speed-of-light limit in the full-fledged type IIA superstring theory. This dependence on the RR potential is not yet accessible to us, since we are only considering the bosonic sector of the fundamental string at the moment. 

In order to decode the physics of the Nambu-Goto action~\eqref{eq:snvs}, it is useful to derive its associated Polyakov formulation, which we construct in the rest of this section.

\subsection{Phase-Space Action} 

To facilitate the derivation of the desired Polyakov action, it is instructive to first consider the phase-space formulation. 
In terms of the worldsheet coordinates $\sigma^\alpha = (\tau\,, \,\sigma)$\,, the phase-space action is~\cite{Batlle:2016iel} 
\be \label{eq:psf0}
    S^{}_\text{p.s.} = \int d^2 \sigma \, \biggl\{ P^{}_\mu \, \p_\tau X^\mu - \frac{\chi}{2 \, T} \Bigl[ P^{}_i \, P^{}_i - T^2 \, \bigl( \p_\sigma X^0 \bigr)^2 \Bigr] - \rho \, P^{}_\mu \, \p_\sigma X^\mu \biggr\}\,.
\ee
Here, $P_\mu$ are the conjugate momenta associated with the generalized coordinates $X^\mu$. 
The Lagrange multipliers $\chi$ and $\rho$ impose the Hamiltonian constraints. 
Integrating out $P_\mu$\,, $\rho$\,, and $\chi$ in Eq.~\eqref{eq:psf0} gives back the Nambu-Goto action \eqref{eq:snvs}. 

We now discuss how to obtain Eq.~\eqref{eq:psf0} from taking the Galilean limit of the phase-space action of the conventional string in the hatted notation. Define $\hat{G}^{}_{\alpha\beta} = \p^{}_\alpha \hat{X}^\mu \, \p^{}_\beta \hat{X}^{}_\mu$\,. The phase-space action for the conventional string is
\be \label{eq:relpsf0}
    \hat{S}^{}_\text{p.s.} = \int d^2 \sigma \, \biggl[ \hat{P}^{}_\mu \, \p_\tau \hat{X}^\mu - \frac{\hat{\chi}}{2 \, T} \Bigl( \hat{P}^{}_\mu \, \hat{P}^\mu + \hat{T}^2 \, \hat{G} \, \hat{G}^{\tau\tau} \Bigr) - \hat{\rho} \, \hat{P}^{}_\mu \, \p_\sigma \hat{X}^\mu \biggr],
\ee
where $\hat{G}^{\tau\tau}$ is a component of the inverse $\hat{G}^{\alpha\beta}$ of $\hat{G}_{\alpha\beta}$\,.
Integrating out $\hat{P}_\mu$\,, $\hat{\rho}$\,, and $\hat{\chi}$ in Eq.~\eqref{eq:relpsf0} gives back the Nambu-Goto action~\eqref{eq:relngsa}. 
The dependence on the speed of light in the conjugate momentum $\hat{P}^{}_\mu$ induces the following reparametrization: 
\be
 	\hat{P}^{}_\mu = \bigl( \omega^{-1} \, P^{}_0\,, \, P^{}_i \bigr)\,.
\ee
Together with the reparametrization~\eqref{eq:txomega} for $\hat{T}$ and $\hat{X}^\mu$\,, and requiring that Eq.~\eqref{eq:psf0} arise from the $\omega \rightarrow \infty$ limit of Eq.~\eqref{eq:relpsf0}, we find the following prescriptions for $\hat{\chi}$ and $\hat{\rho}$\,:
\be \label{eq:chirho0}
    \hat{\chi} = \omega^{-1} \, \chi + O\bigl(\omega^{-3}\bigr)\,,
        \qquad%
    \hat{\rho} = \rho + O\bigl(\omega^{-2}\bigr)\,.
\ee
These relations between the Lagrange multipliers imposing the Hamiltonian constraints in the phase-space action will play an important role for us to formulate the string worldsheet geometry, which naturally acquires a nonrelativistic structure. As we will reveal later in Section~\ref{sec:wstnrs}, this nonrelativistic geometry corresponds to a singular topology, which is the gauge-independent structure of the worldsheet. 

\subsection{Nonrelativistic Parametrization of the Worldsheet} \label{sec:nonrelws}

When the conventional string is concerned, in order to pass from the phase-space action \eqref{eq:relpsf0} to its Polyakov action, we introduce the worldsheet zweibein field $\hat{e}_\alpha{}^a$\,, $a=0\,,1$ and write
\be \label{eq:hchie0}
    \hat{\chi} = \frac{\hat{e}}{\hat{h}_{\sigma\sigma}}\,,
        \qquad%
    \hat{\rho} = \frac{\hat{h}_{\tau\sigma}}{\hat{h}_{\sigma\sigma}}\,,
        \qquad%
    \hat{e} = \det \bigl( \hat{e}_\alpha{}^a \bigr)\,,
        \qquad%
    \hat{h}_{\alpha\beta} = \hat{e}_\alpha{}^a \, \hat{e}_\beta{}^b \, \eta_{ab}\,.
\ee
Further integrating out $\hat{P}_\mu$ in Eq.~\eqref{eq:relpsf0} yields the standard Polyakov action,
\be \label{eq:relpf00}
    \hat{S}_\text{P} = - \frac{T}{2} \int d^2 \sigma \, \sqrt{-\hat{h}} \, \hat{h}^{\alpha\beta} \, \p^{}_\alpha \hat{X}^\mu \, \p^{}_\beta \hat{X}^{}_\mu\,, 
\ee
with the worldsheet metric $\hat{h}_{\alpha\beta}$\,. In order to achieve the rescalings in Eq.~\eqref{eq:chirho0}, it is natural to take the following ansatze:
\be
    \hat{e}_\alpha{}^0 = \omega^{z} \, e_\alpha{}^0\,, 
        \qquad
    \hat{e}_\alpha{}^1 = e_\alpha{}^1\,,
\ee
where the exponent $z$ is to be determined. Note that $e^{}_\alpha{}^a$ will be the zweibein field that encodes the worldsheet geometry after the $\omega \rightarrow \infty$ limit is performed. Matching $\hat{\chi}$ and $\hat{\rho}$ in Eqs.~\eqref{eq:chirho0} and~\eqref{eq:hchie0}, we find $|z| = 1$\,. In the $\omega \rightarrow \infty$ limit, the worldsheet metric
\be
    \hat{h}^{}_{\alpha\beta} = - \omega^{\pm2} \, e^{}_\alpha{}^0 \, e^{}_\beta{}^0 + e^{}_\alpha{}^1 \, e^{}_\beta{}^1
\ee
becomes singular. Therefore, the resulting worldsheet geometry does \emph{not} admit any metric description and is only appropriately described by the vielbein fields $e_\alpha{}^0$ and $e_\alpha{}^1$\,. In this sense, the string worldsheet now acquires a nonrelativistic geometry. We discuss both the $z=\pm1$ choices below:
\begin{enumerate}[(1)]

\item

\emph{Galilean Parametrization of the Worldsheet.} When $z=1$\,, we have
\be \label{eq:wsgge}
    \chi = \frac{e}{\bigl(e_\sigma{}^0 \bigr)^2} \,,
        \qquad%
    \rho = \frac{e_\tau{}^0}{e_\sigma{}^0}\,,
        \qquad%
    e = \det e_\alpha{}^a\,,
\ee
and
\be \label{eq:wsgg0}
    \hat{h}_{\alpha\beta} = - \omega^{2} \, e^{}_\alpha{}^0 \, e^{}_\beta{}^0 + e^{}_\alpha{}^1 \, e^{}_\beta{}^1\,.
\ee
The place where $\omega$ shows up in the worldsheet metric implies that the worldsheet speed of light in this parametrization coincides with the target space speed of light. 
The worldsheet becomes Galilean in the $\omega \rightarrow \infty$ limit, with the vielbein fields related to each other via the Galilean boost,
\be \label{eq:glb0}
    \delta_\text{g} e^0 = 0\,,
        \qquad%
    \delta_\text{g} e^1 = v \, e^0\,,
\ee
where $v$ is the worldsheet boost velocity. We now abuse the notation and define $t = e^0$ and $x = e^1$\,, which are actually one-forms, such that we could recast Eq.~\eqref{eq:glb0} in the familiar form of Galilean transformation,
\be \label{eq:txvt}
    t \rightarrow t\,,
        \qquad%
    x \rightarrow x + v \, t\,.
\ee
This Galilean boost arises from sending the worldsheet speed of light $c^{}_\text{w}$ to infinity in following Lorentz transformation:  
\be \label{eq:lrzb0}
    t \rightarrow \gamma \, \left( t + \frac{v}{c^{2}_\text{w}} \, x \right),
        \qquad%
    x \rightarrow \gamma \, \bigl( x + v \, t \bigr)\,,
        \qquad%
    \gamma = \left( 1 - \frac{v^2}{c^{2}_\text{w}} \right){\phantom{\bigg)}}^{\hspace{-4.5mm}-\frac{1}{2}}\!.
\ee
Intriguingly, the target space Galilean limit induces the Galilean limit on the worldsheet. In particular, the target space and worldsheet speed of light coincide with each other, \emph{i.e.}~$c^{}_\text{w} = c$\,. 

\item

\emph{Carrollian Parametrization of the Worldsheet.} When $z=-1$\,, we have
\be \label{eq:wscge0}
    \chi = \frac{e}{\bigl(e_\sigma{}^1 \bigr)^2} \,,
        \qquad%
    \rho = \frac{e_\tau{}^1}{e_\sigma{}^1}\,,
\ee
and
\be \label{eq:wscg0}
    \hat{h}_{\alpha\beta} = - \omega^{-2} \, e^{}_\alpha{}^0 \, e^{}_\beta{}^0 + e^{}_\alpha{}^1 \, e^{}_\beta{}^1\,.
\ee
In the $\omega \rightarrow \infty$ limit, the vielbein fields are related to each other via the following boost that essentially swaps the roles of space and time in~\eqref{eq:glb0}:
\be \label{eq:wsgb0}
    \delta^{}_\text{c} e^0 = \beta \, e^1,
        \qquad%
    \delta^{}_\text{c} e^1 = 0\,.
\ee
This is the so-called \emph{Carrollian boost}, which arises from sending both $v$ and $c_\text{w}$ in the Lorentz boost transformation~\eqref{eq:lrzb0} to zero while keeping $\beta \equiv v / c^2_\text{w}$ finite, such that Eq.~\eqref{eq:txvt} becomes $t \rightarrow t + \beta \, x$ and $x \rightarrow x$\,. In this sense, we say that the worldsheet geometry is Carrollian, where the ``space'' is absolute while the ``time'' is relative. This is in contrast to the Galilean case, where the ``time'' is absolute while the ``space'' is relative.

\end{enumerate}
After Wick rotating the worldsheet time, the Galilean parametrization from~\eqref{eq:wsgg0} and the Carrollian parametrization from~\eqref{eq:wscg0} are equivalent to each other on the two-dimensional Euclidean manifold, up to a conformal factor. However, the Schild (or transverse) gauge 
\be
    \chi = 1\,, 
        \qquad%
    \rho = 0\,,
\ee
appears to be more natural in the Carrollian parametrization~\eqref{eq:wscge0} of the worldsheet: in this parametrization, the Schild gauge corresponds to the conformal gauge $e^{}_\alpha{}^a \propto \delta_\alpha^a$\,. In contrast, the Schild gauge is associated with setting $e^{}_\alpha{}^b \, \epsilon^{}_b{}^a = \delta_\alpha^a$ in the Galilean parametrization~\eqref{eq:wsgge} of the worldsheet. Here, the Levi-Civita symbol $\epsilon_{ab}$ is defined via $\epsilon^{}_{01} = 1$ and the indices are raised (or lowered) by a Minkowski metric. 

We will mostly use the Carrollian worldsheet in the rest of the paper. However, we will resort to the Galilean worldsheet when we discuss the Spin Matrix theory later in Section~\ref{sec:smtnh}. At least at the classical level, it is straightforward to go between the Galilean and Carrollian parametrizations of the worldsheet by using the mapping
\be \label{eq:ctg}
	e^0 \rightarrow i \, e^1\,,
		\qquad%
	e^1 \rightarrow i \, e^0\,,
\ee
under which Eqs.~\eqref{eq:wsgg0} and \eqref{eq:wscg0} are mapped to each other up to a conformal factor. 

\subsection{Polyakov Action} 

Finally, we are ready to derive the Polyakov action for the non-vibrating string. Plug the Carrollian parametrization~\eqref{eq:wscge0} into the phase-space action \eqref{eq:psf0}, and then integrate out $P_i$\,, we find the following equivalent form: 
\be \label{eq:m0tsp}
    S^{}_\text{P} = \frac{T}{2} \int d^2 \sigma \, e \ls \lr \frac{\p^{}_\sigma X^0}{e^{}_\sigma{}^1} \rr^{\!2} + \Bigl( e^\alpha{}^{}_0 \, \p_\alpha X^i \Bigr)^2 + 
    \frac{2 \, P^{}_0}{T \, e^{}_\sigma{}^1} \,  e^\alpha{}^{}_0 \, \p^{}_\alpha{} X^0 \rs\,, 
\ee
where $e^\alpha{}_a = e^{-1} \, \epsilon^{\alpha\beta} \, e_\beta{}^b \, \epsilon_{ab}$\,. Here, the Levi-Civita symbol $\epsilon^{\alpha\beta}$ is defined via $\epsilon^{01} = 1$\,. 
The action \eqref{eq:m0tsp} is not yet manifestly covariant with respect to the worldsheet diffeomorphisms, which can be fixed by redefining the Lagrange multiplier $P^{}_0$ as 
\be \label{eq:redp0}
    P_0 = \frac{T}{2} \ls \lambda \, e_\sigma{}^1 - e_\sigma{}^0 \, \biggl( e^\alpha{}_1 \, \p_\alpha X^0 + \frac{\p_\sigma X^0}{e_\sigma{}^1} \biggr) \rs.
\ee
Here, $P_0$ is traded with $\lambda$ that plays the role of the new Lagrange multiplier\,\footnote{The presence of the Lagrange multiplier in spirit resembles the magnetic limit that leads to non-Lorentzian theories with constraints in~\cite{Bergshoeff:2022qkx}.}.
Plugging Eq.~\eqref{eq:redp0} into Eq.~\eqref{eq:m0tsp}, we derive the Polyakov action for the non-vibrating string,
\begin{align} \label{eq:m0tspf0}
    S^{}_\text{P} & = \frac{T}{2} \int d^2 \sigma \, e \, \biggl[ \Bigl( e^\alpha{}^{}_1 \, \p^{}_\alpha X^0 \Bigr)^2 + \Bigl( e^\alpha{}^{}_0 \, \p^{}_\alpha X^i \Bigr)^2 + \lambda \, e^\alpha{}^{}_0 \, \p^{}_\alpha X^0 \biggr]\,.
\end{align}
This action is invariant under the spacetime Galilean boost \eqref{eq:glbt} and the worldsheet Carrollian boost~\eqref{eq:wsgb0}, if supplemented with the additional infinitesimal transformation 
\be \label{eq:btl}
    \delta \lambda = - 2 \, \Lambda^{}_{i} \, e^\alpha{}^{}_0 \, \p^{}_\alpha X^i + 2 \, \beta \, e^\alpha{}^{}_1 \, \p^{}_\alpha X^0 \,.
\ee
The Polyakov action~\eqref{eq:m0tspf0} is a central result of the current paper. The discussions through the rest of this paper will hinge on this worldsheet action. 

The Polyakov action~\eqref{eq:m0tspf0} can be reproduced by taking a direct Galilean limit of the conventional Polyakov action~\eqref{eq:relpf00}. Using the Galilean reparametrization~\eqref{eq:txomega} of the target space and the Carrollian reparametrization~\eqref{eq:wscg0} of the worldsheet metric, we expand the Polyakov string action \eqref{eq:relpf00} with respect to a large $\omega$ as
\begin{align} \label{eq:ephsp0}
\begin{split}
    \hat{S}_\text{P} & = \frac{T}{2} \! \int \! d^2 \sigma \, e \! \ls - \, \omega^2 \, \Bigl( e^\alpha{}^{}_0 \, \p^{}_\alpha X^0 \Bigr)^{\!2} \! + \Bigl( e^\alpha{}^{}_1 \, \p^{}_\alpha X^0 \Bigr)^{\!2} \! + \Bigl( e^\alpha{}^{}_0 \, \p^{}_{\alpha} X^i \Bigr)^2 \! - \omega^{-2} \, \Bigl( \p^{}_{\alpha} X^i \Bigr)^2 \rs\,.
\end{split}
\end{align}
Using the Hubbard-Stratonovich transformation to integrate in an auxiliary field $\lambda$\,, we write Eq.~\eqref{eq:ephsp0} equivalently as
\begin{align} \label{eq:hsprrw0}
\begin{split}
    \hat{S}_\text{P} \rightarrow \frac{T}{2} \int d^2 \sigma \, e \, \biggl[ \Bigl( e^\alpha{}^{}_1 \, \p^{}_\alpha X^0 \Bigr)^{\!2} + \Bigl( e^\alpha{}^{}_0 \, \p^{}_\alpha X^i \Bigr)^{\!2} + \lambda \, e^\alpha{}^{}_0 \, \p^{}_\alpha X^0 \biggr]& \\[4pt]
    + \frac{T}{2 \, \omega^2} \int d^2 \sigma \, e \ls \frac{\lambda^2}{4} - \Bigl( e^\alpha{}^{}_1 \, \p^{}_{\alpha} X^i \Bigr)^{\!2} \rs&.
\end{split}
\end{align}
In the $\omega \rightarrow \infty$ limit, the Polyakov action~\eqref{eq:m0tspf0} is reproduced. Intuitively, the uncanceled $\omega^2$ divergent term in Eq.~\eqref{eq:ephsp0} induces strong back-reactions on the worldsheet gravity, which is responsible for why the worldsheet develops nonrelativistic features. 

\subsection{Worldsheet Topology: Nodal Riemann Spheres} \label{sec:wstnrs}

So far, we have only discussed the matter sector~\eqref{eq:m0tspf0} of the non-vibrating string. The complete (bosonic) sigma model also includes a dilaton term that couples to the worldsheet topology. In Euclidean time, and before the Galilean limit is performed,  the dilaton term in conventional string theory is
\be \label{eq:grs}
    \hat{S}_\text{gr.} = \frac{1}{4\pi} \int d^2 \sigma \, \sqrt{\hat{h}} \, R (\hat{h}) \, \hat{\Phi}\,. 
\ee
Here, $R(\hat{h})$ is the Ricci scalar defined with respect to the worldsheet metric $\hat{h}_{\alpha\beta} = \hat{e}_\alpha{}^a \, \hat{e}_\beta{}^b \, \eta^{}_{ab}$ and $\hat{\Phi}$ is the background dilaton field. We have already performed the Wick rotation $\hat{e}^{\,2} = i \, \hat{e}^{\,0}$. On flat worldsheet, this Wick rotation gives rise to the Euclidean coordinates $\sigma^\alpha = (\sigma^1, \sigma^2)$ with $\sigma^1 = \sigma$ and $\sigma^2 = i \, \tau$\,. 
For simplicity, we focus on the closed string case where the worldsheet is closed. 
We also assume that the dilaton $\hat{\Phi} = \hat{\Phi}_0$ is constant. The Gauss-Bonnet theorem implies 
\be \label{eq:eh}
	\hat{S}_\text{gr.} =  \hat{\chi}^{}_{\text{\scalebox{0.8}{E}}} \, \hat{\Phi}_0 \,,  
		\qquad%
	\hat{\chi}^{}_{\text{\scalebox{0.8}{E}}} = 2 - 2 \, n\,.
\ee
Here, $\hat{\chi}^{}_{\text{\scalebox{0.8}{E}}}$ is the Euler characteristic of the genus-$n$ Riemann surface. 

Next, we consider the Galilean limit. We have learned in Section~\ref{sec:nonrelws} that the implementation of this limit requires reparametrizing the worldsheet metric as in Eq.~\eqref{eq:wscg0}\,\footnote{One may also take the Galilean parametrization from~\eqref{eq:wsgg0}. As we have commented earlier, after Wick rotation, this is equivalent to the Carrollian reparametrization from~\eqref{eq:wscg0} up to a conformal factor.}. We will learn later in Eq.~\eqref{eq:hphirs} that the reparametrization of the dilaton $\hat{\Phi}_0$ in terms of $\omega$ is given by
\be \label{eq:hphi}
	\hat{\Phi}_0 = \Phi_0 - \frac{3}{2} \, \ln \omega\,.
\ee
The effective dilaton in non-vibrating string theory is taken to be constant here. The shift in $\omega$ contributes an overall factor in the associated path integral and does not affect the worldsheet topology directly. In the $\omega \rightarrow \infty$ limit, and dropping the $\omega$ shift in Eq.~\eqref{eq:hphi}, we write the dilaton contribution to the non-vibrating string sigma model generically as
\be
	S_\text{gr.} = \chi^{}_{\text{\scalebox{0.8}{E}}} \, \Phi_0\,.
\ee
The new Euler characteristic $\chi^{}_{\text{\scalebox{0.8}{E}}}$ is not necessarily the same as the original $\hat{\chi}^{}_{\text{\scalebox{0.8}{E}}}$ at finite $\omega$\,, as the infinite $\omega$ limit might well change the genus-$n$ topology. We now discuss the $\omega \rightarrow \infty$ limit in the cases of genus $n=0$\,, $n = 1$\,, and $n > 1$ separately, and show that the resulting topology is described by the nodal Riemann spheres~\cite{Geyer:2015bja, Geyer:2018xwu}. Later around Eq.~\eqref{eq:chie1}, we will see that the new Euler characteristic $\chi^{}_\text{E}$ after performing the $\omega \rightarrow \infty$ limit can be computed in an intrinsic way.

\vspace{3mm}

\noindent $\bullet$~\text{\emph{Tree level.}} We start with the $n=0$ case where, at finite $c$\,, the worldsheet topology is $S^2$\,. The situation here is very similar to the Riemannian case, for which we follow closely and generalize the discussion in~\cite{kreyszig2013differential} (see around Chapter X, Section 69). Cut the worldsheet $\Sigma_2$ into two pieces $\mathcal{S}_1$ and $\mathcal{S}_2$ with sufficiently\,\footnote{The surface $S$ and its boundary curve $\p S$ must be sufficiently smooth: $S$ is required to be of differentiability class $r \geq 3$ while $\p S$ is required to be of differentiability class $r^* \geq 2$.} differentiable boundary curves. We then map conformally $\mathcal{S}_1$ to a flat disk and introduce a (geodesic) polar coordinate system $(r\,, \theta)$ on it, with $r$ the radial direction and $\theta$ the polar angle. Choose the cut such that the rescaling $\omega$ of the Vielbein field in Eq.~\eqref{eq:wscg0} is mapped to the rescaling $r \rightarrow \omega^{-1} \, r$\,. In this case, the $\omega \rightarrow \infty$ limit preserves the topology, and the Gauss-Bonnet theorem implies
\be
	\lim_{\omega \rightarrow \infty} \! \lr \frac{1}{4\pi} \int_{\mathcal{S}^{}_1} d^2 \sigma \, \sqrt{\hat{h}} \, R (\hat{h}) + \frac{1}{2\pi} \int_{\p \mathcal{S}^{}_1} ds \, \hat{\kappa}_\text{g} \rr = 1\,,
\ee
where $\hat{\kappa}_\text{g}$ is the \emph{geodesic curvature} of the curve $\p \mathcal{S}$ at finite $\omega$\,. 
The same equation holds for $\mathcal{S}_2$\,. Since $\Sigma_2 = \mathcal{S}_1 \cup \mathcal{S}_2$ and $\p \mathcal{S}_1 + \p \mathcal{S}_2 = 0$\,, we find
\be
	\lim_{\omega \rightarrow \infty} \frac{1}{4\pi} \int d^2 \sigma \sqrt{\hat{h}} \, R(\hat{h}) = 2\,.
\ee 
We conclude that the topology of the sphere at zero genus is not affected by the $\omega \rightarrow \infty$ limit and the Euler characteristic is still $\chi^{}_{\text{\scalebox{0.8}{E}}} = 2$\,. 

\vspace{3mm}

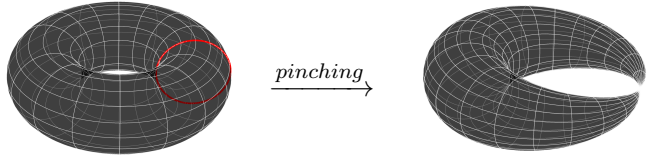
\begin{figure}[t!]
	\centering
	\begin{minipage}{3cm}
	\includegraphics[width=1\textwidth]{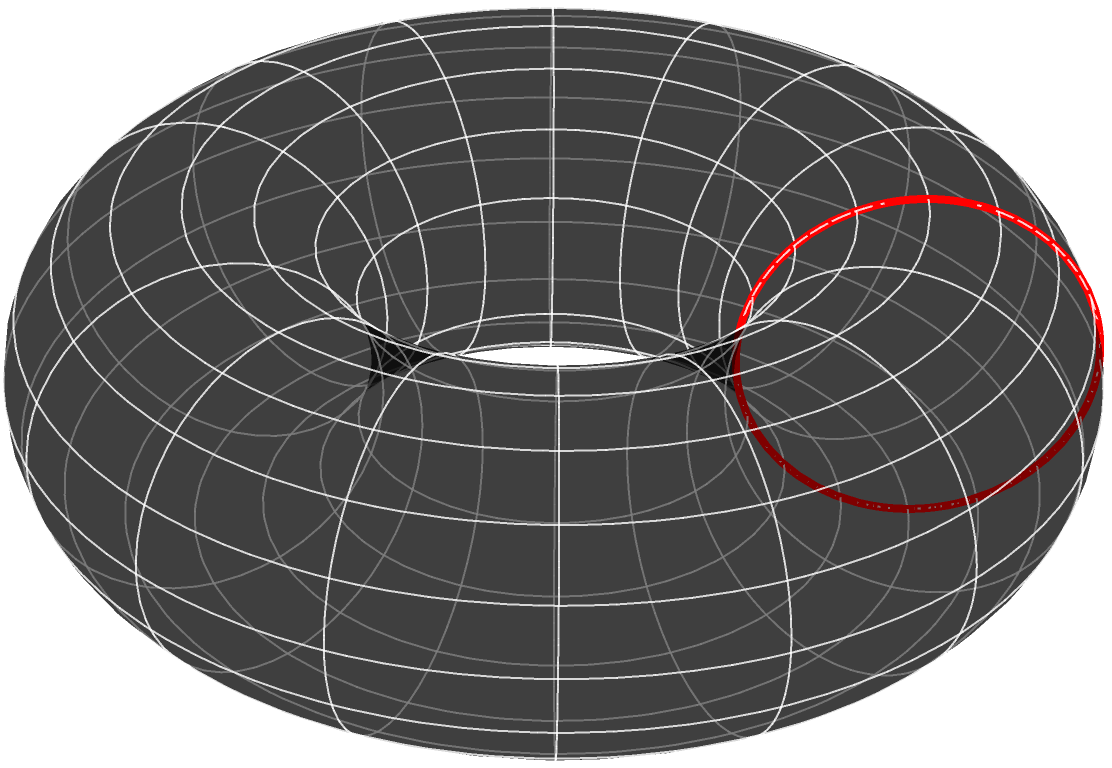}
	\end{minipage}
	\hspace{0.1cm}
	\begin{minipage}{2cm}
	\begin{tikzpicture}
		\node at (0,0) {$\xrightarrow{pinching}$};
	\end{tikzpicture}
	\end{minipage}
	\hspace{-0.1cm}
	\begin{minipage}{3cm}
	\includegraphics[width=1\textwidth]{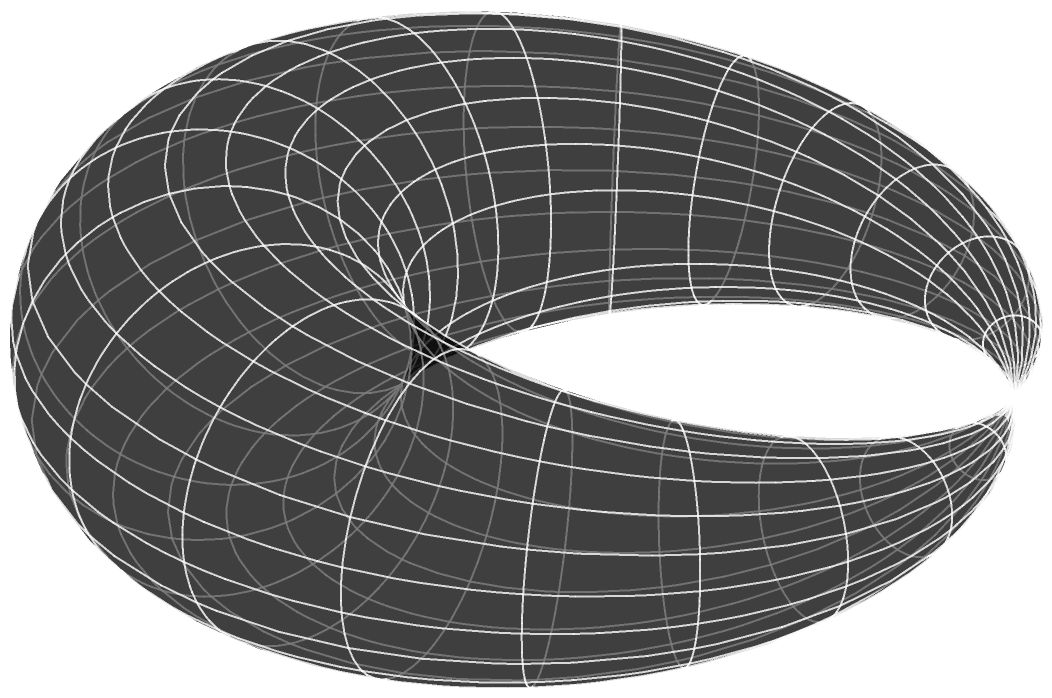}
	\end{minipage}
	\caption{Pinching a torus: the red circle on the torus collapses to zero.}
	\label{fig:nrs}
\end{figure}

\noindent $\bullet$~\text{\emph{One-loop level.}} Next, we consider the genus-one case with $n = 1$ and the worldsheet topology is a two-torus at finite $\omega$\,. The first fundamental form on the two-torus is
\be
	ds^2 = \frac{\Gamma}{\tau_2} \, \Bigl| d\sigma^1 + \tau \, d\sigma^2 \Bigr|^2\,,
\ee
where $\Gamma$ is the area of the torus and $\tau = \tau^{}_1 + i \, \tau^{}_2$ is the modulus. The worldsheet metric is
\be \label{eq:tmzbf}
	h^{}_{\alpha\beta} = e^{}_\alpha{}^a \, e^{}_\beta{}^b \, \delta^{}_{ab}\,, 
		\qquad%
	e^{}_\alpha{}^1 =
	\sqrt{\Gamma}
	\begin{pmatrix}
		0 \\[4pt]
		\sqrt{\tau^{}_2}
	\end{pmatrix},
		\qquad%
	e^{}_\alpha{}^2 =
	\sqrt{\frac{\Gamma}{\tau_2}}
	\begin{pmatrix}
		1 \\[4pt]
		\tau^{}_1
	\end{pmatrix},
\ee
where $e_\alpha{}^a$ is the zweibein field. The rescaling $e_\alpha{}^2 \rightarrow \omega^{-1} \, e_\alpha{}^2$ implied by Eq.~\eqref{eq:wscg0} translates to $\tau^{}_2 \rightarrow \omega \, \tau^{}_2$ and $\Gamma \rightarrow \omega^{-1} \, \Gamma$\,. In this case, the effect of the $\omega \rightarrow \infty$ limit is translated to the limit of the modulus\,\footnote{In the case where the definitions for $e_\alpha{}^1$ and $e_\alpha{}^2$ are swapped in Eq.~\eqref{eq:tmzbf}, we have $\tau \rightarrow \text{real}$ instead of $\tau \rightarrow \infty$\,. These two different choices are related to each other via an SL($2\,,\mathbb{Z})$ transformation. We prefer the choice that leads to $\tau \rightarrow i \, \infty$ as it is within the fundamental domain while $\tau \rightarrow \text{real}$ is not. Another subtlety here is that what we wrote in Eq.~\eqref{eq:inftau} really means $\tau = c_0 + i \, \infty$\,, where $c^{}_0$ is a real number. When $c^{}_0 \neq 0$\,, $\tau$ is outside the fundamental domain. This $\tau = c_0 + i \, \infty$ is of course also related to $\tau = i \, \infty$ without any real part via an SL($2\,,\mathbb{Z}$) transformation.}, 
\be \label{eq:inftau}
	\tau \rightarrow i \, \infty\,. 
\ee
This is equivalent to a \emph{Deligne-Mumford compactification}~\cite{deligne1969irreducibility} of the torus, where a homological cycle is collapsed to a singular point called a \emph{node}. The resulting topology in $\omega \rightarrow \infty$ is a pinched torus, which is equivalent to a sphere with two nodal points being identified with each other. See Fig.~\ref{fig:nrs} for an illustration. A detailed review of the reasoning that underlies this relation between the $\tau \rightarrow i \, \infty$ limit in Eq.~\eqref{eq:inftau} and the pinching of the torus in the language of elliptic curves is given in Appendix~\ref{app:nodalcurve}. We follow~\cite{Geyer:2015bja, Geyer:2018xwu} and refer to the general topologies from identifying pairs of nodes on the sphere as \emph{nodal Riemann spheres}.   

The appearance of nodal spheres here is reminiscent of the situation of the ambitwistor string amplitudes~\cite{Geyer:2015bja, Geyer:2018xwu}. For example, it is shown in~\cite{Geyer:2015bja} that the one-loop moduli space is localized to a discrete set of points selected by the scattering equations that encode the kinematics of particle scatterings. Using the ``global residue theorem," the contour integrals around these discrete points can be rewritten as the contour integral over the boundary of the fundamental domain for the torus moduli space, and the only nontrivial contribution comes from $\tau \rightarrow i \, \infty$\,, just as in Eq.~\eqref{eq:inftau}. This implies that we encounter the same worldsheet topology in both the non-vibrating and ambitwistor string theory. This resemblance is \emph{not} a coincidence: as we will see in Section~\ref{sec:fdlcq}, the non-vibrating string is related to ambitwistor string theory via a timelike T-duality transformation. 

\begin{figure}[t!]
    \centering
    \includegraphics[width=0.22\textwidth]{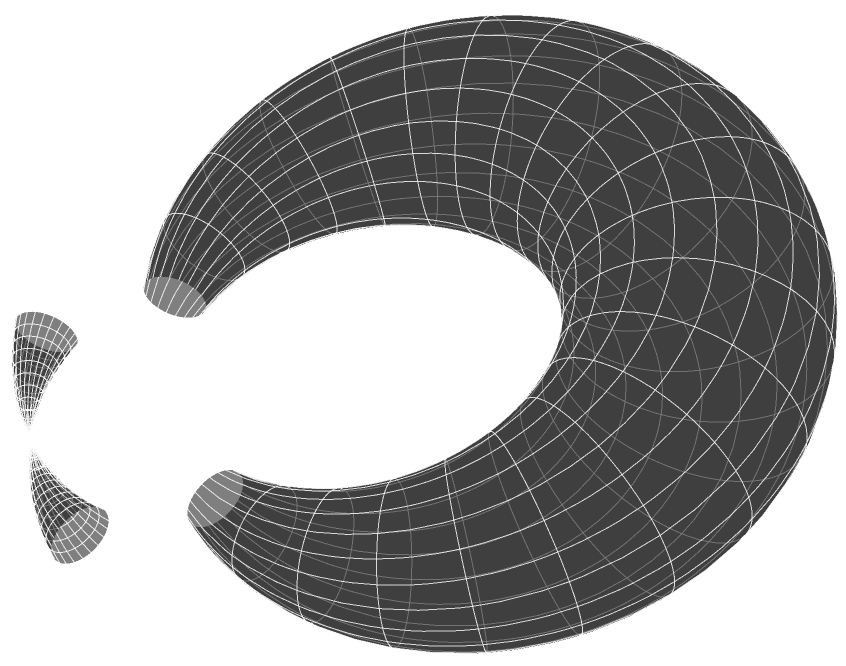}
    \caption{A cut of the pinched torus. The component on the right is Riemannian.}
    \label{fig:cpt}
\end{figure}

It is interesting to note that the nodal Riemann sphere is \emph{almost} a Riemann sphere: if we perform a cut of the worldsheet as in Fig.~\ref{fig:cpt}, then the component on the right is a regular Riemann surface. This suggests that it might be possible to map the non-vibrating string sigma model defined on this patch of the surface to be a conventional conformal field theory. 

It is important to note that there is still a well-defined Euler characteristic after the limit~\eqref{eq:inftau} is performed, under which the Riemann surface becomes singular. In fact, the Euler characteristics of the pinched torus is 
\be \label{eq:chie1}
    \chi^{}_{\text{\scalebox{0.8}{E}}} = 1\,,
\ee
which can be explicitly computed by \emph{e.g.}~using the triangulation given in Fig. 1 of~\cite{brasselet1996intersection}.

\vspace{3mm}

\begin{figure}[b!]
	\centering
	\begin{minipage}{2cm}
	\includegraphics[width=1\textwidth]{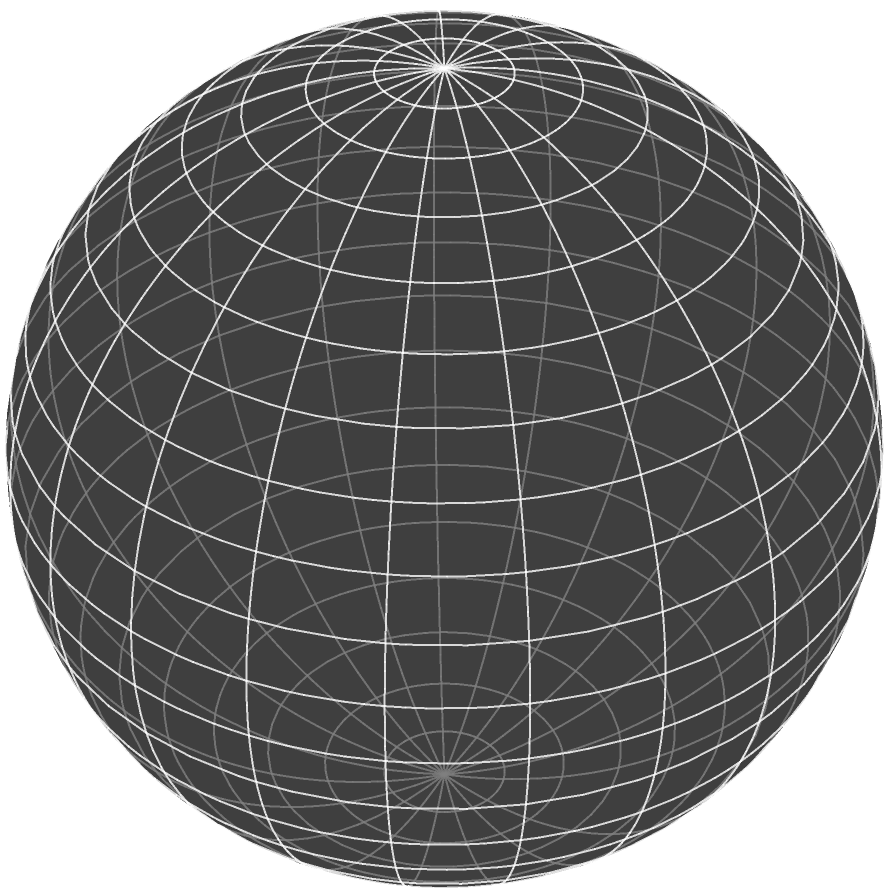}
	\end{minipage}
	\hspace{0.2cm}+\hspace{0.2cm}
	\begin{minipage}{3cm}
	\includegraphics[width=1\textwidth]{pinched_torus.png}
	\end{minipage}
	\hspace{0.2cm}+\hspace{0.2cm}
	\begin{minipage}{4.5cm}
	\includegraphics[width=1\textwidth]{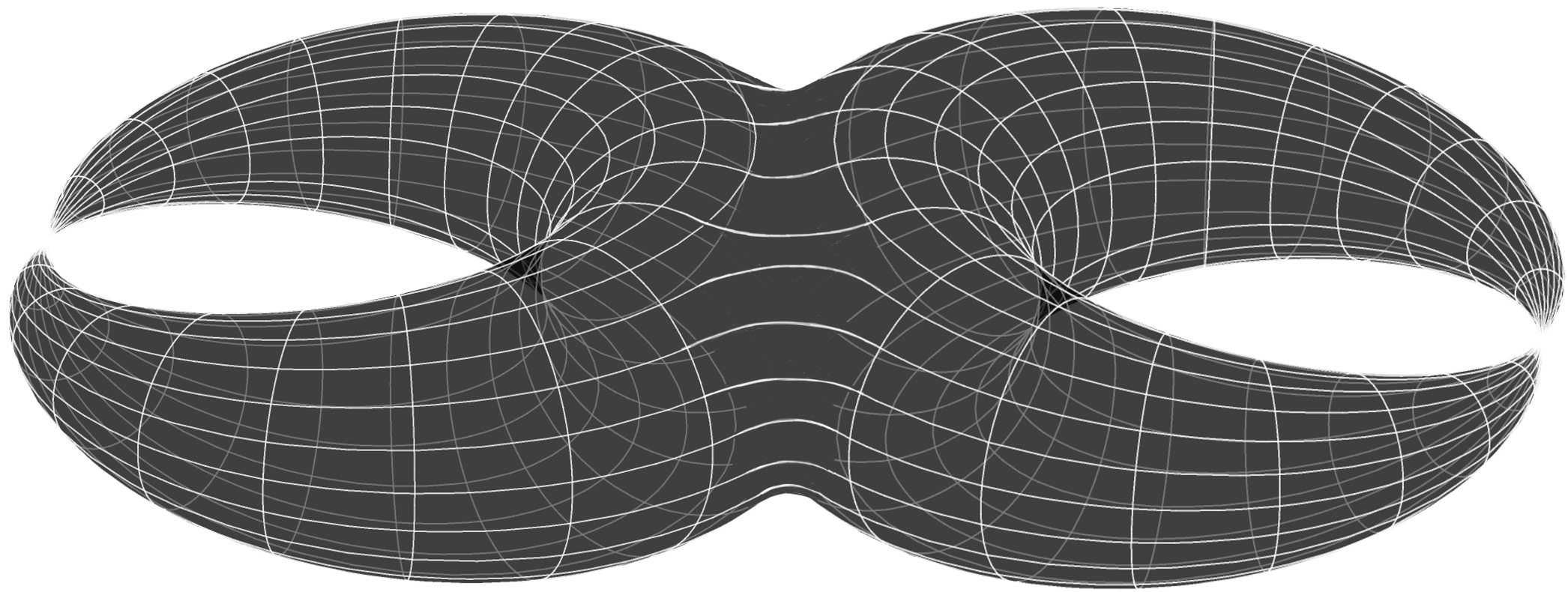}
	\end{minipage}
	\hspace{0.2cm}+\,$\cdots$
	\caption{Sum over the topologies of nodal Riemann spheres.}
	\label{fig:snrs}
\end{figure}

\noindent $\bullet$~\text{\emph{Higher-loop orders.}} The topology in the genus-$n$ case follows directly from our discussion on the genus-one case above via the sewing method: we just obtain a nodal Riemann sphere with $n$ pairs of identified points in the $\omega \rightarrow \infty$ limit from pinching different cycles on the genus-$n$ Riemann surface. Therefore, the path integral for the non-vibrating string is associated with the sum of different topologies described by nodal Riemann surfaces as in Fig.~\ref{fig:snrs}. The Euler characteristics of the $n$-loop string worldsheet is given by 
\be \label{eq:chie}
	\chi^{}_{\text{\scalebox{0.8}{E}}} = 2 - 2 \, n + \text{number of pinches} = 2 - n\,,
\ee
where the number of pinches is equal to the order of string loops. The Euler characteristic~\eqref{eq:chie} follows directly from the result~\eqref{eq:chie1} for the pinched torus via sewing. In the case of ambitwistor string theory, the number of the pairs of identified points on the Riemann sphere corresponds to the loop order of the associated Feynman diagrams~\cite{Geyer:2015bja, Geyer:2018xwu}.   

\vspace{3mm} 

Later in Sections~\ref{sec:fdlcq}$-$\ref{sec:tnddlcq}, we will build a duality web of string sigma models by performing T-duality transformations of the Polyakov action describing the non-vibrating string. It is important to note that such perturbative duality transformations do \emph{not} alter the nature of the worldsheet structure and, therefore, the discussions of the worldsheet properties in this section continue to apply.   

\subsection{Worldsheet Symmetries and Gauge Fixing} \label{sec:wsgf}

We now study the worldsheet symmetries of the non-vibrating string action \eqref{eq:m0tspf0} and their gauge fixing. It is sufficient to consider the flat target space and the gauge fixing in the Carrollian parametrization of the worldsheet. This subsection extends the previous work~\cite{Batlle:2016iel} on non-vibrating strings to the Polyakov formulation. 
See related discussions for the same worldsheet structure in the context of tensionless/ambitwistor string theory~\cite{Isberg:1993av, Bagchi:2013bga, Casali:2016atr, Chen:2023esw}, Spin Matrix theory~\cite{Harmark:2018cdl, Harmark:2019upf, Bidussi:2023rfs}, and tropological sigma models~\cite{Albrychiewicz:2023ngk}. We will visit the connections between non-vibrating string theory and these corners using T-duality later after understanding the gauge fixing. 

We transcribe the non-vibrating string action~\eqref{eq:m0tspf0} as below:
\be \label{eq:mztfscw}
	S = \frac{T}{2} \int d^2 \sigma \, e \, \biggl[ \Bigl( e^\alpha{}^{}_1 \, \p^{}_\alpha X^0 \Bigr)^2 + e^\alpha{}^{}_0 \, e^\beta{}^{}_0 \, \p_\alpha X^i \, \p_\beta X^i + \lambda \, e^\alpha{}^{}_0 \, \p_\alpha X^0 \biggr]\,.
\ee
Under the reparametrization symmetry of the worldsheet coordinates $\sigma^\alpha \rightarrow \sigma^{\prime\alpha} (\tau\,, \, \sigma)$\,, the induced transformations on the zweibein fields $e_\alpha{}^a$ and the worldsheet fields $X^\mu$ and $\lambda$ are:
\be \label{eq:wdiff}
	e'_\alpha{}^a (\tau'\!, \sigma') = \frac{\p \sigma^\beta}{\p \sigma'{}^\alpha (\sigma)} \, e_\beta{}^a (\tau, \sigma)\,,
        \qquad%
    X'{}^\mu (\sigma') = X^\mu(\sigma)\,,
        \qquad%
    \lambda'(\sigma') = \lambda(\sigma)\,.
\ee
In addition, the action is invariant under the local worldsheet gauge symmetries
\begin{subequations} \label{eq:lgsw}
\begin{align} 
	e^0 & \rightarrow h(\tau, \sigma) \, e^0 + b(\tau, \sigma) \, e^1\,, 
        &%
    X^\mu & \rightarrow X^\mu\,, \\[4pt]
	e^1 & \rightarrow h(\tau, \sigma) \, e^1\,, 
        &%
    \lambda & \rightarrow \frac{\lambda}{h} + \frac{b}{h^3} \, \bigl( 2 \, h \, e^\alpha{}_1 - b \, e^\alpha{}_0 \bigr) \, \p_\alpha X^0\,,
\end{align}
\end{subequations}
where $b(\sigma)$ parametrizes the local Carrollian boost that generalizes Eq.~\eqref{eq:wsgb0} to finite transformations and $h(\sigma)$ parametrizes the local Weyl symmetry. 

\vspace{3mm}

\noindent $\bullet$~\text{\emph{Gauge fixing.}}
The infinitesimal version of Eqs.~\eqref{eq:wdiff} and~\eqref{eq:lgsw} are
\begin{subequations} \label{eq:infwt}
\begin{align}
	\delta e^{}_\alpha{}^0 & = \xi^\beta \, \p^{}_\beta e^{}_\alpha{}^0 + e^{}_\beta{}^0 \, \p^{}_\alpha \xi^\beta + \theta \, e^{}_\alpha{}^0 + \beta \, e^{}_\alpha{}^1\,, \label{eq:infwt1} \\[4pt]
	\delta e^{}_\alpha{}^1 & =  \xi^\beta \, \p^{}_\beta e^{}_\alpha{}^1 + e^{}_\beta{}^1 \, \p^{}_\alpha \xi^\beta + \theta \, e^{}_\alpha{}^1\,, \label{eq:infwt2} \\[4pt]
    \delta X^\mu & = \xi^\alpha \, \p_\alpha X^\mu\,, \label{eq:infwt3} \\[4pt]
	\delta \lambda & = \xi^\alpha \p_\alpha \lambda - \theta \, \lambda + 2 \, \beta \, e^\alpha{}_1 \, \p_\alpha X^0\,. \label{eq:infwt4}
\end{align}
\end{subequations}
where $\xi^\alpha$ parametrizes the diffeomorphisms, $\theta$ the dilatation, and $\beta$ the worldsheet Carrollian boost.
We choose the flat gauge $e^{}_\alpha{}^a = \delta_\alpha^a$\,, which implies $\delta e_\alpha{}^a = 0$\,. From Eqs.~\eqref{eq:infwt1} and \eqref{eq:infwt2}, we find
\be
	\theta = - \p_\tau \xi^\tau = - \p_\sigma \xi^\sigma\,,
		\qquad%
	\beta = - \p_\sigma \xi^\tau\,,
		\qquad%
	\p_\tau \xi^\sigma = 0\,,	
\ee
which are solved by
\be
	\xi^\tau = u' (\sigma) \, \tau + v(\sigma)\,,
		\qquad%
	\xi^\sigma = u (\sigma)\,, 
		\qquad%
	\theta = - u' (\sigma)\,,
		\qquad%
	\beta =  - u'' (\sigma) \, \tau - v' (\sigma)\,.
\ee
From Eqs.~\eqref{eq:infwt3} and \eqref{eq:infwt4}, we find
\begin{subequations} \label{eq:rsdg}
\begin{align}
	\delta X^\mu & = \bigl( u' \, \tau + v \bigr) \, \p_\tau X^\mu + u \, \p_\sigma X^\mu\,, \label{eq:dxm} \\[4pt]
	\delta \lambda & = \bigl( u' \, \tau + v \bigr) \, \p_\tau \lambda + \bigl( u \, \lambda \bigr)' - 2 \, \bigl( u'' \, \tau + v' \bigr) \, \p_\sigma X^0\,. \label{eq:dlrs}
\end{align}
\end{subequations}
In this flat gauge, the non-vibrating string action~\eqref{eq:mztfscw} becomes
\be \label{eq:fstmztp}
    S^{}_\text{P} = \frac{T}{2} \int d^2 \sigma \, \Bigl( \p^{}_\sigma X^0 \, \p^{}_\sigma X^0 + \p^{}_\tau X^i \p^{}_\tau X^i + \lambda \, \p^{}_\tau X^0 \Bigr)\,,
\ee
We recalled that the target space symmetry is now nonrelativistic, as the non-vibrating string action \eqref{eq:mztfscw} is invariant under the global transformations
\be
	X^0 \rightarrow \Xi^0\,,
		\qquad%
	X^i \rightarrow \Xi^i + \Lambda^i \, X^0\,,
		\qquad%
	\lambda \rightarrow \lambda - 2 \, \Lambda_i \, \p_\tau X^i\,.
\ee
Here, $\Lambda^i$ parametrizes the target space Galilean boost. The target space Lorentz symmetry in conventional string theory is, however, broken now. 

Following the reparametrizations of the target space data in Eq.~\eqref{eq:txomega} and of the worldsheet metric in Eq.~\eqref{eq:wscg0}, the Galilean limit of string theory can now be rephrased as a rescaling of both the worldsheet and spacetime coordinates in the flat case\,\footnote{Note that the Nambu-Goto action is invariant under the rescalings of the worldsheet coordinates $\tau$ and $\sigma$\,.}: 
\be \label{eq:rsc}
    \hat{X}^0 = \omega^{1/2} \, X^0\,,
        \qquad%
    \hat{X}^{i} = \omega^{-1/2} \, X^{i}\,,
        \qquad%
    \hat{\tau} = \omega^{-1} \, \tau\,,
        \qquad%
    \hat{\sigma} = \sigma\,,
\ee
where we have conveniently absorbed the rescaling~\eqref{eq:txomega} of the string tension into the rescalings of the embedding coordinates. This convention with the string tension remaining untouched has the benefit of making it easier to track the target space physics. This is the convention that we will stick to through the rest of this paper. 
Plugging Eq.~\eqref{eq:rsc} into the conventional string sigma model in conformal gauge,
\be \label{eq:hsfs}
    \hat{S}^{}_\text{P} = - \frac{T}{2} \int d^2 \hat{\sigma} \, \p_\alpha \hat{X}^\mu \, \p^\alpha \hat{X}_\mu\,,
\ee
we find
\be
    \hat{S}^{}_\text{P} = \frac{T}{2} \int d^2 \sigma \, \Bigl[ - \omega^2 \, \bigl( \p_\tau X^0 \bigr)^2 + \p^{}_\sigma X^0 \, \p^{}_\sigma X^0 + \p^{}_\tau X^{i} \p^{}_\tau X^{i} + O\bigl(\omega^{-2}\bigr) \Bigr]\,.
\ee
Using the Hubbard-Stratonovich transformation to integrate in an auxiliary field $\lambda$ as in Eq.~\eqref{eq:hsprrw0} followed by taking the infinite $c$ limit, we recover the M0T string action \eqref{eq:fstmztp}. 

\vspace{3mm}

\noindent $\bullet$~\emph{Tropological sigma models.} Intriguingly, if $X^2 = \cdots X^9 = 0$\,, the M0T string action~\eqref{eq:fstmztp} can be identified with the bosonic part of \emph{tropological sigma models} recently proposed in~\cite{Albrychiewicz:2023ngk}. More specifically, identify $T = e^{-1}$\,, $\tau = \theta$\,, $\sigma = r$\,, $\lambda = 2 \, \bigl( \beta - \p_r \Theta \bigr)$\,, $X^0 = X$\,, and $X^1 = \Theta$ in the M0T string action~\eqref{eq:fstmztp}, we find\,\footnote{The more general sigma model in Eq.~(6.6) from~\cite{Albrychiewicz:2023ngk} can be obtained by truncating the M$p$T string action~\eqref{eq:m0tst} that we will introduce later via T-duality, together with a Wick rotation.}
\be \label{eq:tsm}
	S = \frac{1}{e} \int dr \, d\theta \, \Bigl[ \tfrac{1}{2} \bigl( \p_\theta \Theta - \p_r X \bigr)^2 + \beta \, \p_\theta X + \text{total derivative} \Bigr]\,.
\ee
Up to a boundary term, Eq.~\eqref{eq:tsm} is identical to the bosonic part of the tropological sigma model~(3.21) in \cite{Albrychiewicz:2023ngk}. Such tropological sigma model is partly motivated by the study of Gromov-Witten invariants in topological QFTs, as some of these invariants can be computed by performing the ``tropical" limit of the associated geometries. In the target space, the tropological sigma model realizes the geometric structures associated with tropical localization equations, while the tropicalization of the worldsheet metric essentially gives the Carrollian worldsheet that we have discussed around Eq.~\eqref{eq:wscg0}. The tropical geometry is a useful mathematical concept that naturally appears in a range of different fields including computer science, \emph{e.g.}~the \emph{Floyd-Warshall algorithm}, which is a dynamic-programming method that solves the all-pairs shortest-paths problem on a directed graph~\cite{cormen2022introduction}. It is argued in~\cite{Albrychiewicz:2023ngk} that the nonrelativistic worldsheet of the tropological sigma models shares similarities with the worldsheet of nonequilibrium string perturbation theory, \emph{i.e.} the string theoretical version of the Schwinger-Keldysh formalism~\cite{Horava:2020she, Horava:2020val, Horava:2020apz}.

\vspace{3mm}

\noindent $\bullet$~\text{\emph{Galilean conformal algebra.}}
There is a residual gauge symmetry, \emph{i.e.} a reparametrization of the worldsheet coordinates, that leaves the M0T string action~\eqref{eq:fstmztp} invariant. We start with a general reparametrization of the worldsheet coordinates with $\tau \rightarrow T (\tau, \sigma)$ and $\sigma \rightarrow \Sigma (\tau, \sigma)$\,, together with $\lambda \rightarrow \lambda + \Delta (\tau, \sigma)$\,. Requiring that the action~\eqref{eq:fstmztp} be invariant implies
\be \label{eq:condsrg}
	\p_\tau \Sigma = 0\,,
		\qquad%
	\p_\tau T = \p_\sigma \Sigma\,,
		\qquad%
	\Delta = \frac{1}{\p_\sigma \Sigma} \ls \lambda + 2 \, \p_\sigma X^0 \, \frac{\p_\sigma T}{\p_\sigma \Sigma} - \p_\tau X^0 \lr \frac{\p_\sigma T}{\p_\sigma \Sigma} \rr^2 \rs. 
\ee
Therefore, the residual worldsheet diffeomorphisms are
\be \label{eq:}
	\tau \rightarrow T(\tau, \sigma) = \mathcal{S}' (\sigma) \, \tau + \mathcal{T} (\sigma)\,. 
		\qquad%
	\sigma \rightarrow \Sigma(\tau, \sigma) = \mathcal{S} (\sigma)\,,
\ee
for arbitrary functions $\mathcal{S}$ and $\mathcal{T}$ of the worldsheet spatial coordinate $\sigma$\,. Note that these transformations are accompanied with an appropriate reparametrization $\lambda \rightarrow \lambda + \Delta$ of the Lagrange multiplier $\lambda$\,, where $\Delta$ is given in Eq.~\eqref{eq:condsrg}. 

In the infinitesimal case, we write
\be
	\delta \tau = \xi' (\sigma) \, \tau + \zeta (\sigma)\,,
		\qquad%
	\delta \sigma = \xi (\sigma) \,.
\ee
These transformations are generated by the operators
\be
	L = \xi' (\sigma) \, \tau \, \p_\tau + \xi (\sigma) \, \p_\sigma\,,
		\qquad%
	M = \zeta (\sigma) \, \p_\tau\,.
\ee
The Fourier expansions $\xi = \sum_n k_n \, e^{i \, n \, \sigma}$ and $\zeta = \sum_n z_n \, e^{i \, n \, \sigma}$ imply $L = i \sum_n k_n \, L_n$ and $M = i \sum_n z_n \, M_n$, with
\be
	L_n = e^{i \, n \, \sigma} \, \bigl( - n \, \tau \, \p_\tau + i \, \p_\sigma \bigr)\,,
		\qquad%
	M_n = i \, e^{i \, n \, \sigma} \, \p_\tau\,.
\ee
These generators form the two-dimensional worldsheet Galilean conformal algebra~\cite{Bagchi:2009pe, Bagchi:2009my, Bagchi:2013bga},
\be
	\bigl[ L_n\,, L_m \bigr] = \bigl( n - m \bigr) \, L_{n+m}\,,
		\qquad%
	\bigl[ L_n\,, M_m \bigr] = \bigl( n - m \bigr) \, M_{n+m}\,,
		\qquad%
	\bigl[ M_n\,, M_m \bigr] = 0\,,
\ee
which is isomorphic to the Bondi-Metzner-Sachs (BMS) algebra~\cite{Bondi:1962px, Sachs:1962zza} in three dimensions~\cite{Bagchi:2010zz}. 
At the classical level, there is \emph{no} central extension. See~\cite{Bagchi:2009pe} for central extensions of the Galilean conformal algebra. 

\vspace{3mm}

\noindent $\bullet$~\text{\emph{Equations of motion.}} The equations of motion from varying the worldsheet fields in the gauge-fixed action \eqref{eq:fstmztp} are 
\be
    \p^{}_\tau X^0 = \p^{}_\tau \lambda + 2 \, \p_\sigma^2 X^0 = \p_\tau^2 X^{i} = 0\,,
\ee 
which are solved by
\begin{subequations} \label{eq:solm0teom}
\begin{align}
    X^0 \bigl(\tau, \sigma\bigr) & = f^{}_1 \bigl( \sigma \bigr)\,,
        \qquad%
    \lambda \bigl(\tau, \sigma\bigr) = f^{}_2 \bigl( \sigma \bigr) - 2 \, f^{\prime\prime}_1 \bigl( \sigma \bigr) \, \tau\,, \\[4pt]
    X^{i} \bigl(\tau, \sigma\bigr) & = g_1^{i} (\sigma) + g_2^{i} (\sigma) \, \tau\,.
\end{align}
\end{subequations}
Because the embedding coordinates do \emph{not} satisfy any wave function, the M0T string is called \emph{non-vibrating string} in~\cite{Batlle:2016iel}. This non-vibrating feature is ubiquitous for different fundamental strings related to the M0T string via T-duality.   
Under the residual worldsheet gauge transformation of $X^\mu$ in Eq.~\eqref{eq:rsdg}, we find that $f_1$\,, $g_1^{i}$, and $g_2^{i}$ transform as
\be \label{eq:df1g1g2}
	\delta f_1 = u \, f'_1\,,
		\qquad%
	\delta g_1^{i} = u \, \bigl( g_1^{i} \bigr)' + v \, g_2^{i}\,,
		\qquad%
	\delta g_2^{i} = \bigl( u \, g_2^{i} \bigr)'\,.
\ee
Furthermore, the residual gauge transformation \eqref{eq:dlrs} of $\lambda$ implies
\be \label{eq:df1f2}
	\delta f''_1 = \bigl( u \, f'_1 \bigr)''\,,
		\qquad
	\delta f^{}_2 = \bigl( u \, f^{}_2 \bigr)' - 2 \, \bigl( v \, f_1' \bigr)'\,.
\ee
It is reassuring to observe that the transformations $\delta f^{}_1$ in Eq.~\eqref{eq:df1g1g2} and $\delta f''_1$ in Eq.~\eqref{eq:df1f2} match each other, which ensures that $u = u(\sigma)$ is \emph{not} constrained. We can therefore use $u(\sigma)$ to gauge fix $f^{}_1(\sigma)$ to be $\sigma$ and use $v(\sigma)$ to gauge fix $f^{}_2(\sigma)$ to a constant, \emph{i.e.},
\be \label{eq:solsgf}
	X^0 \! = \sigma\,,
		\qquad%
	X^{i} \!\! = x^{i} (\sigma) + T^{-1} P^{i} (\sigma) \, \tau\,,
		\qquad%
	\lambda = 2 \, T^{-1} P^{}_0\,,
\ee 
where $P^{}_0$ is the conjugate momentum of $X^0$ and $P^{}_{i} (\sigma)$ the conjugate momentum of $X^{i}$. The constant $P^{}_0$ represents the constant effective energy of the string. Using the Hamiltonian constraints imposed by $\chi$ and $\rho$\,, respectively, in the phase-space action~\eqref{eq:psf0}, we find
\be \label{eq:hc}
	P^{}_{i} (\sigma) \, P^{}_i (\sigma) = T^2,  
        \qquad%
    P^{}_0 = - P^{}_{i} (\sigma) \, \frac{dx^{i} (\sigma)}{d\sigma}\,.
\ee
The following Virasoro-like constraints arise from varying the worldsheet zweibein fields $e_\alpha{}^a$ in Eq.~\eqref{eq:mztfscw} and then fixing the gauge:
\begin{subequations} \label{eq:vcs}
\begin{align}
    - \bigl( \p_\sigma X^0 \bigr)^2 + \p_\tau X^{i} \, \p_\tau X^{i} & = 0\,,
        &%
    \lambda \, \p_\tau X^0 & = 0\,, \\[4pt]
    \p_\sigma X^0 \, \p_\tau X^0 & = 0\,, 
        &%
    \lambda \, \p_\sigma X^0 + 2 \, \p_\tau X^{i} \, \p_\sigma X^{i} & = 0\,.
\end{align}
\end{subequations}
These equations are consistent (by definition) with the solutions in Eqs.~\eqref{eq:hc} and \eqref{eq:solsgf}.  

\vspace{3mm}

\noindent $\bullet$~\text{\emph{Massless Galilean system and geometric optics.}} We now focus on the string state with a constant collective momentum, \emph{i.e.} we only keep the zero mode in $P_{i} (\sigma)$\,. For this purpose, we treat $P_{i}$ as a constant. As $P_0$ is also a constant, the second Hamiltonian constraint in Eq.~\eqref{eq:hc} implies that
\be \label{eq:xap}
    x^{i} = X^{i}_0 + V^{i} X^0\,,
\ee
where $X^{i}_0$ marks the position of the string and $V^{i}$ is the center-of-mass velocity of the string in the target space. From Eq.~\eqref{eq:hc}, we find the dispersion relation,
\be \label{eq:nvsm}
    P^{}_{i} \, P^{i} = T^2\,,
        \qquad%
    P^{}_0 = - V^{i} P^{}_{i}\,.
\ee
In order to understand the physics of the zero mode derived in Eq.~\eqref{eq:nvsm}, we first consider the spacetime Galilean boost that acts nontrivially on $x^{i}$ and $P^{}_0$ as  
\be \label{eq:gbgf}
    \delta^{}_\text{\scalebox{0.8}{G}} x^{i} = v^{i} X^0\,,
        \qquad%
    \delta^{}_\text{\scalebox{0.8}{G}} P^{}_0 = - v^{i} P^{}_{i}\,.
\ee
The solution~\eqref{eq:hc} to the Hamiltonian constraints are clearly invariant under the Galilean boost~\eqref{eq:gbgf}. 
Eq.~\eqref{eq:xap} implies that the Galilean boost~\eqref{eq:gbgf} acts on $V^{i}$ as $\delta V^{i} = - v^{i}$. Therefore, we can always perform a Galilean boost to go to the ``rest frame" with $V^{i} = 0$\,, such that the dispersion relation~\eqref{eq:nvsm} becomes
\be \label{eq:dsgo}
    \text{\emph{rest frame:}}
        \qquad%
    P^{}_{i} \, P^{}_i = T^2\,,
        \qquad%
    P^{}_0 = 0\,.     
\ee
Since the excitation has zero energy in the rest frame, it behaves like a phonon but with Galilean symmetry. However, the spatial momentum of the excitation always has a fixed length. This is a massless Galilean system, which can be obtained from a Galilean limit of the Souriau tachyon, where the mass is imaginary~\cite{Batlle:2017cfa, souriau1970structure}. 

The seemingly exotic dispersion relation~\eqref{eq:dsgo} of the massless Galilean system in fact describes the ordinary physics of geometric optics~\cite{Duval:2005ry, Duval:2013aza}. Define the optical length from point $p^{}_1$ to point $p^{}_2$ via a spatial path $X^{i} (s)$ to be
\be \label{eq:sfermat}
    S = \int_{p^{}_1}^{p^{}_2} ds \, n \bigl(X^{i}\bigr)\,,
\ee
where $n = n(x^{i})$ is the index of refraction and $ds = \sqrt{dX^{i} \, dX^{i}}$ is the element of arc length. Fermat's principle states that the physical light ray is selected by minimizing the action~\eqref{eq:sfermat}. The conjugate momentum with respect to $X^{i}$ is
\be
    P_{i} = n \, \frac{dX^{i}}{ds}\,,
\ee
which implies the dispersion relation $P_{i} \, P_{i} = n^2$\,. Moreover, the associated Hamiltonian vanishes, \emph{i.e.}~$P^{}_0 = 0$\,. This precisely matches Eq.~\eqref{eq:dsgo} after identifying the index of refraction $n$ with the string tension $T$\,. 

\section{Relation to Matrix Theory} \label{sec:mtnlsw}

In this section, we start with a brief review of the BFSS Matrix theory~\cite{Banks:1996vh} and show that it arises from the same Galilean limit under which the non-vibrating string also arises. 
Both the non-vibrating string and BFSS Matrix theory (at finite $N$) reside in the corner of type IIA superstring theory that corresponds to M-theory compactified over a lightlike circle, \emph{i.e.}~M-theory in the Discrete Light Cone Quantization (DLCQ)~\cite{Susskind:1997cw, Seiberg:1997ad, Sen:1997we}. The BFSS Matrix theory describes the dynamics of the light excitations in DLCQ M-theory, which are the Kaluza-Klein modes on the lightlike compactification. In~\cite{Seiberg:1997ad}, it is argued that the lightlike compactification in DLCQ M-theory can be viewed as an infinite boost limit of a spatial circle. We will see that this infinite boost limit in eleven dimensions corresponds to a BPS limit in ten-dimensional type IIA superstring theory: the background RR one-form in IIA is fined tuned to cancel the associated background D0-brane tension. This corner of IIA is referred to as \emph{Matrix 0-brane Theory} (M0T) in~\cite{udlstmt}, which contains various D$p$-branes and the fundamental string. We will find that the fundamental string in Matrix 0-brane theory is described by the non-vibrating string action that we have studied in Section~\ref{sec:nvst}. However, unlike conventional string theory, the ``fundamental'' degrees of freedom in M0T are the D0-branes, whose dynamics is encoded by the BFSS Matrix theory. 

\subsection{D0-Branes in the BPS Limit} \label{eq:dzbbosl}

For pedagogical reason, before discussing the BFSS Matrix theory, we consider the infinite speed-of-light limit of a single D0-particle in IIA superstring theory, whose bosonic part is  %
\be \label{eq:hsd0}
	\hat{S}^{}_{D0} = - \frac{1}{\sqrt{\hat{\alpha}'}} \, \biggl( \int d\tau \, e^{-\hat{\Phi}} \sqrt{- \p^{}_\tau \hat{X}^\mu_{\phantom{\dagger}} \, \p^{}_\tau {\hat{X}}_{\mu}} - \int \hat{C}^{(1)} \biggr)\,.
\ee
This action essentially describes a charged relativistic particle, with the RR one-form 
$\hat{C}^{(1)}$ 
playing the role of the gauge potential. Here, $\hat{\alpha}' = \bigl( 2 \pi \, \hat{T} \bigr)^{-1}$ is the Regge slope and $\hat{\Phi}$ is the dilaton field. In order for the resulting theory to be finite in the $\omega \rightarrow \infty$ limit, we are required to rescale the string coupling $\langle \exp \hat{\Phi} \rangle$ and $\hat{C}^{(1)}$ in addition to the rescaling of $\hat{\alpha}'$, with
\be \label{eq:hphirs}
	\hat{\Phi} = \Phi - \tfrac{3}{2} \, \ln \omega\,,
		\qquad%
	\hat{C}^{(1)} = \omega^2 \, dX^0\,,
		\qquad%
	\hat{\alpha}' = \omega \, \alpha'\,,
\ee
with the subleading terms in $\omega$ set to zero for simplicity. Here, $\Phi$ and $\alpha'$ will be the dilaton and effective Regge slope, respectively, after sending $\omega$ to infinity. 
Plugging the ansatz~\eqref{eq:txomega} into the D0-brane action~\eqref{eq:hsd0}, in static gauge with $X^0 = \tau$\,, we find the following finite action at infinite $\omega$\,:
\be \label{eq:sdzic}
	S^{}_\text{D0} = \frac{1}{2 \, \sqrt{\alpha'}} \int d\tau \, e^{-\Phi} \, \p^{}_\tau X^i \, \p^{}_\tau X^i\,.
\ee
This resulting D0-brane theory describes a free Galilean particle. The above infinite speed-of-light limit of this particle case has been discussed in~\cite{Gomis:2000bd}.  

Let us collect the prescriptions that we have found. In the same spirit of what we have discussed around Eq.~\eqref{eq:rsc}, we further rescale the embedding coordinates (as in Eq.~\eqref{eq:rsc}) and the RR one-form such that \emph{no} rescaling of the Regge slope is introduced. We therefore rewrite the embedding coordinates, dilaton, and RR one-form as
\be \label{eq:npresxzigc}
	\hat{X}^0 = \omega^{1/2} \, X^0\,,
		\qquad%
	\hat{X}^i = \omega^{-1/2} \, X^i\,,
		\qquad%
	\hat{\Phi} = \Phi - \tfrac{3}{2} \, \ln \omega\,,
		\qquad%
	\hat{C}^{(1)} = \omega^2 \, e^{-\Phi} \, dX^0\,.
\ee
The relevant $\omega$ limit of type IIA superstring theory is closely related to the ones discussed in~\cite{Gopakumar:2000ep, Harmark:2000ff, Gomis:2000bd, Danielsson:2000gi}. In particular, the prescription~\eqref{eq:npresxzigc} is a special case of the general backgrounds considered in~\cite{udlstmt}, where the full-fledged string theory that arises from this $\omega \rightarrow \infty$ limit of IIA is referred to as Matrix 0-brane theory. 

Now, we generalize the single brane result in Eq.~\eqref{eq:sdzic} to a stack of coinciding D0-branes. The nonabelian effective action~\cite{Myers:1999ps} of a stack of $N$ coinciding relativistic D0-branes is
\begin{align} \label{eq:sd0rel}
    	\hat{S}^{}_\text{D0} = - \frac{1}{\sqrt{\alpha'}} \int d \tau \, e^{-\hat{\Phi}} \, \tr \sqrt{\Bigl( - \p_\tau \hat{X}^\mu_{\phantom{\dagger}} \, \p_\tau \hat{X}_\mu \Bigr) \, \det \Bigl( \delta^i_j + 2 \pi i \, \alpha' \bigl[ \hat{X}^i, \, \hat{X}_j \bigr] \Bigr)} + \frac{1}{\sqrt{\alpha'}} \int \hat{C}^{(1)}\,,
\end{align}
where we only included a non-zero RR one-form potential but set all the other RR potentials to zero. Here, $X^i$ are scalars in the adjoint representation of $U(N)$\,. Since we have taken the convention where the Regge slope is not rescaled, the hat in $\hat{\alpha}'$ is {dropped. After} the $\omega \rightarrow \infty$ limit is performed, we take the dilaton field $\Phi$ to be constant and define the string coupling $g^{}_s = e^\Phi$\,. {Choosing the static gauge $X^0 = \tau$, and plugging} Eq.~\eqref{eq:npresxzigc} into the D0-brane action~\eqref{eq:sd0rel}, we find in the infinite $\omega$ limit that
\be \label{eq:sdz}
	S^{}_\text{D0} = \frac{1}{2 \, g^{}_s \, \sqrt{\alpha'}} \int d\tau \, \tr \Bigl[ \p^{}_\tau X^i \, \p^{}_\tau X^i + \bigl(\pi \, \alpha'\bigr)^2 \, \bigl[ X^i\,, \, X^j \bigr]^2 \Bigr]\,.
\ee
Note that, in Eq.~\eqref{eq:sd0rel}, an $\omega^2$ divergence from the square-root term is canceled by the RR one-form potential. The appropriate decoupling limit of IIA is thus more than a na\"{i}ve infinite speed-of-light limit: the RR one-form is also fine tuned to cancel the D0-brane tension. In this sense, the Galilean limit introduced in Section~\ref{sec:nvst} is now refined to be a BPS limit, which preserves half of the supersymmetry for the D0-brane states\,\footnote{In the original Galilean limit, there is \emph{no} central charge. However, upon the coupling to the RR potential, the Galilei algebra now acquires an extension to the Bargmann algebra with a central charge.}. In retrospective, the necessity of including a critical electric potential to cancel the particle mass in the infinite $\omega$ limit is required by supersymmetry. In the next subsection, we will see that the resulting D0-brane action~\eqref{eq:sdz} is the bosonic sector of the BFSS Matrix theory defined by the Hamiltonian~\eqref{eq:hmt}.

\subsection{Review of BFSS Matrix Theory} \label{sec:rbfssmt}

We start with a brief review of BFFS Matrix theory and refer the readers to~\cite{Taylor:2001vb} for further details and references. Historically, in the context of M-theory, Matrix theory arises from the attempt of quantizing the supermembrane~\cite{deWit:1988wri}, which is described by a three-dimensional sigma model that maps the supermembrane to the eleven-dimensional target spacetime~\cite{Bergshoeff:1987cm}. We denote the membrane worldvolume coordinates by $\sigma^\alpha = (\tau, \, \sigma^1, \, \sigma^2)$ and the target space coordinates by $X^\text{M}$\,, $\text{M} = 0\,, 1\,, \cdots\,, 10$\,. In general, this sigma model is not power-counting renormalizable. The Hamiltonian formalism of the membrane simplifies in the light-cone gauge, where the target space light-cone direction $X^- = \bigl( X^0 - X^{10} \bigr) / \sqrt{2}$ is identified with the worldvolume time $\tau$\,. 
However, even in the light-cone gauge, the equations of motion from varying the embedding coordinates $X^\text{M}$ are still nonlinear, which is different from the situation in string theory and makes the membrane theory difficult to solve. This difficulty of quantizing the supermembrane motivates one to discretize the membrane surface by using a ``matrix regularization," where functions on the membrane surface are regularized to be $N \times N$ matrices~\cite{Goldstone:1982, hoppe1987phd}. It was later shown in \cite{deWit:1988wri} that the quantized supermembrane in the matrix regularization is described by the Hamiltonian~\cite{baake1985fierz, flume1985quantum, Claudson:1984th},
\be \label{eq:hmt}
     H = \frac{R}{2} \, \tr \Bigl( P_i \, P_i - \tfrac{1}{2} \, \bigl[ X^i, \, X^j \bigr] \, \bigl[ X^i, \, X^j \bigr] + \psi^\intercal \, \gamma^i \, \bigl[X^i, \, \psi \bigr] \Bigr)\,,
\ee
where 
$X^i$ and its conjugate momentum $P_i$\,, $i = 1\,, \cdots\,, 9$ are $N \times N$ matrices, $\psi$ is a 16-component Matrix-valued spinor of SO$(9)$\,, and $\gamma^i$ are the SO$(9)$ Dirac matrices in the 16-dimensional representation. Moreover, $R = 2\pi \, \ell^3_{11}$\,, with $\ell^{}_{11}$ the Planck length in eleven dimensions.
The Hamiltonian~\eqref{eq:hmt} describes an $\CN = 16$ supersymmetric quantum mechanical theory with matrix degrees of freedom, which is referred to as the Banks-Fischler-Shenker-Susskind (BFSS) Matrix theory~\cite{Banks:1996vh}. However, the BFFS Matrix theory does \emph{not} describe a single first-quantized supermembrane, which is prone to creating long thin spikes at a cost of negligible energy and is thus unstable \cite{deWit:1988xki}. Instead, the action~\eqref{eq:hmt} describes a ``second-quantized" theory with a continuous spectrum. This important multi-particle interpretation was given in~\cite{Banks:1996vh}. Heuristically, this is because the continuous large $N$ limit of the discretized membrane surface in the matrix regularization does not necessarily lead to a single membrane anymore. We have already seen the bosonic sector of the BFSS Matrix theory in Eq.~\eqref{eq:sdz} (up to rescalings and in string unit), which arises from taking the BPS limit of a stack of D0-branes.  

The BFSS Matrix theory is related to the low-energy description of the bound state of $N$ D0-branes in type IIA superstring theory~\cite{Townsend:1995af}. The action~\eqref{eq:hmt} is identical to the dimensional reduction of ten-dimensional SYM theory to (0+1)-dimensions~\cite{Witten:1995im}. The ten-dimensional SYM arises from a zero Regge slope (field theory) limit of a stack of $N$ D9-branes in type IIA superstring theory. On the other hand, this dimensional reduction is essentially equivalent to performing T-duality transformations along all the spatial directions of the stack of D9-branes, which gives rise to a bound state of $N$ D0-branes. Therefore, in the T-dual frame of the zero-Regge slope limit of the D9-branes, the bound state of D0-branes is naturally described by the nonrelativistic quantum mechanical system~\eqref{eq:hmt}. 

Based on the above observations, it is argued in~\cite{Susskind:1997cw, Seiberg:1997ad, Sen:1997we} that the Matrix theory~\eqref{eq:hmt} at finite $N$ describes M-theory in spacetime with a lightlike compactification, where $N$ is the Kaluza-Klein (KK) momentum number in the lightlike circle. Such KK modes in the M-theory circle correspond to the bound D0-brane states in type IIA superstring theory. Compactifying M-theory over a lightlike circle is referred to as the Discrete Light Cone Quantization (DLCQ) of M-theory. In the large $N$ limit, the lightlike circle decompactifies in DLCQ M-theory, which leads to M-theory in the infinite momentum frame. It is therefore conjectured in~\cite{Banks:1996vh} that the large $N$ limit of the BFSS Matrix theory describes M-theory in asymptotically flat spacetime. 

\subsection{Lightlike Compactification of M2-Brane} \label{sec:zblttast}

We have discussed the relation between the BFSS Matrix theory and DLCQ M-theory in the previous subsection. We have also seen earlier in Section~\ref{eq:dzbbosl} that both the BFSS Matrix theory and non-vibrating string arise from the same infinite speed-of-light limit and thus live in Matrix 0-brane theory. It is therefore natural to ask the question: how the non-vibrating string is related to DLCQ M-theory? This requires us to compactify the M2-brane over a lightlike circle\,\footnote{Compactification of the M2-brane over a lightlike circle has been considered in~\cite{Kluson:2021pux}, but the results there are different from what we present here.}. 

We start with reviewing how DLCQ M-theory is defined in~\cite{Seiberg:1997ad}, via an infinite boost limit of a spatial circle. We start with a spatial compactification in $X^{10}$ with proper radius $R_0$\,, \emph{i.e.}
$X^{10} \sim X^{10} + 2 \pi \, R^{}_0$\,,
and then perform a large boost along $X^{10}$ such that
\be
    X^0 \rightarrow \gamma \, \bigl( X^0 + v \, X^{10} \bigr)\,,
        \qquad%
    X^{10} \rightarrow \gamma \, \bigl( X^{10} + v \, X^0 \bigr)\,.
\ee
Here, $v$ is the boost velocity and $\gamma = 1 / \sqrt{1 - v^2}$ is the Lorentz factor. At large $\gamma$\,, we define the ``almost'' lightlike directions,
\be \label{eq:amllc}
    X^+ = X^0 + X^{10}\,, 
        \qquad%
    X^- = X^0 - v \, X^{10}\,.
\ee
Here, $X^-$ becomes lightlike when the boost velocity $v \rightarrow 1$\,, \emph{i.e.}~in the infinite boost limit with $\gamma \rightarrow \infty$\,. It follows that
\be
    X^+ \rightarrow 2 \, \gamma \, X^+ + O\bigl(\gamma^{-1}\bigr)\,,
        \qquad%
    X^- \rightarrow \frac{X^0}{\gamma} + O\bigl(\gamma^{-2}\bigr)\,.
\ee
In the double scaling limit $\gamma \rightarrow \infty$ and $R_0 \rightarrow 0$\,, while keeping $R = 2 \, \gamma \, R^{}_0$ fixed, both $X^\pm$ become lightlike and $X^+$ picks up a periodicity, with 
\be
    X^+ \sim X^+ + 2 \pi \, R\,.
\ee
The other lightlike direction $X^-$ remains noncompact. Note that this infinite Lorentz boost that we have performed is more than just a change of frame, as compactifying the spatial $X^{10}$ direction over a circle already breaks Lorentz symmetry. In this sense, this ``infinite boost'' limit should be viewed as a decoupling limit: from the ten-dimensional perspective, this is the BPS limit of type IIA superstring theory.  

In the DLCQ of string theory, there also exists an alternative definition of the lightlike circle via T-duality. The lightlike circle in DLCQ string theory is T-dual to a spatial circle in the so-called ``nonrelativistic string theory,'' which is unitary and UV-complete~\cite{Gomis:2000bd, Oling:2022fft}. Nonrelativistic string theory arises from a BPS limit of conventional string theory that zooms in on the fundamental string, where the $B$-field becomes critical and cancels the string tension. One may therefore in turn use nonrelativistic string theory to define DLCQ string theory~\cite{Bergshoeff:2018vfn}. See Section~\ref{sec:sdnrst} for further discussions on nonrelativistic string theory. 

\subsubsection{Double dimensional reduction}

We now make the connection between DLCQ M-theory and the non-vibrating string. We start with a single M2-brane in eleven-dimensional M-theory. We have defined the worldvolume coordinates on the M2-brane manifold to be $\sigma^\text{m}$\,, $\text{m} = 0\,, 1\,, 2$\,, and the embedding coordinates that map the membrane worldvolume to the target space to be $X^\text{M}$\,, $\text{M} = 0\,, 1\,, \cdots\,, 10$\,. 
The classical dynamics of the M2-brane is described by the three-dimensional worldvolume action in~\cite{Bergshoeff:1987cm}, whose bosonic sector is given by
\be \label{eq:sm2}
    S^{}_\text{M2} = - T^{}_\text{M2} \int d^3 \sigma \, \sqrt{-\det \Bigl( \p^{}_\text{m} X^\text{M}_{\phantom{\dagger}} \, \p^{}_\text{n} X_\text{M} \Bigr)}\,.
\ee 
The non-vibrating string action arises from wrapping the M2-brane over the lightlike circle $X^+ = X^0 + X^{10}$ in DLCQ M-theory. In practice, we take $X^+ = \sigma^2$\,. This procedure is referred to as the \emph{double dimensional reduction}, as we dimensionally reduce simultaneously the brane direction on the worldvoume and an ambient target space direction. The resulting fundamental string action is
\be \label{eq:nvstd}
    S^{}_\text{F1} = - T \int d^2 \sigma \, \sqrt{-\det 
    \begin{pmatrix}
        0 &\,\, \p^{}_\beta \tilde{X}^0 \\[4pt]
        \p^{}_\alpha \tilde{X}^0 &\,\, \p^{}_\alpha X^i \, \p^{}_\beta X^i
    \end{pmatrix}}\,,
        \qquad%
    \alpha = 0\,, \, 1\,,
\ee 
where $\tilde{X}^0$ is proportional to the other lightlike direction $X^-$\,. 
Upon identifying $\tilde{X}^0$ with the time direction in the ten-dimensional target space, we find that the action~\eqref{eq:nvstd} is identical to the non-vibrating string action~\eqref{eq:snvs}. 

Next, we discuss how the fundamental string action~\eqref{eq:nvstd} arises from the infinite boost limit in M-theory. We start with the M2-brane action~\eqref{eq:sm2} in the almost lightlike coordinates defined in Eq.~\eqref{eq:amllc}, which takes the following form:
\be \label{eq:sm2nc}
	S^{}_\text{M2} = - T^{}_\text{M2} \int d^3 \sigma \sqrt{- \det \Bigl( \tfrac{1}{\omega^2} \, \p_\text{m} X^+_{\phantom{\dagger}} \p_\text{n} X^+_{\phantom{\dagger}} + 2 \, \p_\text{(m} X^+_{\phantom{\dagger}} \p_\text{n)} \tilde{X}^0_{\phantom{\dagger}} + \p_\text{m} X^i_{\phantom{\dagger}} \, \p_\text{n} X^i_{\phantom{\dagger}} \Bigr)},
\ee 
where the symmetrization of the indices is defined via ${T}_\text{(mn)} = \frac{1}{2} \bigl( {T}_\text{mn} + {T}_\text{nm} \bigr)$ and
\begin{subequations}
\begin{align}
	\frac{1}{\omega^2} & = 2 \gamma \, \Bigl( \gamma - \sqrt{\gamma^2 - 1} \Bigr) - 1 = \frac{1}{4 \, \gamma^2} + O \bigl( \gamma^{-4} \bigr)\,, \label{eq:omegagamma} \\[4pt]
	\tilde{X}^0 & = \gamma \, \Bigl( \sqrt{\gamma^2 - 1} - \gamma \Bigr) \, X^- = - \frac{1}{2} \, X^- + O \bigl( \gamma^{-2} \bigr)\,.
\end{align}
\end{subequations}
We then perform the lightlike compactification and wrap the M2-brane around the $X^+$ direction by setting $X^+ = \sigma^2$\,. The M2-brane action~\eqref{eq:sm2nc} gives rise to the F1-string action
\be \label{eq:f1sg}
    S^{}_\text{F1} = - T \int d^2 \sigma \, \sqrt{- \det
    \begin{pmatrix}
        \omega^{-2} &\,\, \p_\beta \tilde{X}^0 \\[4pt]
        \p_\alpha \tilde{X}^0 &\,\, \p_\alpha X^i \, \p_\beta X^i
    \end{pmatrix}}\,,
        \qquad%
    \alpha\,, \, \beta = 0\,, \, 1\,.
\ee
In the infinite-boost limit $\gamma \rightarrow \infty$\,, we have $\omega \rightarrow \infty$ and $\tilde{X}^0 \rightarrow - X^- / 2$ kept finite, and this action reduces to the non-vibrating string action~\eqref{eq:nvstd}. 
At finite $\gamma$\,, the exotic expression~\eqref{eq:f1sg} is equivalent to the familiar form of the Nambu-Goto string action below, reparametrized as in Eq.~\eqref{eq:npresxzigc}:
\be
    S^{}_\text{F1} = - T \int d^2 \sigma \, \sqrt{-\det \Bigl( - \omega \, \p_\alpha \tilde{X}^0 \, \p_\beta \tilde{X}^0 + \tfrac{1}{\omega} \, \p_\alpha X^i \, \p_\beta X^i \Bigr)}\,,
\ee
This is the Galilean reparametrization of the Nambu-Goto string that we have introduced in Section~\ref{sec:nga}, and the $\omega \rightarrow \infty$ limit corresponds to the infinite speed-of-light limit. Moreover, Eq.~\eqref{eq:omegagamma} implies that $\omega = 2 \, \gamma + O(\gamma^{-1})$\,. Therefore, the parameter $\omega$ used to define the decoupling limit in IIA corresponds to the Lorentz factor $\gamma$ associated with the large boost along $X^{10}$ in M-theory. 

\subsubsection{Direct dimensional reduction}

In contrast to the double dimensional reduction, it is also possible to consider a direct dimensional reduction of the M2-brane, which gives rise to the D2-brane in type IIA superstring theory~\cite{Townsend:1995af, Townsend:1996xj}. In the direct dimensional reduction, we place the M2-brane orthogonally to the compact circle $X^+$\,. Gauging the isometry along $X^+$ in the M2-brane action~\eqref{eq:sm2nc}, we rewrite it equivalently as
\begin{align} \label{eq:sm2nc2}
\begin{split}
	S'_\text{M2} = & - T^{}_\text{M2} \int d^3 \sigma \sqrt{-\det \Bigl( \omega^{-2} \, D_\text{m} X^+ D_\text{n} X^+ + 2 \, D_\text{(m} X^+_{\phantom{\dagger}} \p_\text{n)} \tilde{X}^0 + \p_\text{m} X^i \, \p_\text{n} X^i \Bigr)} \\[4pt]
	& + T^{}_\text{M2} \int d^3 \sigma \, A \wedge d v\,,
\end{split}
\end{align}
where $D^{}_\text{m} X^+ = \p^{}_\text{m} X^+ + v^{}_\text{m}$ and $v = v^{}_\text{m} \, d\sigma^\text{m}$\,. The action~$S'_\text{M2}$ recovers the diffeomorphism invariance $\delta X^+ = \xi$\,, which has to be supplemented with the gauge transformation $\delta v = - d\xi$\,. Integrating out the worldvolume one-form $A$ in $S'_\text{M2}$ imposes that $v$ is exact, \emph{i.e.}~$dv = 0$\,, which implies that $v$ is pure gauge. Fixing the gauge such that $v = 0$\,, $S'_\text{M2}$ reduces to the original M2-brane action~\eqref{eq:sm2nc}. Instead, integrating out $v$ in $S'_\text{M2}$ gives rise to the following D2-brane action in the ten-dimensional target space:
\begin{align} \label{eq:hdt}
\begin{split}
	\hat{S}^{}_\text{D2} = & - T_\text{D2} \int d^3 \sigma \, \omega^{3/2} \, e^{-\Phi} \sqrt{- \det \Bigl( - \omega \, \p_\text{m} X^0 \, \p_\text{n} X^0 + \omega^{-1} \, \p_\text{m} X^i \, \p_\text{n} X^i \Bigr)} \\[4pt]
	& + \frac{T_\text{D2}}{2} \int d^3 \sigma \, \epsilon^\text{mnk} \, \Bigl( \omega^2 \, e^{-\Phi} \, \p_\text{m} X^0 \Bigr) \, F_\text{nk}\,,
\end{split}
\end{align}
where $F = dA$ is the $U(1)$ gauge field strength and we have recovered the dependence on the dilaton $\hat{\Phi}$ that we ignored before. Here, the Levi-Civita $\epsilon^{\alpha\beta\gamma}$ is defined via $\epsilon^{012} = 1$\,. Note that the eleventh dimension $X^+$ in M-theory is now traded with the $U(1)$ gauge field $A_\alpha$ on the D2-brane. Comparing with the conventional D2-brane action in IIA, 
\begin{align}
	\hat{S}^{}_\text{D2} = - T_\text{D2} \int d^3 \sigma \, e^{-\hat{\Phi}} \sqrt{- \det \Bigl( \p^{}_\text{m} \hat{X}^\mu_{\phantom{\dagger}} \, \p^{}_\text{n} \hat{X}^{}_\mu + F^{}_\text{mn} \Bigr)} 
	+ T_\text{D2} \int  \hat{C}^{(1)} \wedge F\,,
\end{align} 
we rediscover the full set of prescriptions in Eq.~\eqref{eq:npresxzigc}. 

In the $\omega \rightarrow \infty$ limit of IIA, we are led to Matrix 0-brane theory, which contains various extended objects including the fundamental string and D$p$-branes. We have shown that this fundamental string is the non-vibrating string, while the D0-brane dynamics is described by the BFSS Matrix theory. The D2-brane in Matrix 0-brane theory arises from the $\omega \rightarrow \infty$ limit of the action~\eqref{eq:hdt}, which results in a finite action,
\begin{align} \label{eq:ncdt}
    S^{}_\text{D2} & = - \frac{T^{}_\text{D2}}{2} \int d^3 \sigma \, \frac{\det \!
	\begin{pmatrix}
		0 &\,\, \p_\text{n} X^0 \\[4pt]
		\p^{}_\text{m} X^0 &\,\, \p^{}_\text{m} X^i \, \p_\text{n} X^i
	\end{pmatrix}
    	+ \bigl( \star \CF \bigr)^{\!\text{m}} \, \p^{}_\text{m} X^i \, \p^{}_\text{n} X^i \, \bigl( \star \CF \bigr)^\text{n}}{e^{\Phi}  \, \bigl( \star \CF \bigr)^{\!\text{k}} \, \p^{}_\text{k} X^0}\,, 
\end{align}
where we have recovered the dependence on the $B$-field in $\CF = F + B$ and defined the Hodge dual $\bigl( \star \CF \bigr)^\text{m} = \frac{1}{2} \, \epsilon^\text{mnk}_{\phantom{\dagger}} \CF^{}_\text{nk}$\,. This exotic-looking D2-brane theory turns out to be T-dual to four-dimensional \emph{Non-Commutative Yang-Mills} (NCYM)~\cite{Gopakumar:2000na, Gopakumar:2000ep}. In fact, the gauge theory on the a stack of $N$ D2-branes in the infinite $\omega$ limit is three-dimensional NCYM, where the two spatial directions on the D2-brane do \emph{not} commute with each other. We demonstrate this below for the single D2-brane described by Eq.~\eqref{eq:ncdt}. In static gauge with $X^\text{m} = \sigma^\text{m}$\,, we have
\be \label{eq:b12f12}
	\bigl( \star \CF \bigr)^{\!\text{k}}_{\phantom{\dagger}} \, \p^{}_\text{k} X^0 = B^{}_{12} + F_{12}\,. 
\ee
Treating the gauge strength $F_{12}$ as a small fluctuation, we have to require that $B_{12}$ do not vanish for the D2-brane action to be well defined. This magnetic $B$-field acts as the source of the spatial noncommutativity on the D2-brane. For simplicity, we set the other components in the $B$-field to zero. 

The noncommutative behavior can be made manifest by transforming the closed string data to the effective background fields seen by the open strings. This is done by using by applying the Seiberg-Witten map~\cite{Seiberg:1999vs}. In particular, the open-string background field measuring the noncommutativity between the D2-brane worldvolume coordinates in type IIA superstring theory is $[\sigma^\text{m}, \sigma^\text{n}] \sim \hat{\Theta}^\text{mn}$\,, where
\be \label{eq:nct0}
    \hat{\Theta}^\text{mn} = - \Bigl[ \bigl( \hat{G} + \hat{B} \bigr)^{-1} \hat{B} \, \bigl( \hat{G} + \hat{B} \bigr)^{-1} \Bigr]^\text{mn}.
\ee
Here, $\hat{G}^{}_\text{mn}$ and $\hat{B}^{}_\text{mn}$ are the pullbacks of the target space metric and $B$-field in the target space to the three-dimensional worldvolume. Note that the above expression is only valid before the $\omega \rightarrow \infty$ limit is performed, which implies that we have to refer back to the limiting prescription that we have derived in Eq.~\eqref{eq:npresxzigc}. Recasting the reparametrization of the embedding coordinates in Eq.~\eqref{eq:npresxzigc} as a reparametrization of the background fields, we write the line element in the target space as
\be
	ds^2 = d \hat{X}^\mu \, d \hat{X}^{}_\mu = \hat{G}^{}_{\mu\nu} \, d X^\mu \, d X^\nu,
		\qquad%
	\hat{G}^{}_{\mu\nu} = - \omega \, \delta_\mu^0 \, \delta_\nu^0 + \omega^{-1} \, \delta_\mu^i \, \delta_\nu^i\,.
\ee	
Then, in the static gauge $X^\text{m} = \sigma^\text{m}$\,, we find that the pullback of the metric on the D2-brane worldvolume is
\be
    \hat{G}^{}_\text{mn} = 
    \begin{pmatrix}
        -\omega &\,\,\,\, 0 &\,\, 0 \\[4pt]
        0 &\,\,\,\, \omega^{-1} &\,\, 0 \\[4pt]
        0 &\,\,\,\, 0 &\,\, \omega^{-1}
    \end{pmatrix}.
\ee
Also note that, as shown in the reparametrization~\eqref{eq:npresxzigc}, $\hat{B}$ does not contain any divergence at large $\omega$\,. From the discussion around Eq.~\eqref{eq:b12f12}, we learned that the minimal configuration for the background $B$-field is
\be
	\hat{B}^{}_{\text{mn}} = 
	\begin{pmatrix}
        0 &\,\,\,\, 0 &\,\, 0 \\[4pt]
        0 &\,\,\,\, 0 &\,\, B^{}_{12} \\[4pt]
        0 &\,\,\,\, - B^{}_{12} &\,\, 0
    \end{pmatrix}.
\ee
The $ \omega \rightarrow \infty$ limit of the noncommutativity in Eq.~\eqref{eq:nct0} gives $\hat{\Theta}^{\text{mn}} \rightarrow \Theta^{\text{mn}}$\,, with
\be
    \Theta^{\text{mn}} = 
    B_{12}^{-1}
    \begin{pmatrix}
        \,0 &\,\,\, 0 &\,\,\, 0\, \\[2pt]
        \,0 &\,\,\, 0 &\,\,\, 1\, \\[2pt]
        \,0 &\,\,\, -1 &\,\,\, 0\,
    \end{pmatrix},
\ee
which implies $[\sigma^1,\, \sigma^2] \sim B_{12}^{-1}$\,, \emph{i.e.}, the inverse of the magnetic $B$-field $B_{12}$ controls the noncommutativity between the spatial directions on the D2-brane. Together with the $\omega$-reparmetrization of $\hat{\Phi}$ from Eq.~\eqref{eq:npresxzigc}, we find that the open string coupling $\hat{g}_\text{o}$ in~\cite{Seiberg:1999vs} becomes
\be
    \hat{g}^2_\text{o} = e^{\hat{\Phi}} \, \sqrt{\frac{\det \bigl( \hat{G}_{\alpha\beta} + \hat{B}_{\alpha\beta} \bigr)}{\det \hat{G}_{\gamma\delta}}} = \omega^{-1/2}  \, g_\text{o}^2 + O(\omega^{-5/2})\,,
        \qquad%
    g_\text{o}^2 = e^\Phi \, |B_{12}|\,,
\ee
which shows that, in the $\omega \rightarrow \infty$ limit, the effective gauge coupling $g_\text{o}$ in NCYM is controlled by the vacuum expectation value $e^{\Phi/2} \sqrt{|B_{12}|}$\,.

Using the fundamental string and D-branes as probes, we have now demonstrated that the BPS limit discussed in Section~\ref{eq:dzbbosl} of type IIA superstring theory corresponds to the infinite boost limit along a spatial compactification in M-theory. It also follows that Matrix 0-brane theory arises from the null compactification of M-theory.

\section{Spacelike T-Duality} \label{sec:fdlcq}

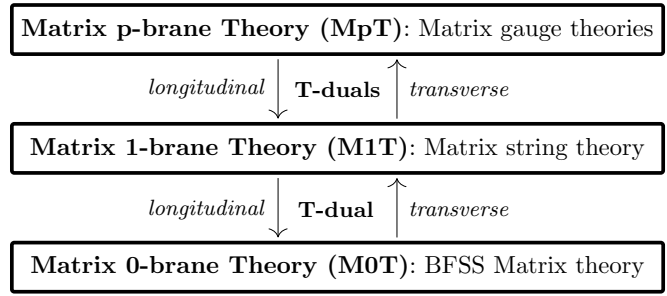
\begin{figure}[b!]
\centering
\begin{adjustbox}{width=.52\textwidth}
\hspace{-8mm}
\begin{tikzpicture}
	\begin{scope}[scale=.85]

        	\path[every node/.style=draw, rounded corners=1, line width=1.5pt, minimum width=265pt, minimum height=20pt, font = \small]   
                (11,1.5) node {\textbf{Matrix 0-brane Theory~(M0T)}: BFSS Matrix theory}
                (11,3.5) node {\textbf{Matrix 1-brane Theory~(M1T)}: Matrix string theory}
                (11,5.5) node {\textbf{Matrix p-brane Theory (MpT)}: Matrix gauge theories}
                ;

        	\path  
                (11,4.5) [font=\footnotesize] node {\textbf{T-duals}}
                (10.97,2.47) [font=\footnotesize] node {\textbf{T-dual}}
                (8.8,2.47) [font=\footnotesize] node {\emph{longitudinal}}
                (13.05,2.47) [font=\footnotesize] node {\emph{transverse}}
                (8.8,4.5) [font=\footnotesize] node {\emph{longitudinal}}
                (13.05,4.5) [font=\footnotesize] node {\emph{transverse}}
                ;

        \draw [{[length=1.3mm]<}-{[length=1.3mm]}] (10,4) -- (10,5);
        \draw [{[length=1.3mm]<}-{[length=1.3mm]}] (10,2) -- (10,3);
        \draw [{[length=1.3mm]}-{[length=1.3mm]>}] (12,4) -- (12,5);
        \draw [{[length=1.3mm]}-{[length=1.3mm]>}] (12,2) -- (12,3);

\end{scope}
\end{tikzpicture}
\end{adjustbox}
\caption{T-duality relations between different Matrix $p$-brane Theories (M$p$Ts) and the associated Matrix (gauge) theories on the critical D$p$-branes. In M$p$T, the RR ($p+1$)-form is taken to be critical. The spacetime geometry in M$p$T develops a ($p$\,+1)-dimensional sector longitudinal to the critical background D$p$-brane, and a (9-$p$)-dimensional sector transverse to the same background brane. These sectors are related via a $p$-brane Galilean boost.}
\label{fig:tdmpt}
\end{figure}

In this section, we consider T-duality transformations of the non-vibrating string action~\eqref{eq:fstmztp} in Matrix 0-brane Theory (M0T). We have shown that M0T arises from the critical RR one-form limit of type IIA superstring theory. In practice, this limit is defined by sending $\omega$ to infinity in IIA reparametrized as in Eq.~\eqref{eq:npresxzigc}. In this decoupling limit, all light excitations except the D0-branes described by the BFSS Matrix theory are decoupled. Studying the T-duality transformations of the non-vibrating string theory will allow us to probe a zoo of T-dual decoupling limits of type II string theories, all connected to DLCQ M-theory. In this section, we focus on the spacelike T-duality transformations of M0T, which maps M0T to \emph{Matrix $p$-brane Theory} (M$p$T). The light excitations in M$p$T are the D$p$-branes, which are described by the associated Matrix gauge theory~\cite{udlstmt}. A summary of the results in this section is given in Figure~\ref{fig:tdmpt}. 

\subsection{Strings in Matrix \texorpdfstring{$p$}{p}-Brane Theory} \label{sec:sfspt0}

In conformal gauge, we transcribe the Polyakov formulation of the M0T string action~\eqref{eq:m0tspf0} as follows:
\be \label{eq:fstmztpmzt}
    S^{}_\text{M0T} = \frac{T}{2} \int d^2 \sigma \, \Bigl( \p^{}_\sigma X^0 \, \p^{}_\sigma X^0 + \p^{}_\tau X^{i} \p^{}_\tau X^{i} + \lambda \, \p^{}_\tau X^0 \Bigr)\,.
\ee
We compactify the spatial directions $X^u$, $u = 1\,, \, \cdots\,, \, p$ over individual circles, and then perform a T-duality transformation along each of these circles. This can be done by gauging the isometries $X^u$ and then rewriting Eq.~\eqref{eq:fstmztpmzt} equivalently as~\cite{Rocek:1991ps, Alvarez:1994dn}
\begin{align} \label{eq:pamzt}
\begin{split}
    S^{}_\text{gauged} = \frac{T}{2} \int d^2 \sigma \, \Bigl( \p^{}_\sigma X^0 \, \p^{}_\sigma X^0 & + \p^{}_\tau X^{A'} \, \p^{}_\tau X^{A'} + \lambda \, \p^{}_\tau X^0 \\[4pt]
    & +  D^{}_\tau X^u \, D^{}_\tau X^u - 2 \, \epsilon^{\alpha\beta} \, \tilde{X}^u \, \p_\alpha v^u_\beta \Bigr)\,.
\end{split}
\end{align}
where $A' = p+1\,, \cdots\,, 9$ and $D_\alpha X^u = \p_\alpha X^u + v^u_\alpha$\,. This action preserves the U$(1)$ gauge symmetry 
$\delta X^u = \xi^u$ and $\delta v^u_\alpha = - \p_\alpha \xi^u$\,.
The to-be T-dual coordinates $\tilde{X}^u$ are Lagrange multipliers imposing that $v^u_\alpha$ is pure gauge. Integrating out $\tilde{X}^u$ sets $\epsilon^{\alpha\beta} \, \p_\alpha v^u_\beta = 0$\,, which can be solved locally by $v^u_\alpha = \p_\alpha \Phi^u$ and it implies $D_\alpha X^u  = \p_\alpha (X^u + \Phi^u)$\,. After absorbing $\Phi^u$ into the definition of $X^u$, we recover the original action~\eqref{eq:fstmztpmzt}. 
Instead of integrating out $\tilde{X}^u$\,, integrating out $v^u_\tau$ in the gauged action~\eqref{eq:pamzt} gives the dual theory,
\begin{align}
\begin{split}
    \tilde{S}^{}_\text{P} & = - \frac{T}{2} \int d^2 \sigma \, \biggl[ \Bigl( - \p_\sigma X^0 \, \p_\sigma X^0 + \p_\sigma \tilde{X}^u \, \p_\sigma \tilde{X}_u \Bigr) - \p^{}_\tau X^{A'} \, \p^{}_\tau X_{A'} \\
    & \hspace{3.42cm} - \Bigl( \lambda \, \p_\tau X^0 + 2 \, v^u_\sigma \, \p_\tau \tilde{X}_u \Bigr) - 2 \, \p_\tau X^u \, \p_\sigma \tilde{X}_u \biggr]\,,
\end{split}
\end{align}
where $v_\sigma^u$ stays as a Lagrange multiplier imposing the constraint $\p_\tau \tilde{X}_u = 0$\,. 
Define $\lambda_0 = \lambda$ and $\lambda^u = 2 \, v^u_{\sigma} + 2 \, \p^{}_\sigma X^u$, we find the dual action
\begin{align} \label{eq:m1tst0}
    \tilde{S} & = \frac{T}{2} \int d^2 \sigma \, \Bigl( \p^{}_\sigma X^0 \, \p^{}_\sigma X^0 - \p^{}_\sigma \tilde{X}^u \, \p^{}_\sigma \tilde{X}^u + \p^{}_\tau X^{A'} \p^{}_\tau X^{A'} 
    + \lambda^{}_0 \, \p^{}_\tau X^0 + \lambda^{}_u \, \p^{}_\tau \tilde{X}^u \biggr)\,,
\end{align}
where we have ignored the topological term $\int dX^u \wedge d\tilde{X}^u$ that encodes the winding Wilson lines~\cite{Alvarez:1996up}. Throughout the rest of the paper, we will always ignore this topological term for simplicity. 
Drop the tilde in $\tilde{X}^u$ and define $A = (0\,, \, u)$\,, we write the dual action as
\begin{align} \label{eq:m0tst}
    \tilde{S} & = \frac{T}{2} \int d^2 \sigma \, \Bigl( - \p^{}_\sigma X^A \, \p^{}_\sigma X_A + \p^{}_\tau X^{A'} \p^{}_\tau X^{A'} 
    + \lambda^{}_A \, \p^{}_\tau X^A \biggr)\,, 
\end{align}
Here, $A = 0\,, \, \cdots, \, p$ and $A' = p+1\,, \, \cdots, \, 9$\,. 
The target space of the dual action~\eqref{eq:m0tst} adopts a codimension-$(p+1)$ foliated structure and, upon decompactifying all the circles, admits a spacetime $p$-brane Galilean boost,
\be
	\delta^{}_{\text{G}} X^A = 0\,,
		\qquad%
	\delta^{}_{\text{G}} X^{A'} = \Lambda_A{}^{A'} \, X^A,
		\qquad%
	\delta^{}_{\text{G}} \lambda_A = - 2 \, \Lambda_{AA'} \, \p_\tau X^{A'},
\ee
which naturally generalizes the usual Galilean boost in the particle case. 

The worldsheet action~\eqref{eq:m0tst} describes the fundamental string coupled to the target space geometry with different foliation structures for different $p$'s. Therefore, T-dualizing along spatial circles in M0T leads to other corners, which arise from distinct decoupling limits of type II superstring theories. Following~\cite{udlstmt}, we refer to the corner that is T-dual to M0T compactified over a $p$-torus as \emph{Matrix $p$-brane Theory} (M$p$T), where the light degrees of freedom are the wrapped D$p$-branes described by Matrix gauge theories. We will comment more on the relation to Matrix gauge theories in Section~\ref{eq:rmgtsltd}. We therefore refer to the fundamental string described by the action~\eqref{eq:m0tst} as the M$p$T string. 

\subsection{Nambu-Goto Action and Decoupling Limit}  \label{sec:ngadl}

In Section~\ref{sec:sfspt0}, we have derived the M$p$T string sigma model~\eqref{eq:m0tst} in the Polyakov action using T-duality transformations. Undoing the conformal gauge, this Polyakov action becomes
\begin{align} \label{eq:hspmpt20}
	S^{}_\text{M$p$T} = & \frac{T}{2} \int d^2 \sigma \, e \, \Bigl( - e^\alpha{}^{}_1 \, e^\beta{}^{}_1 \, \p_{\alpha} X^A \, \p_{\beta} X^{}_A + e^\alpha{}^{}_0 \, e^\beta{}^{}_0 \, \p^{}_\alpha X^{A'} \p^{}_\beta X^{A'} + \lambda^{}_A \, e^\alpha{}^{}_0 \, \p^{}_\alpha X^A \Bigr)\,.
\end{align}
The string worldsheet is nonrelativistic, with the topology that we have discussed in Section~\ref{sec:wstnrs}. One natural question is whether one could write down the analog of the Nambu-Goto M0T string action~\eqref{eq:nvstd} for the M$p$T string, which is insensitive to the worldsheet geometry. Moreover, being able to reproduce both the Polyakov and Nambu-Goto action from the same decoupling limiting will provide us with an important self-consistency check.  

In order to obtain the Nambu-Goto formulation, we integrate out the Lagrange multiplier $\lambda_A$ in the M$p$T string action~\eqref{eq:hspmpt20}, which gives rise to the constraint $e^\alpha{}^{}_0 \, \p^{}_\alpha X^A = 0$\,. This constraint is solved by
\be \label{eq:solet}
	e^\alpha{}^{}_0 = \Gamma \, \epsilon^{\alpha\beta} \, \p^{}_\beta X^0\,,
		\qquad%
	\p_\alpha X^A = \Gamma^A \, \p_\alpha X^0\,.
\ee
Note that $\Gamma^0 = 1$ and the second equation in Eq.~\eqref{eq:solet} can be regarded as the general solution to the constraint equation,
\be \label{eq:exab}
	\epsilon^{\alpha\beta} \, \p_\alpha X^A \, \p_\beta X^B = 0\,.
\ee
Plugging Eq.~\eqref{eq:solet} back into the Polyakov action~\eqref{eq:hspmpt20} yields
\be \label{eq:mptng}
	S^{}_\text{M$p$T} = - \frac{T}{2} \int d^2 \sigma \ls \chi \, \Gamma^A \, \Gamma_A + \frac{1}{\chi} \det \! 
	\begin{pmatrix}
		0 &\,\, \p^{}_\beta X^0 \\[4pt]
		\p^{}_\alpha X^0 &\,\, \p^{}_{\alpha} X^{A'} \p^{}_{\beta} X^{A'}
	\end{pmatrix} 
	\rs,
\ee
where $\chi = \Gamma^{-1} \, e^\alpha{}^{}_1 \, \p^{}_\alpha X^0$\,. 
Finally, integrating out $\chi$ gives rise to the Nambu-Goto-like action:
\be \label{eq:ngrwmpt}
	S^{}_\text{NG} = - T \int d^2 \sigma \, \sqrt{\Gamma^A \, \Gamma_A \det \!
	\begin{pmatrix}
		0 &\,\, \p^{}_\beta X^0 \\[4pt]
		\p^{}_\alpha X^0 &\,\, \p^{}_{\alpha} X^{A'} \p^{}_{\beta} X^{A'}
	\end{pmatrix}}\,.
\ee
This result is not completely satisfying: the action is not directly a functional of the embedding coordinates; instead, it depends on $\Gamma^A$ that comes from solving a constraint equation~\eqref{eq:exab}. Although the combination $\Gamma^A \, \Gamma^{}_A$ in Eq.~\eqref{eq:ngrwmpt} can be replaced with
\be
	\Gamma^A \Gamma^{}_A = \frac{\p_\alpha X^A \, \p^\alpha X^{}_A}{\p_\beta X^0 \, \p^\beta X^0}\,,
\ee
the action is not manifestly invariant under the Lorentz boost transformation $\delta X^A = \Lambda^A{}_B X^B$ within the longitudinal sector with the index $A$. Moreover, the action still has to be supplemented with the constraint~\eqref{eq:exab}. Below, via a limiting procedure, we will find a more self-contained Nambu-Goto action, where the action is manifestly invariant under the longitudinal Lorentz boosts and the constraint~\eqref{eq:exab} is imposed by a Lagrange multiplier. 

The limit of type II superstring theory that leads to M$p$T has been given in~\cite{udlstmt}. This limit is also closely related to the ones that lead to open D$p$-brane theory in~\cite{Gopakumar:2000ep} and the Galilean D$p$-brane in~\cite{Gomis:2000bd}. When the fundamental string is concerned, we simply need to T-dualize the reparametrizations of the embedding and worldsheet coordinates in Eq.~\eqref{eq:npresxzigc}. T-duality acts trivially on the worldsheet coordinates, but it reverts the scaling of the T-dualized embedding coordinates. The dual reparametrization of the embedding coordinates is given by~\cite{udlstmt}
\be \label{eq:mptsps}
	\hat{X}^A = \omega^{1/2} \, X^A\,,
		\qquad%
	\hat{X}^{A'} = \omega^{-1/2} \, X^{A'}.
\ee
Together with the Carrollian reparametrization~\eqref{eq:wscg0} of the worldsheet metric,  
\be \label{eq:repahhab}
	\hat{h}^{}_{\alpha\beta} = - \omega^{-2} \, e^{}_\alpha{}^0 \, e^{}_\beta{}^0 + e^{}_\alpha{}^1 \, e^{}_\beta{}^1\,,
\ee
we expand the conventional Polyakov action
\be \label{eq:relpf0}
    \hat{S}_\text{P} = - \frac{T}{2} \int d^2 \sigma \, \sqrt{-\hat{h}} \, \hat{h}^{\alpha\beta} \, \p^{}_\alpha \hat{X}^\mu \, \p^{}_\beta \hat{X}^{}_\mu
\ee
with respect to a large $\omega$ as
\begin{align} \label{eq:hspmpt0}
\begin{split}
	\hat{S}^{}_\text{P} = - \frac{T}{2} \int d^2 \sigma \, e \, \biggl[ & - e^\alpha{}^{}_0 \, e^\beta{}^{}_0 \, \Bigl( \omega^2 \, \p^{}_{\alpha} X^A \, \p^{}_{\beta} X^{}_A + \p^{}_{\alpha} X^{A'} \, \p^{}_{\beta} X^{A'} \Bigr) \\[4pt]
	& + e^\alpha{}^{}_1 \, e^\beta{}^{}_1 \, \Bigl( \p^{}_{\alpha} X^A \, \p^{}_{\beta} X^{}_A + \omega^{-2} \, \p^{}_{\alpha} X^{A'} \, \p^{}_{\beta} X^{A'} \Bigr) \biggr] \,.
\end{split}
\end{align}
Introducing the Lagrange multipliers $\lambda^{}_A$\,, we rewrite~\eqref{eq:hspmpt0} equivalently as
\begin{align} \label{eq:hspmpt2}
\begin{split}
	\hat{S}^{}_\text{P} = & - \frac{T}{2} \int d^2 \sigma \, e \, \Bigl( e^\alpha{}^{}_1 \, e^\beta{}^{}_1 \, \p_{\alpha} X^A \, \p_{\beta} X^{}_A - e^\alpha{}^{}_0 \, e^\beta{}^{}_0 \, \p^{}_\alpha X^{A'} \p^{}_\beta X^{A'} - \lambda^{}_A \, e^\alpha{}^{}_0 \, \p^{}_\alpha X^A \Bigr) \\[4pt]
    	& - \frac{T}{2} \int d^2 \sigma \, \omega^{-2} \, e \, \Bigl( e^\alpha{}^{}_1 \, e^\beta{}^{}_1 \, \p^{}_\alpha X^{A'} \p^{}_\beta X^{A'} + \tfrac{1}{4} \, \lambda^{}_A \, \lambda^A \Bigr) \,. 
\end{split}
\end{align}
In the $\omega \rightarrow \infty$ limit, the M$p$T string sigma model~\eqref{eq:hspmpt20} is recovered. 

In order to derive the analogous M$p$T Nambu-Goto action, we now apply the same limiting prescription to the conventional Nambu-Goto action
\be \label{eq:sngt}
	\hat{S}^{}_\text{NG} = - T \int d^2 \sigma \, \sqrt{-\det \Bigl( \p^{}_\alpha \hat{X}^\mu_{\phantom{\dagger}} \, \p^{}_\beta \hat{X}^{}_\mu \Bigr)}\,.
\ee 
We start with rewriting this Nambu-Goto action equivalently as
\be \label{eq:mptngv0}
	\hat{S}^{}_\text{NG} = - \frac{T}{2} \int d^2 \sigma \, \biggl[ v - \frac{1}{v} \det \bigl( \p^{}_\alpha \hat{X}^\mu_{\phantom{\dagger}} \, \p^{}_\beta \hat{X}^{}_\mu \bigr) \biggr]\,.
\ee
Integrating out $v$ gives back the original action~\eqref{eq:sngt}. 
Using the reparametrization~\eqref{eq:mptsps} of the embedding coordinates, we find
\begin{align}
\begin{split}
	\det \bigl( \p^{}_\alpha \hat{X}^\mu_{\phantom{\dagger}} \, \p^{}_\beta \hat{X}^{}_\mu \bigr) = \omega^2 \, \det \bigl( \p_{\alpha} X^A \, \p_{\beta} X^{}_A \bigr) & + \epsilon^{\alpha\beta} \, \epsilon^{\gamma\delta} \, \bigl( \p_{\alpha} X^A \, \p_{\gamma} X^{}_A \bigr) \, \bigl( \p^{}_{\beta} X^{A'} \p^{}_{\delta} X^{A'} \bigr) \\[4pt]
	& + \omega^{-2} \, \det \bigl( \p_{\alpha} X^{A'} \, \p_{\beta} X^{}_{A'} \bigr)\,,
\end{split}
\end{align}
where 
\be
	\det \bigl( \p_{\alpha} X^A \, \p_{\beta} X^{}_A \bigr) = \Bigl( \epsilon^{\alpha\beta} \, \p^{}_\alpha X^A \, \p^{}_\beta X^B \Bigr) \, \Bigl( \epsilon^{\gamma\delta} \, \p^{}_\gamma X^C \, \p^{}_\delta X^D \Bigr) \, \eta^{}_{AC} \, \eta^{}_{BD}\,.
\ee
Introducing an auxiliary antisymmetric tensor $\lambda^{}_{AB}$\,, we rewrite the action~\eqref{eq:mptngv0} as
\begin{align} \label{eq:sng}
	\hat{S}^{}_\text{NG} = - \frac{T}{2} \int d^2 \sigma \, \biggl\{ v & - \frac{1}{v} \, \Bigl[ \epsilon^{\alpha\beta} \, \epsilon^{\gamma\delta} \, \bigl( \p_{\alpha} X^A \, \p_{\gamma} X^{}_A \bigr) \, \bigl( \p^{}_{\beta} X^{A'} \p^{}_{\delta} X^{A'} \bigr) + \omega^{-2} \, \det \bigl( \p_{\alpha} X^{A'} \, \p_{\beta} X^{}_{A'} \bigr) \Bigr] \notag \\[4pt]
	& + \epsilon^{\alpha\beta} \, \p^{}_\alpha X^A \, \p^{}_\beta X^B \,  \lambda^{}_{AB} + \frac{v}{4 \, \omega^2} \, \lambda^{}_{AB} \, \lambda^{AB} \biggr\}\,.
\end{align}
In the $\omega \rightarrow \infty$ limit, integrating out $v$ gives the following finite action:
\be \label{eq:ngf}
	S^{}_\text{NG} = - T \int d^2 \sigma \, \sqrt{- \epsilon^{\alpha\beta} \, \epsilon^{\gamma\delta} \, \bigl( \p_{\alpha} X^A \, \p_{\gamma} X^{}_A \bigr) \, \bigl( \p^{}_{\beta} X^{A'} \p^{}_{\delta} X^{A'} \bigr)} - T \int d X^A \, \wedge dX^B \,  \lambda^{}_{AB}\,.
\ee
This action is now manifestly invariant under the longitudinal Lorentz boost $\delta X^A = \Lambda^A{}_B \, X^B$, supplemented with $\delta \lambda_{AB} = \Lambda_A{}^C \, \lambda_{CB} - \Lambda_B{}^C \, \lambda_{CA}$\,. The invariance under the $p$-brane Galilean boost $\delta_\text{\scalebox{0.8}{G}} X^{A'} = \Lambda^{A'}{}_A X^A$ is guaranteed on-shell by the constraint $dX^A \wedge dX^B = 0$ imposed by the Lagrange multiplier $\lambda^{}_{AB}$\,. The off-shell invariance under the boost therefore requires a non-trivial transformation of $\lambda^{}_{AB}$\,.  
When the longitudinal index $A$ is one-dimensional, the Lagrange multiplier term in the action drops out, and the action coincides with the Nambu-Goto formulation~\eqref{eq:snvs} of the M0T string. 

The Nambu-Goto action~\eqref{eq:ngf} is equivalent to Eq.~\eqref{eq:ngrwmpt} that we derived earlier from integrating out the auxiliary fields in the Polyakov formulation~\eqref{eq:hspmpt20} of the M$p$T string: integrating out the Lagrange multiplier $\lambda_{AB}$ imposes the constraint $\epsilon^{\alpha\beta} \, \p^{}_\alpha{}X^A \, \p^{}_\beta{}X^B = 0$\,, which is solved by $\p^{}_\alpha X^A = \Gamma^A \, \p^{}_\alpha X^0$ as in Eq.~\eqref{eq:solet}; plugging this solution back into the Nambu-Goto action~\eqref{eq:ngf} gives back Eq.~\eqref{eq:ngrwmpt}.
%

\subsection{Relation to Matrix Gauge Theories} \label{eq:rmgtsltd}

We have seen in Section~\ref{sec:mtnlsw} that, in M0T, the fundamental string is the non-vibrating string, while the light excitations are the D0-branes described by BFSS Matrix theory. We have also seen that M0T arises from the critical RR one-form limit of type IIA superstring theory, which means that a BPS limit is taken on a background D0-brane. 
Under T-dualities, this critical RR one-form is eventually dualized to be the critical RR ($p$+1)-form. This corner of type II superstring theory that arises from such a critical RR ($p$+1)-form limit is referred to as Matrix $p$-brane Theory (M$p$T) in~\cite{udlstmt}. We have found in this paper that the fundamental string in M$p$T is described by the action~\eqref{eq:mptsps}. The name M$p$T is motivated by the fact that the critical RR ($p$+1)-form limit of a stack of D$p$-branes leads to Matrix gauge theory~\cite{Obers:1998fb}, \emph{i.e.}, the BFSS Matrix theory compactified over a vanishing $p$-torus~\cite{Fischler:1997kp}, which describes the light excitations in M$p$T. For example, when $p=1$\,, the Matrix gauge theory is Matrix string theory~\cite{Dijkgraaf:1997vv, Motl:1997th}; when $p=3$, the Matrix gauge theory is $\CN=4$ SYM. See further details in \cite{udlstmt, longpaper}. Here, we content ourselves with performing the M$p$T limit of a single D$p$-brane, from which we will be able to provide a qualitative understanding of why the RR potential is required to be critical~\cite{Gomis:2000bd, Gomis:2005bj} (see also~\cite{Kamimura:2005rz} for dual D-branes in this context). We start with the D$p$-brane action in conventional type II superstring theories, focusing on the bosonic part in flat target space but with a nontrivial background RR ($p+1$)-form potential,  
\be \label{eq:hsdp}
    	\hat{S}_{\text{D}p} = - T_{p} \int d^{p+1} \sigma \, e^{-\hat{\Phi}} \, \sqrt{-\det \Bigl( \p_\alpha \hat{X}^\mu_{\phantom{\dagger}} \, \p_\beta \hat{X}_\mu + F_{\alpha\beta} \Bigr)} + T_{p} \int \hat{C}^{(p+1)}\,.
\ee
Here, $F = dA$ is the $U(1)$ gauge field strength on the brane and $\hat{\Phi}$ is the dilaton field. We have set all the other RR potentials except the $(p+1)$-form $\hat{C}^{(p+1)}$ to zero. Under the rescalings of the embedding coordinates in Eq.~\eqref{eq:mptsps}, in the static gauge with $X^A = \sigma^A$, $A = 0\,, \cdots, \, p$\,, we find at large $\omega$\,, 
\begin{align} \label{eq:dbirs}
	\sqrt{-\det \Bigl( \p_\alpha \hat{X}^\mu_{\phantom{\dagger}} \, \p_\beta \hat{X}_\mu + F_{\alpha\beta} \Bigr)} = \omega^{\frac{p-3}{2}} \biggl[ \, \omega^2 + \tfrac{1}{2} \Bigl( \p^\alpha X^{A'} \, \p_\alpha X^{A'} \! + F^{\alpha\beta} \, F_{\alpha\beta} \Bigr) + O \bigl( \omega^{-2} \bigr) \biggr].
\end{align}
Under the additional reparametrization~\cite{Gopakumar:2000ep, Gomis:2000bd, udlstmt},
\be \label{eq:cphicp}
	\hat{\Phi} = \Phi + \tfrac{p-3}{2} \, \ln \omega\,,
		\qquad%
	\hat{C}^{(p+1)} \rightarrow - \omega^2 \, e^{-\Phi} \, dX^0 \wedge \cdots dX^{p}\,,
\ee
we find that the $\omega \rightarrow \infty$ limit of the D$p$-brane action~\eqref{eq:hsdp} gives rise to the following finite D$p$-brane action~\cite{Gomis:2005bj, Kluson:2019uza, Ebert:2021mfu}:
\be \label{eq:sdpmpt}
	S^{}_{\text{D}p} = - \frac{T_p}{2} \int d^{p+1} \sigma \, e^{-\Phi} \, \Bigl( \p^\alpha X^{A'} \, \p_\alpha X^{A'} + F^{\alpha\beta} \, F_{\alpha\beta} \Bigr)\,.
\ee
Here, $\Phi$ is the effective dilaton field in M$p$T. It is shown explicitly in~\cite{udlstmt, longpaper} that the nonabelian version of Eq.~\eqref{eq:sdpmpt} gives rise to Matrix gauge theory. Note that the background D$p$-brane charge is fine tuned to cancel to effective brane tension in this critical RR $(p+1)$-form limit. This is ultimately a BPS limit.

\section{Timelike T-Duality}

\begin{figure}[t!]
\centering
\begin{adjustbox}{width=0.7\textwidth}
\hspace{-8mm}
\begin{tikzpicture}
\begin{scope}[scale=.85]
 
        	\path[every node/.style=draw, rounded corners=1, line width=1.5pt, minimum width=135pt, minimum height=20pt, font = \small]   
	(11,1.5) node {\textbf{M0T}: BFSS Matrix theory}
	(11,3.5) node {\textbf{MpT}: Matrix gauge theories}
	(19,1.07) [align=center] node[minimum height = 40pt] {\hspace{4mm} IKKT Matrix theory\\
	\hspace{4mm} tensionless string theory\\
	\hspace{4mm} ambitwistor string theory}
	(19,3.5) node {\textbf{M(-p\,-1)T}/Carrollian string}
	;

    	\path  
        (10.97,2.47) [font=\footnotesize] node {\textbf{T-duals}}
        (18.97,2.47) [font=\footnotesize] node {\textbf{T-duals}}
        (16.5,1.09)[font=\small] node[rotate=90] {\textbf{M(-1\hspace{-0.5mm})T}} 
        (9.4,2.5) [font=\footnotesize] node {\emph{long.}}
        (12.72,2.5) [font=\footnotesize] node {\emph{trans.}}
		(17.4,2.5) [font=\footnotesize] node {\emph{long.}}
        (20.72,2.5) [font=\footnotesize] node {\emph{trans.}}
        (15,1.8) [font=\footnotesize] node {\emph{timelike}}
       	(15,1.25) [font=\footnotesize] node {\emph{T-dual}}
        (15,3.8) [font=\footnotesize] node {\emph{timelike}}
        (15,3.25) [font=\footnotesize] node {\emph{T-dual}}
        ;

        \draw [{[length=1.3mm]<}-{[length=1.3mm]}] (10,2) -- (10,3);
        \draw [{[length=1.3mm]}-{>[length=1.3mm]}] (12,2) -- (12,3);        
        \draw [{[length=1.3mm]<}-] (18,2) -- (18,3);
        \draw [-{[length=1.3mm]>}] (20,2) -- (20,3);

        \draw [{<[length=1.3mm]}-{>[length=1.3mm]}] (14,1.5) -- (16,1.5);
       	\draw [{<[length=1.3mm]}-{>[length=1.3mm]}] (14,3.5) -- (16,3.5);
 
\end{scope}
\end{tikzpicture}
\end{adjustbox}
\caption{A road map for the timelike T-dual relation between M$p$T and M(-$p$\,-1)T and the spacelike T-duality relation between M$p$Ts with $p<0$.}
\label{fig:tltd}
\end{figure}
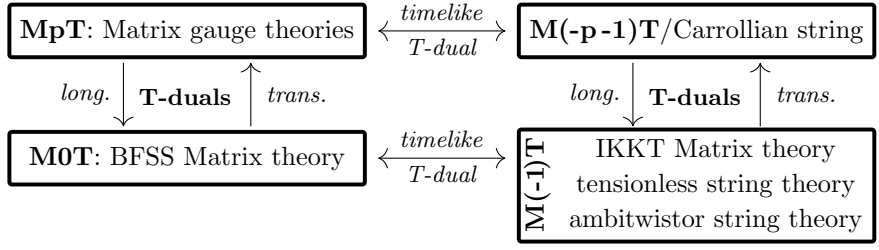

The second class of theories that we consider arise from the T-duality transformation along a target space timelike isometry in Matrix $p$-brane Theory (M$p$T). T-duality in a timelike direction was first introduced in \cite{Hull:1998vg}, where it is shown that type IIA (IIB) superstring theory maps to the IIB${}^*$ (IIA${}^*$) theory. These starred theories are distinct from the conventional type II theories as certain background fields are effectively complexified. In such type II${}^*$ theories, the type II D-branes are mapped to Euclidean branes, which are localized in time and are also known as S(pacelike)-branes in~\cite{Gutperle:2002ai}. In particular, in the IIB${}^*$ theory, the spacelike D3-branes are argued to be holographically dual to de Sitter space~\cite{Hull:1998vg, Hull:2001ii}. In this subsection, we will provide a worldsheet perspective for how this timelike T-duality relation between type II and II${}^*$ theories works in the critical RR limits~\cite{udlstmt}. We will show how the M$p$T strings connect to tensionless, ambitwistor, and Carrollian string theories at the worldsheet level. See Fig.~\ref{fig:tltd} for a summary of the T-dual relations in this section. 

\subsection{Tensionless String and IKKT Matrix Theory} \label{eq:ikktts}

We start with considering the timelike T-duality transformation of the M0T string sigma model~\eqref{eq:fstmztp} in conformal gauge. For this purpose, we gauge the timelike isometry $X^0$ and rewrite Eq.~\eqref{eq:fstmztp} as
\begin{align} \label{eq:pamzttl}
    	S^{}_\text{gauged} = \frac{T}{2} \int d^2 \sigma \, \Bigl( D^{}_{\!\sigma} X^0 \, D^{}_{\!\sigma} X^0 & + \p^{}_\tau X^i \, \p^{}_\tau X^i + \lambda \, D^{}_\tau X^0 
    	- 2\, \tilde{X}^0 \, \epsilon^{\alpha\beta} \, \p_\alpha v_\beta \Bigr)\,,
\end{align}
where $D_\alpha X^0 = \p_\alpha X^0 + v_\alpha$ and $i = 1\,, \, \cdots, \, 9$\,. 
The Lagrange multiplier $\tilde{X}^0$ imposes the condition that $v_\alpha$ is pure gauge. The equations of motion from varying $v_\alpha$ are
$\lambda = 2 \, \p_\sigma \tilde{X}^0$ and $v^{}_\sigma = - \p_\tau \tilde{X}^0 - \p_\sigma X^0$\,. 
Plugging these equations into the action~\eqref{eq:pamzttl}, we find the dual action (up to a topological term associated with winding Wilson lines)
\be \label{eq:smot00}
	\tilde{S} = \frac{T}{2} \int d^2 \sigma \, \Bigl( - \p_\tau \tilde{X}^0 \, \p_\tau \tilde{X}^0 + \p^{}_\tau X^i \, \p^{}_\tau X^i \Bigr) \,.
\ee
Dropping the tildes, we write the dual action as
\be \label{eq:smot0}
	S_\text{M(-1)T} = \frac{T}{2} \int d^2 \sigma \, \p_\tau X^\mu \, \p_\tau X_\mu\,,
		\qquad%
	\mu = 0\,, \, \cdots, \, 9\,.
\ee
This is the string sigma model in Matrix (-1)-brane Theory (M(-1)T), where the target space is Lorentzian. as indicated by the name M(-1)T, the light excitations in this corner are D(-1)-branes, \emph{i.e.} D-instantons~\cite{udlstmt}. The action~\eqref{eq:smot0} can be thought of as the $p = -1$ extension of the M$p$T string sigma model~\eqref{eq:m0tst}. It is therefore also a straightforward analog that the M$(-1)$T string must arise from the $\omega \rightarrow \infty$ limit of the conventional Polyakov action reparametrized as in Eq.~\eqref{eq:hspmpt0} but with $p = -1$\,. We now demonstrate this limiting procedure explicitly. In the usual Polyakov string action,
\be
	\hat{S}_\text{P} = - \frac{T}{2} \int d^2 \sigma \, \sqrt{-\hat{h}} \, \hat{h}^{\alpha\beta} \, \p_\alpha \hat{X}^\mu \, \p_\beta \hat{X}^{}_\mu\,, 
\ee
we take
\be \label{eq:tsrp}
	\hat{X}^{\mu} = \omega^{-1/2} \, X^{\mu},
		\qquad%
	\hat{h}^{}_{\alpha\beta} = - \omega^{-2} \, e^{}_\alpha{}^0 \, e^{}_\beta{}^0 + e^{}_\alpha{}^1 \, e^{}_\beta{}^1\,.
\ee
In the $\omega \rightarrow \infty$ limit, we find
\be \label{eq:smmote}
	S_\text{M(-1)T} = \frac{T}{2} \int d^2 \sigma \, e \, e^\alpha{}_0 \, e^\beta{}_0 \, \p_\alpha X^\mu \, \p_\beta X_\mu\,,
\ee
which in the conformal gauge $e_\alpha{}^a \propto \delta_\alpha^a$ reduces to Eq.~\eqref{eq:smot0}. 

\vspace{3mm}

\begin{figure}[t!]
	\centering
	\includegraphics[width=0.4\textwidth]{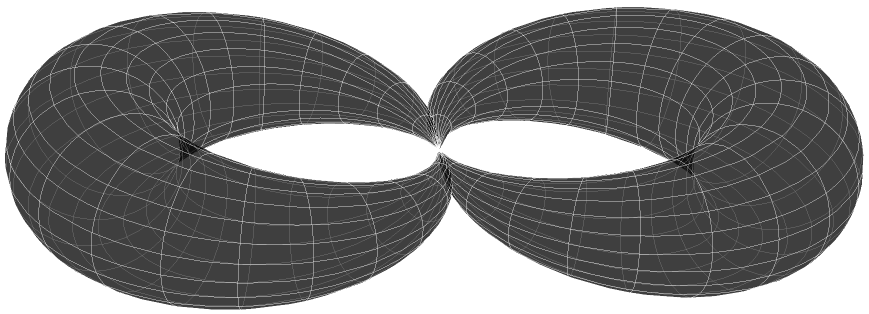} 
    \caption{A more general Deligne-Mumford compactification of Riemann surface.}
    \label{fig:sf}
\end{figure}

\vspace{3mm}

\noindent $\bullet$~\text{\emph{Tensionless string theory.}} It is known that the M(-1)T string action~\eqref{eq:smot0} arises from a tensionless limit of conventional string theory. Upon the identification $V^\alpha = \sqrt{e} \, e^\alpha{}_0$ in the M(-1)T string action~\eqref{eq:smmote}, we find
\be \label{eq:mmosilst}
    S^{}_\text{M(-1)T} = \frac{T}{2} \int d^2 \sigma \, V^\alpha \, V^\beta \, \p^{}_\alpha X^\mu \, \p^{}_\beta X_\mu\,. 
\ee
This action is identical to the Isberg-Lindstr\"{o}m-Sundborg-Theodoris (ILST) tensionless string action~\cite{Lindstrom:1990qb, Isberg:1993av}. In particular, this is Eq.~(18) in~\cite{Isberg:1993av}. See \emph{e.g.}~\cite{Sundborg:2000wp, Bagchi:2013bga, Bagchi:2015nca, Bagchi:2020fpr, Bagchi:2021rfw, Chen:2023esw} for more recent developments on tensionless strings. ILST tensionless string is also referred to as \emph{null} string in the literature~\cite{Schild:1976vq}. 

In order to understand why the M(-1)T limit defined by Eq.~\eqref{eq:tsrp} can be viewed as a tensionless string limit, we consider how the M(-1)T limit is applied to the conventional Nambu-Goto action,
\be
	\hat{S}^{}_\text{NG} = - T \int d^2 \sigma \, \sqrt{- \det \Bigl( \p^{}_\alpha \hat{X}^\mu_{\phantom{\dagger}} \, \p^{}_\beta \hat{X}^{}_\mu \Bigr)}\,.
\ee
Plugging the prescription~\eqref{eq:tsrp} for the embedding coordinates, we find
\be \label{eq:ngns}
    \hat{S}^{}_\text{NG} = - \frac{T}{\omega} \int d^2 \sigma \, \sqrt{-\det \Bigl( \p^{}_\alpha {X}^\mu_{\phantom{\dagger}} \, \p^{}_\beta {X}^{}_\mu \Bigr)}\,.
\ee
It is therefore manifest that the reparametrization in $\omega$ here can be equivalently obtained by replacing the original string tension $\hat{T}$ with $T / \omega$\,, while keeping the embedding coordinates untouched. In this alternative way of thinking, the $\omega \rightarrow \infty$ limit sets the original string tension $\hat{T}$ to zero. However, in the resulting M$(-1)$T string, we still have a finite effective tension $T$\,. 

In order to facilitate the $\omega \rightarrow \infty$ limit in the Nambu-Goto action~\eqref{eq:ngns}, we rewrite it as
\be \label{eq:sngvo}
    \hat{S}^{}_\text{NG} = - \frac{T}{2} \int d^2 \sigma \, \biggl[ \, \omega^{-2} \, v - v^{-1} \det \Bigl( \p^{}_\alpha {X}^\mu_{\phantom{\dagger}} \, \p^{}_\beta {X}^{}_\mu \Bigr) \biggr]\,,
\ee
where $v$ is an auxiliary field. Integrating $v$ out gives back the original action~\eqref{eq:ngns}. 
In the $\omega \rightarrow \infty$ limit, Eq.~\eqref{eq:sngvo} gives rise to the Nambu-Goto formulation for the M(-1)T string,
\be \label{eq:ngav}
    S^{}_\text{NG} = \frac{T}{2} \int d^2 \sigma \, v^{-1} \det \Bigl( \p^{}_\alpha {X}^\mu_{\phantom{\dagger}} \, \p^{}_\beta {X}^{}_\mu \Bigr)\,.
\ee
The associated phase-space formulation is
\be \label{eq:psammot}
    S_\text{p.s.} = \int d^2 \sigma \, \biggl( P^{}_\mu \, \p_\tau X^\mu - \frac{\chi}{2 T} \, P^{}_\mu \, P^\mu - \rho \, P_\mu \, \p_\sigma X^\mu \biggr)\,.
\ee
After integrating out $P^{}_\mu$ and $\rho$\,, this phase-space action gives back the Nambu-Goto formulation~\eqref{eq:ngav}, with $v = \chi \, \tau^{}_{\sigma\sigma}$\,. Finally, plugging the Carrollian parametrization~\eqref{eq:wscge0} of the worldsheet into the phase-space action followed by integrating out $P^{}_\mu$\,, we recover the original Polyakov action~\eqref{eq:smmote}. This proves that the Nambu-Goto formulation~\eqref{eq:ngav} is equivalent to the Polyakov formulation~\eqref{eq:smmote}.

\vspace{3mm}

\noindent $\bullet$~\text{\emph{Type IIB${}^*$ superstring theroy.}} The complete decoupling limit of string theory that leads to M(-1)T has been given in~\cite{udlstmt}, which we review below to be self-contained. Na\"{i}vely, one would expect that M(-1)T is a special case of M$p$T with $p = -1$\,. It is natural to guess the decoupling limit of type IIB superstring theory that leads to M(-1)T by extending the M$p$T prescriptions in Eqs.~\eqref{eq:mptsps} and \eqref{eq:cphicp}, which we transcribe below:
\begin{subequations}
\begin{align}
	\hat{X}^A & = \sqrt{\omega} \, X^A,  		
		&%
	\hat{\Phi} & = \Phi + \frac{p-3}{2} \, \ln \omega\,, \\[4pt]		
	\hat{X}^{A'} & = \frac{X^{A'}}{\sqrt{\omega}}\,,	
		&%
	 \hat{C}^{(p+1)} & = \frac{\omega^2}{e^\Phi} \, dX^0 \wedge \cdots \wedge dX^p\,,
\end{align}
\end{subequations}
where $A = 0\,, \, \cdots, \, p$ and $A' = p+1\,, \, \cdots, \, 9$\,. The $\omega \rightarrow \infty$ limit of the relevant type II superstring theory gives rise to M$p$T. Setting $p = -1$ basically kills the $X^A$ sector. This leads us to conjecture that the M(-1)T prescription is given by
\begin{align} \label{eq:rpmmot}
	\hat{X}^\mu & = \frac{X^{\mu}}{\sqrt{\omega}}\,,		
		&%
	\hat{\Phi} & = \Phi - 2 \, \ln \omega\,, 	
		&%
	 \hat{C}^{(0)} & = \frac{\omega^2}{e^\Phi}\,,
\end{align}
where the RR zero-form $\hat{C}^{(0)}$ becomes critical. 
However, this na\"{i}ve extrapolation is \emph{not} quite right. The associated caveat can be made manifest by \emph{e.g.}~applying the prescription~\eqref{eq:rpmmot} to a single probe D1-brane in type IIB superstring theory, which is described by the action
\be \label{eq:d1ba}
	\hat{S}_\text{D$1$} = - T_1 \int d^2 \sigma \, e^{-\hat{\Phi}} \, \sqrt{- \det \Bigl( \p_{\alpha} \hat{X}^\mu_{\phantom{\dagger}} \, \p^{}_\beta \hat{X}_\mu + \CF^{}_{\alpha\beta} \Bigr)} + T_1 \int \hat{C}^{(0)} \, \CF\,.
\ee  
Here, $\CF = B + dA$\,, with $B$ the background Kalb-Ramond field and $A$ the U(1) gauge potential on the D1-brane. We have set all the other background fields to zero. Plugging the prescription~\eqref{eq:rpmmot} into the D1-brane action leads to a leftover $\omega$ divergence,
\be
    \hat{S}^{}_\text{D1} = \bigl( 1 - i \bigr) \, \omega^2 \, \int d^2\sigma \, e^{-\Phi} \, \CF^{}_{01} + O\bigl(\omega^0\bigl)\,.
\ee
We have chosen the branch with $\CF^{}_{01} > 0$\,\footnote{The existence of different branches is due to the ambiguity in defining the reparametrizations~\eqref{eq:rpmmot}. The two branches marked by $\CF^{}_{01} > 0$ and $\CF^{}_{01} < 0$ are associated with the brane and the anti-brane, respectively, and they are mapped to each other via the SL($2\,,\mathbb{Z}$) duality similarly as in~\cite{Bergshoeff:2022iss}.}. This divergence vanishes identically if one replaces in Eq.~\eqref{eq:rpmmot} \emph{either} the $\hat{\Phi}$ prescription with 
\be \label{eq:complphi}
    \hat{\Phi} = \Phi + \frac{i\pi}{2} - 2 \, \ln \omega\,,
\ee
\emph{or} the $\hat{C}^{(0)}$ prescription with $\hat{C}^{(0)} = i \, \omega^2 \, e^{-\Phi}$ (but \emph{not} both). The resulting theory is Euclidean NCYM on spacelike one-brane that is T-dual to 4D NCYM on D3-brane in M1T. This necessity of introducing an extra factor $i$ in the dilaton reparametrization can also be understood by using the standard Buscher rule for the dilaton transformation~\cite{Buscher:1987qj},
\be
    \tilde{\Phi} = \Phi - \frac{1}{2} \ln k^2\,, 
\ee
where $k_\mu$ is a timelike Killing vector, \emph{i.e.} $k^2 < 0$\,. Hence, with a specific choice of the branch,
\be
    \tilde{\Phi} = \Phi + \frac{i\pi}{2} - \frac{1}{2} \ln |k^2|\,.
\ee
The appearance of the factor $i$ in the dilaton term is a general feature of timelike T-duality. As we have briefly discussed at the beginning of this section, the timelike T-duality maps type II superstring theories to the so-called type II${}^*$ theories, where the latter admit S(spacelike)-branes that satisfy the Dirichlet boundary condition in time~\cite{Hull:1998vg}. 

\vspace{3mm}

\noindent $\bullet$~\text{\emph{IKKT Matrix theory.}} The dynamics of M(-1)T, which is of type IIB${}^*$\,, is captured by the D(-1)-instantons from T-dualizing the BFSS Matrix theory on the D0-branes in M0T along a timelike isometry. This procedure gives rise to the Ishibashi-Kawai-Kitazawa-Tsuchiya (IKKT) Matrix theory on the D-instantons in M(-1)T~\cite{Ishibashi:1996xs}, 
\be
    S_\text{IKKT} = - \frac{1}{2 \, g^2} \, \tr \Bigl( \tfrac{1}{2} \bigl[ A_\mu\,, A_\nu \bigr]^2 + \bar{\Psi} \, \Gamma^\mu \bigl[ A_\mu\,, \Psi \bigr] \Bigr)\,,
\ee
where the vector $A_\mu$ and the Majorana-Weyl spinor $\Psi$ are $N \times N$ matrices. Moreover, $\Gamma^\mu$ is the Weyl-projected $16 \times 16$ Dirac matrices in ten dimensions. 
The IKKT Matrix theory was originally proposed as a nonperturbative regularization of the Green-Schwarz IIB superstring sigma model in the Schild formulation. 

The string worldsheet topology in M(-1)T is the same as in M0T, which we have discussed in Section~\ref{sec:wstnrs}. They are nodal Riemann spheres. The existence of the pinching on the M$0$T string worldsheet might imply that the fundamental strings interact with each other via instantons at the pinching points. It is therefore suggestive to consider more general Deligne-Mumford compactifications of Riemann surfaces (see \emph{e.g.} Fig.~\ref{fig:sf}), and it would be interesting to see whether there is any connection to some version of string field theory that involves instantons. The conjectured string field-theoretical dynamics may be ultimately encoded by IKKT Matrix theory. Moreover, the brane objects other than the D-instanton in M(-1)T also acquire well-defined effective actions from taking the $\omega$ limit of conventional D$p$-brane actions using the prescription~\eqref{eq:rpmmot}, but with the dilaton $\hat{\Phi}$ modified as in Eq.~\eqref{eq:complphi}. Here, the background RR zero-form becomes critical. See also~\cite{Bagchi:2020ats} for discussions on a D-instanton state in the context of tensionless string theory.

\subsection{Ambitwistor String Theory} \label{sec:ast}

We now discuss the gauge fixing of the phase-space action~\eqref{eq:psammot}, \emph{i.e.}
\be \label{eq:psammotn}
    S_\text{p.s.} = \int d^2 \sigma \, \biggl( P^{}_\mu \, \p_\tau X^\mu - \frac{\chi}{2 T} \, P^{}_\mu \, P^\mu - \rho \, P_\mu \, \p_\sigma X^\mu \biggr)\,.
\ee
We have noted that the Schild gauge with $\chi = 1$ and $\rho = 0$ matches the conformal gauge $e^{}_\alpha{}^a \propto \delta_\alpha^a$ in the Polyakov action with the Carrollian parametrization~\eqref{eq:wscge0} of the worldsheet. In this Schild gauge, we reproduce the M(-1)T string action~\eqref{eq:smot0} upon integrating out $P_\mu$ in the phase-space action. In Section~\ref{eq:ikktts}, we have shown that the M(-1)T string action describes the ILST tensionless string~\cite{Lindstrom:1990qb, Isberg:1993av}. This Schild gauge is the gauge choice that we have considered so far. However, if one chooses the ambitwistor string gauge with~\cite{Casali:2016atr, Siegel:2015axg} 
\be \label{eq:ag}
    \chi = 0\,,
        \qquad%
    \rho = 1
\ee
in the phase-space action~\eqref{eq:psammotn}, we are led to the chiral worldsheet action 
\be \label{eq:ambia}
    S^{}_\text{ambi.} = \int d^2\sigma \, P^{}_\mu \, \bar{\p} X^\mu\,,
        \qquad%
    \bar{\p} = \p^{}_\tau - \p^{}_\sigma\,,
\ee
which is supplemented with the constraint $P^{}_\mu \, P^\mu = 0$\,. This is the defining action of the bosonic part of \emph{ambitwistor string theory}~\cite{Mason:2013sva}.  

At the classical level that we have just discussed, ambitwistor string theory appears to be tensionless string theory in an unconventional gauge~\eqref{eq:ag}. At the quantum level, however, this relation between ambitwistor and tensionless string theory develops another layer: the ambitwistor string arises from an exotic choice of vacuum. To illustrate how ambitwistor string arises at the quantum level, we consider the mode expansion of the embedding coordinates in the closed string sector. Denote the Fourier modes as $\alpha_n$ and $\tilde{\alpha}_n$\,, which correspond to the left-moving and right-moving waves of the close string, respectively. These modes satisfy the commutation relations
\be
    \bigl[ \alpha^{}_n\,, \alpha^{}_m \bigr] = \bigl[ \tilde{\alpha}^{}_n\,, \tilde{\alpha}^{}_m \bigr] = n \, \delta^{}_{m+n, \, 0}\,.
\ee
In conventional string theory, the vacuum $|0\rangle$ is defined by $\alpha^{}_n |0\rangle = \tilde{\alpha}^{}_{n} |0\rangle = 0$ for any $n > 0$\,. In contrast, the vacuum $|0\rangle_\text{ambi.}$ in ambitwistor string theory is defined by
\be
    \alpha^{}_n |0\rangle_\text{ambi.} = \tilde{\alpha}^{}_{-n} |0\rangle_\text{ambi.} = 0\,,
        \qquad%
    n > 0\,. 
\ee
See~\cite{Casali:2016atr, Hwang:1998gs} for further details. 

The moduli space of an ambitwistor string amplitude is localized to be a set of discrete points that solve the scattering equations~\cite{Cachazo:2013gna}, which encode the kinematics of particle scatterings in the CHY formalism of QFTs~\cite{Cachazo:2013hca}. We review how the scattering equation arises from the ambitwistor string theory below, following closely the original work~\cite{Mason:2013sva}. At tree level, the scattering equation can be obtained by considering $N$ insertions of plane-wave vertex operators $e^{i k^{(i)} \cdot X}$, $i = 1\,, \cdots, N$, acting as point-like sources coupled to the ambitwistor string,
\be
    S_\text{sourced} = \int  d^2 \sigma \, \biggl[ P^{}_\mu \, \bar{\p} X^\mu - i \, \sum_{i=1}^N k^{(i)} \! \cdot \! X  \, \delta^{(2)} \big(\sigma - \sigma^{(i)} \bigr) \biggr]\,.
\ee
Here, $\sigma^{(i)}$ refers to the location of the $i$-th inserted vertex operator.
Integrating out $X^\mu$ in the associated path integral gives $\bar{\p} P_\mu = i \, \sum_{i} k^{(i)}_\mu \, \delta^{(2)} (\sigma - \sigma^{(i)})$\,. In the tree-level case, where the worldsheet is conformally a sphere, the unique solution is
\be
    P_\mu (\sigma) = \frac{d\sigma}{2\pi} \sum_{i=1}^N \frac{k^{(i)}_\mu}{\sigma - \sigma^{(i)}}\,.
\ee
After integrating over the worldsheet, the Hamiltonian constraint $P_\mu P^\mu = 0$ then implies the scattering equation
\be \label{eq:scatteringequation}
    \sum_{j\neq i} \frac{k^{(i)} \cdot k^{(j)}}{\sigma^{(i)} - \sigma^{(j)}} = 0\,.
\ee
Here, we have used the on-shell condition $k^{(i)} \cdot k^{(i)} = 0$\,. This scattering equation originally comes from the Gross-Mende limit~\cite{Gross:1987ar} of the Koba-Nielsen factor in the conventional string amplitude, where the condition arises from the saddle point evaluation. 

We now return to the Polyakov formulation~\eqref{eq:smmote} of the M(-1)T string, 
\be \label{eq:smmotetrs}
	S_\text{M(-1)T} = \frac{T}{2} \int d^2 \sigma \, e \, e^\alpha{}_0 \, e^\beta{}_0 \, \p_\alpha X^\mu \, \p_\beta X_\mu\,.
\ee
The ambitwistor string gauge~\eqref{eq:ag} is now translated to be
\be \label{eq:agpf}
    e^{}_\tau{}^1 = e^{}_\sigma{}^1 = 1\,,
        \qquad%
    e^{}_\tau{}^ 0 = e^{}_\sigma{}^0 = 0\,,
\ee
which makes the worldsheet singular as the determinant $e = \det e_\alpha{}^a =0$\,. This singular behavior can be regularized by using the Hohm-Siegel-Zwiebach (HSZ) gauge~\cite{Hohm:2013jaa}, 
\be \label{eq:agpfhsz}
    e^{}_\tau{}^1 = e^{}_\sigma{}^1 = 1\,,
        \qquad%
    e^{}_\tau{}^ 0 = \chi/2\,,
        \qquad%
    e^{}_\sigma{}^0 = -\chi/2\,,
\ee
with a finite $\chi$\,. 
This HSZ gauge choice can be achieved by fixing the worldsheet gauge symmetries~\eqref{eq:infwt} accordingly, which yields the residue symmetries, 
\be \label{eq:rsas}
    \delta \chi = \xi^\alpha \, \p_\alpha \chi + \frac{1}{2} \, \chi \, \bar{\p} \bar{\xi} + \theta \, \chi\,,
        \qquad%
    \beta = - \frac{1}{4} \, \chi \, \p \bar{\xi}\,, 
        \qquad%
    \theta = - \p \xi\,,
        \qquad%
    \bar{\p} \xi = 0\,.
\ee
Here, $\p = \p_\tau + \p_\sigma$\,, $\xi = \xi^\tau + \xi^\sigma$, and $\bar{\xi} = \xi^\tau - \xi^\sigma$\,.
Further fixing $\bar{\xi}$ such that $\chi = 0$ brings the HSZ gauge~\eqref{eq:agpfhsz} to the ambitwistor string gauge~\eqref{eq:agpf}, while Eq.~\eqref{eq:rsas} implies
\be
    \beta = 0\,,
        \qquad%
    \theta = -\p \xi\,,
        \qquad%
    \bar{\p} \xi = 0\,.
\ee
From~\eqref{eq:infwt3}, we find the residual gauge transformation
$\delta X^\mu = \frac{1}{2} \, \xi \, \p X^\mu$\,,
under which the ambitwistor string action~\eqref{eq:ambia} is invariant on-shell. 

We have seen that ambitwistor string action can be derived from ILST tensionless string theory by taking the ambitwistor string gauge choice. This singular gauge seems to be mostly benign at the classical level. However, as we have discussed, at the quantum level, the ambitwistor string is associated with a rather distinct vacuum, where the creation and annihilation operators are interchanged~\cite{Casali:2016atr, Bagchi:2020fpr}. Therefore, ambitwistor string theory is physically distinct from the ILST tensionless, rather than simply a gauge choice. 

\subsection{Carrollian String Theory} \label{sec:cst}

It is natural to also consider T-duality transformations along spacelike circles in the M(-1)T string action~\eqref{eq:smot0}. We will show that this procedure leads to strings in target space equipped with Carroll-like geometry, where a collection of spacelike directions are absolute while the rest directions, which include a timelike direction, transform nontrivially under a Carroll-like boost. See~\cite{udlstmt, longpaper} for the target space perspective of such Carrollian string theories. 

\vspace{3mm}

\subsubsection{Spacelike T-duality of Matrix (-1)-brane theory} 

We start with the M(-1)T string action~\eqref{eq:smot0} in conformal gauge, which we transcribe below:
\be
	S_\text{M(-1)T} = \frac{T}{2} \int d^2 \sigma \, \p_\tau X^\mu \, \p_\tau X_\mu\,.
\ee
Compactify the spatial directions $X^{A'}$, with $A' = 1\,, \cdots, \, q$ for a positive integer $q$\,. We dualize each of these circles by gauging the isometries, which results in the gauged action,
\be
	S_\text{gauged} = \frac{T}{2} \int d^2 \sigma \, \Bigl( \p_\tau X^A \, \p_\tau X_A + D_\tau X^{A'} D_\tau X^{A'} - 2 \, \tilde{X}^{A'} \epsilon^{\alpha\beta} \p_\alpha v^{A'}_\beta \Bigr)\,,
\ee
where $D^{}_\tau X^{A'} = \p^{}_\tau X^{A'} + v^{A'}_\tau$. Here, $A = 0\,, \, q+1\,, \, \cdots, \, 9$\,. The Lagrange multiplier $\tilde{X}^{A'}$ guarantees that $v^{A'}_\alpha$ are pure gauge. Integrating out $v^{A'}_\tau$ gives rise to the dual action,
\be \label{eq:mmpt0}
	\tilde{S} = \frac{T}{2} \int d^2 \sigma \, \Bigl( - \p_\sigma \tilde{X}^{A'} \p_\sigma \tilde{X}^{A'} + \p_\tau X^A \, \p_\tau X_A + \lambda^{}_{A'} \, \p_\tau \tilde{X}^{A'} \Bigr)\,.
\ee
Dropping the tildes, we write the dual action as
\be \label{eq:mmpt}
	S_\text{M(-$q$-1)T} = \frac{T}{2} \int d^2 \sigma \, \Bigl( - \p_\sigma X^{A'} \p_\sigma X^{A'} + \p_\tau X^A \, \p_\tau X_A + \lambda^{}_{A'} \, \p_\tau X^{A'} \Bigr)\,,
\ee
which generalizes M$p$T to $p = - q - 1 < -1$. M(-1)T can be thought of as the special case with $q = 0$\,. The target space now develops a codimension-(10-$q$) foliation structure. The action~\eqref{eq:mmpt} is invariant under the Carroll-like boost transformation,
\be \label{eq:pbranecb}
    \delta^{}_\text{\scalebox{0.8}{C}} X^A = \Lambda^A{}_{A'} \, X^{A'}\,,
        \qquad%
    \delta^{}_\text{\scalebox{0.8}{C}} X^{A'} = 0\,,
        \qquad%
    \delta^{}_\text{\scalebox{0.8}{C}} \lambda^{A'} = - 2 \, \Lambda^A{}^{}_{A'} \, \p_\tau X^{A'}\,,
\ee
where the spatial sector with the index $A'$ is absolute. 
In the case where $p = -10$\,, \emph{i.e.}~$q = 9$\,, we have $A = 0$ and $A' = 1\,, \cdots, 9$\,, and Eq.~\eqref{eq:pbranecb} implies that 
\be
    \delta X^0 = \Lambda^0{}_{\!A'} \, X^{A'}\,,
        \qquad%
    \delta X^{A'} = 0\,. 
\ee
This is the conventional Carrollian boost for Carrollian particles~\cite{levy1965nouvelle, sen1966analogue, Duval:2014uoa}, which we refer to as the 0-brane Carrollian boost. Note that, however, the fundamental degrees of freedom are associated with nine-dimensional S(pacelike)-branes in M(-10)T. Along these lines, we refer to the transformation~\eqref{eq:pbranecb} as the (9-$q$)-brane Carrollian boost, in which case the $q$-dimensional S-branes capture the fundamental degrees of freedom. 

We have noted that the fundamental degrees of freedom in M(-1)T are captured by the D-instantons, whose dynamics is described by the IKKT Matrix theory. As we have hinted earlier, these D-instantons are now T-dualized to S-branes~\cite{Hull:1998vg, Gutperle:2002ai}. Such T-duals of IKKT Matrix theory on a vanishing $q$-torus give rise to new Matrix theory on a stack of $q$-dimensional S-branes describing the fundamental degrees of freedom in M(-$q$-1)T, where $q > 0$\,. Such S-branes are localized in time but extending along $q$ spatial directions. It would be interesting to study in the future how Carrollian field theories could be defined on certain brane configurations in such M$p$T with $p < -1$\,. The relation to IKKT Matrix theory on D-instantons via spacelike T-dualities suggest that such Carrollian QFTs acquire nonperturbative features, which may help us understand the pathological behaviors that arise from the perturbative quantization of Carrollian field theories~\cite{Figueroa-OFarrill:2023qty, deBoer:2023fnj}. 

The studies of such Carrollian string theories might eventually be relevant to flat space holography, where the asymptotic flat spacetime symmetries constitute the BMS group. There are two different approaches towards the construction of flat space holography~\cite{Susskind:1998vk, Polchinski:1999ry}, which have different proposals for what the boundary field-theoretical descriptions of the bulk quantum gravity in four-dimensional asymptotically flat spacetime might be: the \emph{celestial holography} program~\cite{deBoer:2003vf, Pasterski:2016qvg} proposes that the holographic dual of the bulk four-dimensional quantum gravity is a two-dimensional celestial conformal field theory, while the \emph{Carrollian holography} program~\cite{Dappiaggi:2005ci, Bagchi:2016bcd} proposes that the holographic dual is a three-dimensional conformal Carrollian field theory. The relation between these two different proposals was initiated in~\cite{Donnay:2022aba}. It is therefore natural to suspect that it might be possible to embed Carrollian holography within the top-down string-theoretical framework proposed in this paper. The close interplay between Matrix theory and the relevant decoupling limits~\cite{udlstmt} that lead to Carrollian string theories may thus be useful for the study of flat space holography.  

\subsubsection{Decoupling limits in type II\texorpdfstring{${}^*$}{star} superstring theory} 

For self-containedness, we briefly review the decoupling limit of string theory that leads to M$p$T with $p < -1$, which is discussed in~\cite{udlstmt}.   
In flat spacetime, M$p$T with $p < -1$ arises from the $\omega \rightarrow \infty$ limit of type II superstring theory with the following reparametrizations of the embedding coordinates $\hat{X}^\mu$, dilaton $\hat{\Phi}$\,, and RR $q$-form $\hat{C}^{(q)}$: 
\begin{subequations} \label{eq:lpmptnp}
\begin{align}
    	\hat{X}^A & = \frac{X^A}{\sqrt{\omega}}\,,
    		\qquad%
	 \hat{X}^{A'} = \sqrt{\omega} X^{A'}, \label{eq:xaxap} \\[2pt]
    	\hat{\Phi} & = \Phi + \frac{i \pi}{2} + \frac{q-4}{2} \, \ln \omega\,, \\[6pt]
    	\hat{C}^{(q)} & = \omega^2 \, e^{-\Phi} \, dX^1 \wedge \cdots \wedge dX^q,
        		\qquad\quad\,\,%
    	p + q = - 1\,,
\end{align}
\end{subequations}
Here, $A = 0\,, \, q+1\,, \, \cdots, \, 9$ and $A' = 1\,, \, \cdots,\, q$\,. The other RR potentials $\hat{C}^{(r)} = 0$\,, $r\neq q$ and the $B$-field are set to zero. The imaginary term~in the reparametrization of the dilaton $\hat{\Phi}$ indicates that the resulting M$p$T theory is of type II${}^*$\,. Here, it is the RR $q$-form coupled to the $q$-dimensional S-branes in the II${}^*$ theory that becomes critical. This also implies that, in the case of tensionless string theory where $p = -1$\,, the associated decoupling limit is defined via the following reparametrization of the IIB background fields:
\begin{align} \label{eq:tll}
	\hat{X}^\mu & = \frac{X^\mu}{\sqrt{\omega}}\,, 
		\qquad%
	\hat{\Phi} = \Phi + \frac{i \pi}{2} - 2 \, \ln \omega\,, 
		\qquad%
	\hat{C}^{(0)} = \frac{\omega^2}{e^{\Phi}}\,.
\end{align}
Here, the prescriptions for $\hat{X}^\mu$ and $\hat{C}^{(0)}$ match the ones in Eq.~\eqref{eq:rpmmot} while the dilaton $\hat{\Phi}$ matches Eq.~\eqref{eq:complphi}.
The background RR zero-form potential associated with the D-instanton is brought to its critical value in M(-1)T. This is consistent with that the light excitations in M(-1)T are the D-instantons, whose dynamics is described by the IKKT Matrix theory. 

\vspace{3mm}

\noindent $\bullet$~\emph{Carrollian string actions revisited.} Plugging the first line of the reparametrization~\eqref{eq:lpmptnp} into the conventional Polyakov string action leads to the Carrollian string action~\eqref{eq:mmpt}. This derivation is very similar to what we have discussed in Section~\ref{sec:ngadl}, where we found the M$p$T string with $p \geq 0$ from a decoupling limit. We repeat this similar derivation here for completeness, starting with the conventional Polyakov formulation,
\be \label{eq:relpf02}
    \hat{S}_\text{P} = - \frac{T}{2} \int d^2 \sigma \, \sqrt{-\hat{h}} \, \hat{h}^{\alpha\beta} \, \p^{}_\alpha \hat{X}^\mu \, \p^{}_\beta \hat{X}^{}_\mu\,.
\ee
Following Eq.~\eqref{eq:repahhab} from Section~\ref{sec:ngadl}, we reparametrize the worldsheet metric $\hat{h}_{\alpha\beta}$ as
\be \label{eq:wsrp}
	\hat{h}^{}_{\alpha\beta} = - \omega^{-2} \, e^{}_\alpha{}^0 \, e^{}_\beta{}^0 + e^{}_\alpha{}^1 \, e^{}_\beta{}^1\,,
\ee
Together with the reparametrization~\eqref{eq:xaxap} of the embedding coordinates, we find
\begin{align} \label{eq:hspmpt0c}
\begin{split}
	\hat{S}^{}_\text{P} = - \frac{T}{2} \int d^2 \sigma \, e \, \biggl[ & - e^\alpha{}^{}_0 \, e^\beta{}^{}_0 \, \Bigl( \omega^2 \, \p^{}_{\alpha} X^{A'} \, \p^{}_{\beta} X^{A'} + \p^{}_{\alpha} X^{A} \, \p^{}_{\beta} X_{A} \Bigr) \\[4pt]
	& + e^\alpha{}^{}_1 \, e^\beta{}^{}_1 \, \Bigl( \p^{}_{\alpha} X^{A'} \, \p^{}_{\beta} X^{A'} + \omega^{-2} \, \p^{}_{\alpha} X^{A} \, \p^{}_{\beta} X_{A} \Bigr) \biggr] \,.
\end{split}
\end{align}
Rewriting the $\omega^2$ divergence using the Hubbard-Stratonovich transformation, we find
\begin{align} \label{eq:hspmpt2c}
\begin{split}
	\hat{S}^{}_\text{P} = & - \frac{T}{2} \int d^2 \sigma \, e \, \Bigl( e^\alpha{}^{}_1 \, e^\beta{}^{}_1 \, \p_{\alpha} X^{A'} \p_{\beta} X^{A'} - e^\alpha{}^{}_0 \, e^\beta{}^{}_0 \, \p^{}_\alpha X^{A} \, \p^{}_\beta X_{A} - \lambda^{}_{A'} \, e^\alpha{}^{}_0 \, \p^{}_\alpha X^{A'} \Bigr) \\[4pt]
    	& - \frac{T}{2} \int d^2 \sigma \, \omega^{-2} \, e \, \Bigl( e^\alpha{}^{}_1 \, e^\beta{}^{}_1 \, \p^{}_\alpha X^{A} \, \p^{}_\beta X_{A} + \tfrac{1}{4} \, \lambda^{}_{A'} \, \lambda^{A'} \Bigr) \,,
\end{split}
\end{align}
where we have integrated in the auxiliary field $\lambda_{A'}$\,. In the $\omega \rightarrow \infty$ limit, we find the Carrollian string action, 
\begin{align} \label{eq:hspmpt2cc}
\begin{split}
	S^{}_\text{P} = & \frac{T}{2} \int d^2 \sigma \, e \, \Bigl( - e^\alpha{}^{}_1 \, e^\beta{}^{}_1 \, \p_{\alpha} X^{A'} \p_{\beta} X^{A'} + e^\alpha{}^{}_0 \, e^\beta{}^{}_0 \, \p^{}_\alpha X^{A} \, \p^{}_\beta X_{A} + \lambda^{}_{A'} \, e^\alpha{}^{}_0 \, \p^{}_\alpha X^{A'} \Bigr)\,.
\end{split}
\end{align}
This same action also arises from undoing the conformal gauge in Eq.~\eqref{eq:mmpt}, which we derived from T-dualizing the M(-1)T string. Integrating out the auxiliary worldsheet geometric data in this Carrollian string as in Section~\ref{sec:ngadl} leads to the Nambu-Goto formulation, 
\be \label{eq:ngfcs}
	S^{}_\text{NG} = - T \int \! d^2 \sigma \, \sqrt{- \epsilon^{\alpha\beta} \, \epsilon^{\gamma\delta} \bigl( \p_{\alpha} X^A \, \p_{\gamma} X^{}_A \bigr) \, \bigl( \p^{}_{\beta} X^{A'} \p^{}_{\delta} X^{A'} \bigr)} - T \int  d X^{A'} \wedge dX^{B'}  \lambda^{}_{A'B'}\,.
\ee
Here, $\lambda^{}_{A'B'}$ is an anti-symmetric two-tensor. 
Note that the discussion here about the fundamental string action is in form the same as in Section~\ref{sec:ngadl}, but with the roles played by $X^A$ and $X^{A'}$ swapped. We reiterate that M$p$T with $p < 0$ is of type II${}^*$. This make it fundamentally different from M$p$T with $p \geq 0$\,, where the latter is of type II. 

\vspace{3mm}

\noindent $\bullet$~\emph{Spacelike branes.} We have noted earlier that the light excitations in M$p$T with $p < 0$ are conceptually distinct from the ones in M$p$T with $p \geq 0$\,. When $p \geq 0$\,, the light excitations are the D$p$-branes, which arise from the critical RR $(p+1)$-form limit of the D$p$-branes in type II superstring theory. In contrast, when $p < -1$\,, the light excitations become $q$-dimensional S-branes satisfying $p + q = - 1$, which arise from bringing the RR $q$-form to its critical value as in Eq.~\eqref{eq:lpmptnp} in the type II${}^*$ theory. To demonstrate how such a decoupling limit works for $p < 0$\,, we start with the D$q$-brane coupled to a non-trivial RR $q$-form in type II superstring theory, 
\be \label{eq:hsdp0}
    	\hat{S}_{\text{D}q} = - T_{q} \int d^{q+1} \sigma \, e^{-\hat{\Phi}} \, \sqrt{-\det \Bigl( \p_\alpha \hat{X}^\mu_{\phantom{\dagger}} \, \p_\beta \hat{X}_\mu + F_{\alpha\beta} \Bigr)} + T_{q} \int \hat{C}^{(q+1)}\,.
\ee
Recall that the complexified dilaton in Eq.~\eqref{eq:lpmptnp} implies that we are transferred to the type II${}^*$ theory, where the D-brane is mapped to an S-brane. In static gauge, we align the S-brane with the spatial sector with the index $A'$, by setting $X^{A'} = \sigma^{A'}$, with $A' = 1\,, \, \cdots, \, q$\,. Plugging the reparametrization~\eqref{eq:lpmptnp} into the above brane action, we find the following finite action in the $\omega \rightarrow \infty$ limit:
\begin{align}
	S_{\text{S}q} = - \frac{T_{q}}{2} \int d^{q+1} \sigma \, \Bigl( \p^\alpha X^{A} \, \p_\alpha X_A \! + F^{\alpha\beta} \, F_{\alpha\beta} \Bigr)\,.
\end{align}
This action describes the $q$-dimensional S-brane in M(-\,$q$\,-1)T. It would be interesting to understand the Matrix theory describing a stack of coinciding S-branes, which is supposed to encode the dynamics of the associated Carrollian string theory. 

\subsubsection{Ambitwistor string gauge} 

As in Section~\ref{sec:ast}, it is also possible to impose the ambitwistor string gauge in the M$p$T string sigma model~\eqref{eq:mmpt}. For this purpose, we start with discussing the phase-space formulation of Carrollian string theory, which can be derived from the Nambu-Goto formulation~\eqref{eq:ngfcs} and takes the following form:
\begin{align} \label{eq:psanp}
    S^{}_\text{p.s.} = \int d^2 \sigma \, \biggl[ P^{}_\mu \, \p^{}_\tau X^\mu - \frac{\chi}{2 \, T} \Bigl( P^A P^{}_A + T^2 \, \p^{}_\sigma X^{A'} \p^{}_\sigma X^{A'} \Bigr) - \rho \, P^{}_\mu \, \p^{}_\sigma X^\mu \biggr].
\end{align}
Recall that $A = 0\,, \, q+1\,, \, \cdots, \, 9$ and $A' = 1\,, \, \cdots, \, 9$\,. The ``particle'' case with $q = 9$ and ``string'' case with $q = 8$ of this phase-space action have appeared in~\cite{Cardona:2016ytk}. 
Integrating out $P_\mu$\,, $\chi$\,, and $\rho$ in this phase-space action leads to the Nambu-Goto action~\eqref{eq:ngfcs}. 
As expected, plugging the Carrollian parametrization~\eqref{eq:wscge0} of the worldsheet into this phase-space action followed by integrating out $P_\mu$ reproduces the Polyakov action~\eqref{eq:mmpt}. 

Choosing the ambitwistor string gauge~\eqref{eq:ag} with $\chi = 0$ and $\rho = 1$ in the phase-space action~\eqref{eq:psanp}, we find the following chiral action:
\be \label{eq:chiralcarroll}
    S_\text{chiral} = \frac{T}{2} \int d^2 \sigma \, P^{}_\mu \, \bar{\p} X^\mu\,,
        \qquad%
    P^A \, P^{}_A + \tfrac{1}{4} \, T^2 \, \p X^{A'} \, \p X^{A'} = 0\,.
\ee
This chiral action is manifestly Carroll invariant. The target space Carroll-like boost acts nontrivially on both $X^\mu$ and the conjugate momentum $P_\mu$\,, 
\be \label{eq:cbxp}
    \delta^{}_\text{\scalebox{0.8}{C}} X^A = \Lambda^A{}_{A'} \, X^{A'}\,,
        \qquad%
    \delta^{}_\text{\scalebox{0.8}{C}} X^{A'} = 0\,,
        \qquad%
    \delta^{}_\text{\scalebox{0.8}{C}} P_A = 0\,,
        \qquad%
    \delta^{}_\text{\scalebox{0.8}{C}} P_{A'} = - \Lambda^A{}_{A'} \, P_{A}\,,
\ee
where $P_A$ and $P_{A'}$ transform as $\p_A$ and $\p_{A'}$, respectively. 
Under the transformation~\eqref{eq:cbxp}, the chiral action and the constraint in Eq.~\eqref{eq:chiralcarroll} are invariant. 
Similarly, it is also possible to write the chiral action associated with the M$p$T string action~\eqref{eq:hspmpt20} with $p \geq 0$\,. In the resulting chiral string theory, the chiral action in Eq.~\eqref{eq:chiralcarroll} remains unchanged but the constraint there is now replaced with
\be \label{eq:congpp}
    P^{}_{A'} \, P^{}_{A'} + \tfrac{1}{4} \, T^2 \, \p X^A \, \p X^{}_A = 0\,.
\ee
The chiral string theory is invariant under the target space Galilei-like boost,
\be
    \delta^{}_\text{\scalebox{0.8}{G}} X^A = 0\,,
        \qquad%
    \delta^{}_\text{\scalebox{0.8}{G}} X^{A'} = \Lambda^{A'}{}_{\!A} \, X^A\,,
        \qquad%
    \delta^{}_\text{\scalebox{0.8}{G}} P_A = - \Lambda^{A'}{}_A \, P_{A'}\,,
        \qquad%
    \delta^{}_\text{\scalebox{0.8}{G}} P_{A'} = 0\,.
\ee
Such chiral string theories in M$p$T with $p \neq -1$ generalize ambitwistor string theory associated with M(-1)T, where the constraint is $P^\mu P_\mu = 0$ in the latter theory. It would be interesting to further understand what these new ambitwistor-like strings might imply for field theories. For example, the constraint in M0T is 
\be
    P^{}_{A'} P^{}_{A'} = \tfrac{1}{4} \, T^2 \, \p X^0 \p X^0\,. 
\ee
Gauge fixing $X^0 = \tau + \sigma$ gives rise to the constraint $P^{}_{A'} P^{}_{A'} = T^2$ as in Section~\ref{sec:wsgf}, where we showed that this is related to the dynamics of geometric optics~\cite{Batlle:2017cfa}. In the presence of spins, this corner may be associated with the dynamics of spinoptics and the optical Hall effect~\cite{Duval:2005ry, Duval:2013aza, bliokh2006conservation, onoda2004hall}. For another example, in M(-2)T, the Hamiltonian constraint becomes 
\be
    - P_0^2 + P^{}_u \, P^{}_u + \tfrac{1}{4} \, T^2 \, \p X^1 \, \p X^1 = 0\,,
        \quad%
    u = 2\,, \, \cdots, \, 9\,.
\ee
Gauge fixing $X^1 = \tau + \sigma$\,, we find the dispersion relation $P_0^2 - P^{}_u \, P^{}_u = T^2$\,, which describes a massive relativistic particle. This new variety of dispersion relations may imply alternative CHY-like formulae that correspond to a larger landscape of field-theoretical observables, \emph{e.g.}~possibly the ones associated with geometric optics and massive particles. 

\subsection{Other T-Dual Relations} \label{sec:otdr}

\noindent $\bullet$~\emph{Timelike T-duality from M$p$T to M(-$p$\,-1)T.} We have discussed the timelike T-duality transformation from M0T to M(-1)T, \emph{i.e.}~tensionless string theory, and then moved on to the spacelike T-duality transformations from M(-1)T to M$p$T with $p < -1$\,, \emph{i.e.}~Carrollian string theories. The road map in Figure~\ref{fig:tltd} also implies that M$p$T and M(-$p$\,-1)T are T-dual to each other along a timelike isometry. We will demonstrate this relation below. 

We start with the M$p$T string action~\eqref{eq:mmpt} with $p < -1$\,. We transcribe such a Carrollian string action below:
\be \label{eq:mmpt2}
	S_\text{M(-$q$-1)T} = \frac{T}{2} \int d^2 \sigma \, \Bigl( - \p_\sigma X^{A'} \p_\sigma X^{A'} + \p_\tau X^A \, \p_\tau X_A + \lambda^{}_{A'} \, \p_\tau X^{A'} \Bigr)\,,
\ee
where $p + q = - 1$\,, $A = 0\,, \, q+1\,, \, \cdots, 9$\,, and $A' = 1\,, \, \cdots, \, q$\,. Gauging the time isometry in $X^0$ leads to the following gauged action:
\begin{align} \label{eq:gaugedscs}
\begin{split}
	S_\text{gauged} = \frac{T}{2} \int d^2 \sigma \, \Bigl( - \p_\sigma X^{A'} \p_\sigma X^{A'} & + \p_\tau X^{\CA'} \, \p_\tau X^{\CA'} + \lambda^{}_{A'} \, \p_\tau X^{A'} \\[2pt]
		& - D_\tau X^0 \, D_\tau X^0 - 2\, \tilde{X}^0 \, \epsilon^{\alpha\beta} \, \p_\alpha v_\beta \Bigr)\,,
\end{split}
\end{align}
where $D_\tau X^0 = \p_\tau X^0 + v_\tau$ and $\CA' = q+1\,, \, \cdots, \, 9$\,. Integrating out $v_\tau$ in this gauged action gives the T-dual theory,
\be \label{eq:dscmqt}
	\tilde{S} = \frac{T}{2} \int d^2 \sigma \, \Bigl( \p_\sigma \tilde{X}^0 \, \p_\sigma \tilde{X}^0 - \p_\sigma X^{A'} \p_\sigma X^{A'} + \p_\tau X^{\CA'} \, \p_\tau X^{\CA'} + \tilde{\lambda}^{}_0 \, \p_\tau \tilde{X}^0 + \lambda^{}_{A'} \, \p_\tau X^{A'} \Bigr)\,,
\ee
where $\tilde{\lambda}_0 = 2 \, v_\sigma$\,. Dropping the tildes in this dual action, we find the M$q$T string action,
\be \label{eq:dscmqtf}
	S^{}_\text{M$q$T} = \frac{T}{2} \int d^2 \sigma \, \Bigl( - \p_\sigma X^{\CA} \, \p_\sigma X^{}_{\CA} + \p_\tau X^{\CA'} \, \p_\tau X^{\CA'} + \lambda^{}_{\CA} \, \p_\tau X^{\CA} \Bigr)\,,
\ee
where $\CA = 0\,, \cdots, \, q$\,. The inverse of this duality transformation maps M$p$T with $p \geq 0$ to M(-$p$\,-1)T. In the special case where $p = 0$\,, this gives the T-dual relation between M0T and M(-1)T that we have introduced in Section~\ref{eq:ikktts}. 

\vspace{3mm}

\noindent $\bullet$~\emph{From ambitwistor to Carrollian string theory.} Another T-dual relation that we have not yet explicitly discussed is the one between two different chiral string theories. This relation is already implied by the T-dual relation between two different M$p$Ts. In the following, as an explicit example, we derive the T-duality transformation from ambitwistor string theory to Carrollian string theory in ambitwistor string gauge. We start with the ambitwistor string action~\eqref{eq:ambia},
\be \label{eq:sambic}
    S^{}_\text{ambi.} = \int d^2 \sigma \, \Bigl( P_\mu \, \bar{\p} X^\mu + e \, P_\mu \, P^\mu \Bigr)\,.
\ee
We have incorporated the constraint $P_\mu \, P^\mu = 0$ by introducing the Lagrange multiplier $e$ as in~\cite{Mason:2013sva}. Split the embedding coordinates as $X^\mu = (X^A, X^{A'})$\,, where $A = 0\,, \, \cdots, \, p$ and $A' = p+1\,, \, \cdots, \, 9$\,. Compactify each of the $X^{A'}$ directions over a spatial circle. We perform T-duality transformations along all the $X^{A'}$ circles. This is done by gauging the isometries $X^{A'}$ in the ambitwistor string action~\eqref{eq:sambic}, which gives the equivalent action
\be \label{eq:spambc}
    S_\text{gauged} = \int d^2 \sigma \, \biggl[ P^{}_A \, \bar{\p} X^A + P^{}_{A'} \, \bar{D} X^{A'} + e \, P^{}_{\mu} \, P^\mu + \tfrac{1}{2} \, T \, \tilde{X}^{A'} \Bigl( \bar{\p} {v}^{A'} - \p \bar{v}^{A'} \Bigr) \biggr]\,,
\ee
where $\bar{D} X^{A'} = \bar{\p} X^{A'} + \bar{v}^{A'}$. 
Integrating out $\bar{v}^{}_{A'}$ imposes the constraint
\be \label{eq:pap}
    P^{}_{A'} = - \tfrac{1}{2} \, T \, \p \tilde{X}^{A'}.
\ee
Redefine $v^{}_{\!A'} = - 2 \, \bigl( \tilde{P}^{}_{\!A'} / T \bigr) - \p X^{A'}$ and plug Eq.~\eqref{eq:pap} into the gauged action, we find that the dual action is
\be
    S^{}_\text{Carroll} = \int d^2 \sigma \, \biggl[ P^{}_A \, \bar{\p} X^A + \tilde{P}^{}_{A'} \, \bar{\p} \tilde{X}^{A'} + e \, \Bigl( P^{}_A \, P^A + \tfrac{1}{4} \, T^2 \, \p \tilde{X}^{A'} \p \tilde{X}^{}_{A'} \Bigr) \biggr].
\ee
This is precisely the chiral string action~\eqref{eq:chiralcarroll}, which is Carrollian string theory in ambitwistor string gauge . 
%

\section{Lightlike T-Duality} \label{sec:tnddlcq}

\begin{figure}[b!]
\centering
\begin{adjustbox}{width=0.7\textwidth}
\hspace{-8mm}
\begin{tikzpicture}
\begin{scope}[scale=.85]
 
        	\path[every node/.style=draw, rounded corners=1, line width=1.5pt, minimum width=135pt, minimum height=20pt, font = \small]   
	(0,1) [align=center] node[minimum height = 30pt] {\textbf{M}atrix (\textbf{\emph{p}}+1)-brane \textbf{T}heory\\
		\emph{in the DLCQ}}
	(4,3.5) [align=center] node[minimum height = 30pt] {\textbf{M}ulticritical \textbf{M}atrix \textbf{\emph{p}}-brane \textbf{T}heory\\
	(\emph{generalized SMT string})}
	(8,1) [align=center] node[minimum height = 30pt] {\textbf{M}atrix (-\textbf{\emph{p}}\,-1)-brane \textbf{T}heory\\
	\emph{in the DLCQ}}
	(4,6) [align=center] node[minimum width=171pt, minimum height = 30pt] {\textbf{M}ulticritical \\ 
	\textbf{M}atrix \textbf{(\emph{p}+1)}-brane \textbf{T}heory}
	;

        \path (4,4.75) [font=\footnotesize] node {\textbf{T-dual}}
        (1.6,2.25) [font=\footnotesize] node {\textbf{T-dual}}
        (6.4,2.25) [font=\footnotesize] node {\textbf{T-dual}}
        (4,2.25) [font=\footnotesize] node {\emph{in lightlike circle}}
        (0.45,2.25) [font=\footnotesize] node {\emph{$X^1$}}
        (7.55,2.25) [font=\footnotesize] node {\emph{$X^0$}}
        (2.85,4.75) [font=\footnotesize] node {\emph{$X^u$}}
        (5.3,4.8) [font=\footnotesize] node {\emph{$X^{A'}$}}
        ;

	\draw [{[length=1.3mm]}-{>[length=1.3mm]}] (3.2,5.3) -- (3.2,4.2);
	\draw [{<[length=1.3mm]}-{[length=1.3mm]}] (4.8,5.3) -- (4.8,4.2);
	\draw [{[length=1.3mm]}-{>[length=1.3mm]}] (0.8,2.8) -- (0.8,1.7);
	\draw [{<[length=1.3mm]}-{[length=1.3mm]}] (2.4,2.8) -- (2.4,1.7);
	\draw [{<[length=1.3mm]}-{[length=1.3mm]}] (5.6,2.8) -- (5.6,1.7);
	\draw [{[length=1.3mm]}-{>[length=1.3mm]}] (7.2,2.8) -- (7.2,1.7);
 
\end{scope}
\end{tikzpicture}
\end{adjustbox}
\caption{T-duality relations between Multicritical Matrix $p$-brane theory (MM$p$T), DLCQ Matrix ($p$\,+1)-brane theory, and DLCQ Matrix (-$p$\,-1)-brane theory. We will see later in Section~\ref{sec:smtnh} that the fundamental string in MM$p$T generalizes the Spin Matrix Theory (SMT) string~\cite{Harmark:2018cdl}. Here, $X^0$ and $X^1$ are longitudinal to the background F1-string, $X^0$ and $X^u$ are longitudinal to the background D$p$-brane, and $X^{A'}$ are transverse to the background F1-D$p$ configuration.}
\label{fig:mmtltd} 
\end{figure}
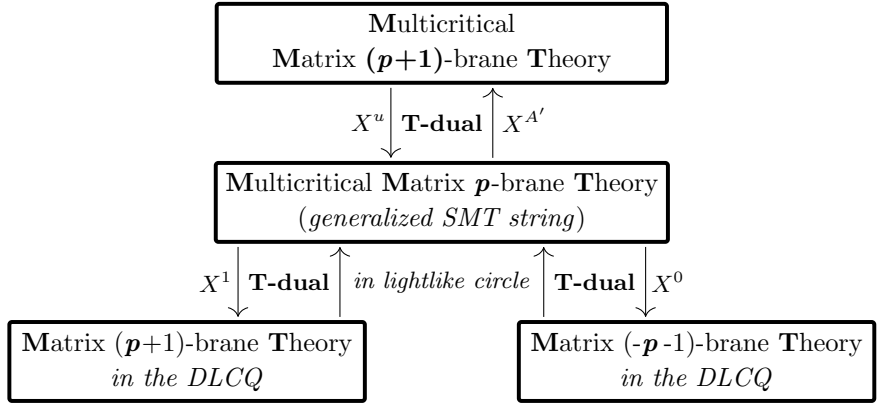

Let us briefly recap what we have done so far. We started with DLCQ M-theory in Section~\ref{sec:mtnlsw}, where M-theory is compactified along a lightlike circle. This procedure leads to the type IIA superstring description that we refer to as Matrix 0-brane theory (M0T), in which the BFSS Matrix theory lives on the D0-branes. We then focused on the fundamental string in M0T, which has a nonrelativistic worldsheet with the topology being the nodal Riemann sphere. Then, we built up a duality web as illustrated in Figs.~\ref{fig:tdmpt} and~\ref{fig:tltd} by performing spacelike and timelike T-duality transformations of the M0T string action~\eqref{eq:m0tspf0}. 

It is also intriguing to observe that a second DLCQ in M$p$T becomes possible, as long as $p \neq 0$\,. This procedure leads to another layer of the duality web that is essentially the DLCQ of the DLCQ of the type II string duality web~\cite{udlstmt, bpslimits}. We now study this duality web from the second DLCQ of M-theory by considering the lightlike T-duality transformation of the M$p$T string sigma model. See Fig.~\ref{fig:mmtltd} for a summary of the results in this section.  

\subsection{The Second DLCQ} \label{sec:sdlcq}

We start with Polyakova formulation~\eqref{eq:m0tst} of the M$p$T string, and write the M($p$+1)T string action with $p\geq0$ as
\begin{align} \label{eq:m1tst2}
    S_{\text{M($p$+1)T}} & = \frac{T}{2} \int d^2 \sigma \, \Bigl( - \p^{}_\sigma X^A \, \p^{}_\sigma X_A + \p^{}_\tau X^{A'} \p^{}_\tau X^{A'} 
    + \lambda^{}_A \, \p^{}_\tau X^A \biggr)\,,
\end{align}
where $A = 0\,, \, \cdots,\, p+1$ and $A' = p+2\,, \, \cdots, \, 9$\,. 
Split the longitudinal coordinates as $X^A = (X^0, \, X^1, \, X^u)$\,, $u = 2\,, \cdots, \, p+1$\,, and then introduce the light-cone variables,
\be
    X^\pm = \tfrac{1}{\sqrt{2}} \, \bigl( X^0 \pm X^1 \bigr)\,,
        \qquad%
    \lambda_\pm = \tfrac{1}{\sqrt{2}} \bigl( \lambda_0 \pm \lambda_1 \bigr)\,.
\ee
We consider M($p$+1)T in the DCLQ, where the lightlike direction $X^+$ is compactified. After gauging this isometry direction along $X^+$\,, the M$p$T string action becomes
\begin{align} \label{eq:gaugedmptll}
\begin{split}
    S^{}_\text{gauged} = \frac{T}{2} \int d^2 \sigma \, \Bigl( - \p^{}_\sigma X^u \, \p^{}_\sigma X^u & + \p^{}_\tau X^{A'} \p^{}_\tau X^{A'} + \lambda_- \, \p_\tau X^- + \lambda^{}_u \, \p^{}_\tau X^u \\[4pt]
    & + 2 \, \p_\sigma X^- \, D_\sigma X^+ + \lambda_+ \, D_\tau X^+ - 2 \, \epsilon^{\alpha\beta} \, \tilde{X} \, \p_\alpha v_\beta \Bigr),
\end{split}
\end{align}
where $D_\alpha X^+ = \p_\alpha X^+ + v_\alpha$\,. Integrating out $\tilde{X}$ gives back the original M($p$+1)T action. Instead, integrating out $v_\tau$ imposes the constraint
$\lambda_+ \! = 2 \, \p_\sigma \tilde{X}$\,,
and the gauged action becomes
\begin{align} \label{eq:gaugedmm}
\begin{split}
    S^{}_\text{gauged} = \frac{T}{2} \int d^2 \sigma \, \Bigl[ - \p^{}_\sigma X^u \, \p^{}_\sigma X^u & + \p^{}_\tau X^{A'} \p^{}_\tau X^{A'} + \lambda_- \, \p_\tau X^- + \lambda^{}_u \, \p^{}_\tau X^u \\[4pt]
    & + 2 \, D_\sigma X^+ \, \bigl( \p_\tau \tilde{X} + \p_\sigma X^- \bigr) - 2 \, \epsilon^{\alpha\beta} \, \p_\alpha \tilde{X} \, \p_\beta X^+ \Bigr]\,.
\end{split}
\end{align}
Define the dual embedding coordinates and Lagrange multipliers, 
\be \label{eq:tildxdef}
	\tilde{X}^0 = - X^-\,,
		\qquad%
	\tilde{X}^1 = \tilde{X},
		\qquad%
	\tilde{\lambda}_0 = - \lambda_-\,,
		\qquad%
	\tilde{\lambda}_1 = 2 \, \bigl( v_\sigma + \p_\sigma X^+ \bigr)\,.
\ee
The dual action is
\begin{align} \label{eq:mmpmot0}
\begin{split}
    	\tilde{S} = \frac{T}{2} \int d^2 \sigma \, \Bigl[ - \p^{}_\sigma X^u \, \p^{}_\sigma X^u & + \p^{}_\tau X^{A'} \p^{}_\tau X^{A'} + {\lambda}_u \, \p_\tau {X}^u \\[4pt]
	& + \tilde{\lambda}_0 \, \p_\tau \tilde{X}^0 + \tilde{\lambda}_1 \, \bigl( \p_\tau \tilde{X}^1 - \p_\sigma \tilde{X}^0 \bigr) \Bigr]\,,
\end{split}
\end{align}
where we have dropped a boundary term.
Dropping the tildes in Eq.~\eqref{eq:mmpmot0}, we write this dual action as
\be \label{eq:mmpmot}
	S_\text{MM$p$T} = \frac{T}{2} \int d^2 \sigma \, \Bigl( - \p^{}_\sigma X^u \, \p^{}_\sigma X^u + \p^{}_\tau X^{A'} \p^{}_\tau X^{A'} + \lambda_A \, \p_\tau X^A - \lambda_1 \, \p_\sigma X^0 \Bigr)\,.
\ee
This action is also found in~\cite{udlstmt} via a decoupling limit of type II superstring theory, which is complementary to the intrinsic derivation that we perform in this paper. 
This dual action defines a class of new theories that is referred to as \emph{Multicritical Matrix $p$-brane Theory} (MM$p$T) in~\cite{udlstmt}. The reason why ``multicritical'' appears in the name will become clear later in Section~\ref{sec:mfl}. 
This action enjoys a more complex set of nonrelativistic boost symmetries in the target space, with the first Galilean boost relating $X^1$ to $X^0$\,,
\be \label{eq:gxzo}
    \delta^{\scalebox{0.6}{(1)}}_\text{G} X^1 = V \, X^{0}\,,
        \qquad%
    \delta^{\scalebox{0.6}{(1)}}_\text{G}
    \lambda_0 = - V \, \lambda_1\,.
\ee
This boost acts trivially on the other worldsheet fields. Here, ${\lambda}_0$ and ${\lambda}_1$ are the conjugate momenta with respect to ${X}^0$ and ${X}^1$, respectively. There is a second Galilei-like boost that relates $X^{A'}$ to $X^0$ and $X^u$,
\begin{align} \label{eq:sndgb}
    \delta^{\scalebox{0.6}{(2)}}_\text{G} X^{A'} = \Lambda^{A'}{}_{\!0} \, {X}^0 + \Lambda^{A'}{}_{\!u} \, \tilde{X}^u\,,
        \quad%
    \delta^{\scalebox{0.6}{(2)}}_\text{G} {\lambda}_0 = - 2 \, \Lambda^{A'}{}_{\!0} \, \p_\tau X_{A'},
        \quad%
    \delta^{\scalebox{0.6}{(2)}}_\text{G} {\lambda}_u = - 2 \, \Lambda^{A'}{}_{\!u} \, \p_\tau X_{A'}.
\end{align}
The remaining worldsheet fields transform trivially under this second Galilean boost. 

Before proceeding, we reexamine the identification~\eqref{eq:tildxdef} between the variables in the two T-dual frames. We recover the tilde notation in the T-dual frame through the following discussion. So far we have been referring to $X^-$ in the original theory as the time direction $\tilde{X}^0$ in the T-dual frame, and it follows that $\tilde{X}$ is a spatial direction that we have identified with $\tilde{X}^1$. Under this choice, the transformation~\eqref{eq:gxzo} receives the interpretation as a Galilean boost. However, at least at the classical level, we also have the option of interpreting $X^-$ as a spatial direction $\tilde{X}^1$ in the T-dual frame, in which case $\tilde{X}^0 = \tilde{X}$ becomes the time direction. Further define $\tilde{\lambda}_0 = 2 \, \bigl( v_\sigma + \p_\sigma X^+ \bigr)$ and $\tilde{\lambda}_1 = \lambda^+$\,. Dropping tildes again, this second choice leads to the action
\begin{align} \label{eq:mmpmotc}
\begin{split}
    S_\text{MM$p$T*} = \frac{T}{2} \int d^2 \sigma \, \Bigl( - \p^{}_\sigma X^u \, \p^{}_\sigma X_u & + \p^{}_\tau X^{A'} \p^{}_\tau X_{A'} + \lambda_A \, \p_\tau X^A - \lambda_0 \, \p_\sigma X^1 \Bigr)\,.
\end{split}
\end{align}
Note that the subscript ``MM$p$T*'' implies that we are now dealing with a type II${}^*$ theory, which we will further elaborate later around Eq.~\eqref{eq:mmpts}. Now, a Carrollian boost relating $X^0$ to $X^1$ replaces the Galilean boost~\eqref{eq:gxzo}, with
\be \label{eq:gxzoc}
    \delta^{\scalebox{0.6}{(1)}}_\text{C} X^0 = V \, X^{1}\,,
        \qquad%
    \delta^{\scalebox{0.6}{(1)}}_\text{C}
    \lambda_1 = - V \, \lambda_0\,.
\ee
The second Galilei-like ``boost''~\eqref{eq:sndgb} now only involves spatial directions, with
\be
    \delta^{\scalebox{0.6}{(2)}}_\text{G} X^{A'} = \Lambda^{A'}{}_{\!1} \, X^1 + \Lambda^{A'}{}_{\!u} \, X^u\,,   
        \qquad%
    \delta^{\scalebox{0.6}{(2)}}_\text{G} \lambda_1 = - \Lambda^{A'}{}_{\!1} \, \p_\tau X_{A'},
        \qquad%
    \delta^{\scalebox{0.6}{(2)}}_\text{G} \lambda_u = - \Lambda^{A'}{}_{\!u} \, \p_\tau X_{A'}.
\ee
It is understood that these transformations act trivially on the rest of the embedding coordinates. 
In the rest of this section, we will mostly focus on the Galilean version of MM$p$T, whose fundamental string is described by the action~\eqref{eq:mmpmot}. Nevertheless, we will show in Section~\ref{sec:gse} that the MM$p$T${}^*$ string~\eqref{eq:mmpmotc} is useful for constructing Galilean CHY-like formulae.

\subsection{Multicritcal Field Limits} \label{sec:mfl}

In~\cite{udlstmt}, it is shown that Multicritical M$p$T (MM$p$T) arises from a decoupling limit of type II superstring theory where both the background $B$-field and RR $(p+1)$-form potential are brought to their critical values. This is why such a decoupling limit is referred to as a multicritical field limit in~\cite{udlstmt}. In flat target space, this decoupling limit is given by the following prescription in type II superstring theory: the embedding coordinates are reparametrized as
\begin{align} \label{eq:repxzouap}
	\hat{X}^0 = \omega \, X^0\,,
        		\qquad%
    	\hat{X}^1 = X^1\,,
        		\qquad%
    	\hat{X}^u = \omega^{1/2} \, X^u\,, 
        		\qquad%
    	\hat{X}^{A'} = \omega^{-1/2} \, X^{A'},
\end{align}  
where $u = 2\,, \, \cdots, \, p+1$ and $A' = p+2\,, \, \cdots, \, 9$\,. Moreover, the background dilaton $\hat{\Phi}$, $B$-field, and RR ($p$+1)-form $\hat{C}^{(p+1)}$ are reparametrized as
\begin{subequations} \label{eq:bphicppo}
\begin{align}
	\hat{B} & = - \omega \, dX^0 \wedge dX^1\,,
		\qquad%
	\hat{\Phi} = \Phi + \tfrac{1}{2} \bigl( p - 2 \bigr) \ln \omega\,, \\[4pt]
	\hat{C}^{(p+1)} & = \omega^2 \, e^{-\Phi} \, dX^0 \wedge dX^2 \wedge \cdots \wedge dX^{p+1}\,.
\end{align}
\end{subequations}
The multicritical decoupling limit is defined by sending $\omega$ to infinity in the reparametrized type II superstring theory. 

\subsubsection{Multicritical fundamental string}

Using the bosonic sector of the fundamental string action that we have been considering, we are able to access the reparametrizations of the embedding coordinates and the $B$-field in the relevant decoupling limit. To test our understanding of the limiting prescriptions in Eqs.~\eqref{eq:repxzouap} and \eqref{eq:bphicppo}, we consider the conventional Polyakov action,
\be \label{eq:hsfsmc}
    \hat{S}^{}_\text{P} = - \frac{T}{2} \int d^2 \hat{\sigma} \, \p_\alpha \hat{X}^\mu \, \p^\alpha \hat{X}_\mu - T \int \hat{B}\,,
\ee
where we have included the $B$-field term. This Chern-Simons term is a total derivative under the prescription in Eq.~\eqref{eq:bphicppo}, when the effective dilaton $\Phi$ is constant. However, this term cannot be dropped when there is string winding in the relevant spatial direction and when a more general configuration of the background fields is considered. A T-duality transformation does not change the worldsheet topology. Therefore, we expect that the worldsheet space and time scale as in Eq.~\eqref{eq:rsc}, with $\hat{\tau} = \omega^{-1} \, \tau$ and $\hat{\sigma} = \sigma$\,. Here, $\hat{\sigma}^\alpha = (\hat{\tau}\,, \, \hat{\sigma})$\,.
Together with the prescriptions of the embedding coordinates in Eq.~\eqref{eq:repxzouap} and of the $B$-field in Eq.~\eqref{eq:bphicppo}, we rewrite the conventional Polyakov action~\eqref{eq:hsfsmc} as
\begin{align} \label{eq:mmptexp}
\begin{split}
    \hat{S}^{}_\text{P} = - \frac{T}{2} \int d^2 \sigma \, \biggl[ \, \omega^3 \, \Bigl( \p_\tau X^0 - \omega^{-2} \, \p_\sigma X^1 \Bigr)^2 \!\! - \omega^2 \, \p_\tau X^u \, \p_\tau X^u - & \, \omega \, \Bigl( \p_\tau X^1 - \p_\sigma X^0 \Bigr)^2 \\[4pt]
    - \Bigl( \p_\tau X^{A'} \, \p_\tau X^{A'} - \p_\sigma X^u \, \p_\sigma X^u \Bigr) + & \, \omega^{-2} \, \p_\sigma X^{A'} \, \p_\sigma X^{A'} \biggr]\,.
\end{split}
\end{align}
Using the Hubbard-Stratonovich transformation for three times, we introduce the Lagrange multipliers $\lambda_A = (\lambda_0\,, \lambda_1\,, \lambda_u )$ to rewrite Eq.~\eqref{eq:mmptexp} equivalently as
\begin{align} \label{eq:expmmptf}
\begin{split}
    \hat{S}^{}_\text{P} \rightarrow - \frac{T}{2} \int d^2 \sigma \, \biggl[ & \p_\sigma X^u \, \p_\sigma X^u - \p_\tau X^{A'} \, \p_\tau X^{A'} \\[2pt]
    & - \lambda_0 \, \Bigl( \p_\tau X^0 - \omega^{-2} \, \p_\sigma X^1 \Bigr) - \lambda_1 \, \Bigl( \p_\tau X^1 - \p_\sigma X^0 \Bigr) - \lambda_u \, \p_\tau X^u \\[4pt]
    & - \tfrac{1}{4 \, \omega^3} \, \lambda_0^2 + \tfrac{1}{4 \, \omega} \, \lambda_1^2 + \tfrac{1}{4 \, \omega^2} \, \lambda^u \, \lambda^u + \tfrac{1}{\omega^2} \, \p_\sigma X^{A'} \, \p_\sigma X^{A'} \biggr]\,, 
\end{split}
\end{align}
where the finite terms are in form the same as in Eq.~\eqref{eq:mmpmot}. In the $\omega \rightarrow \infty$ limit, we find the MM$p$T action,
\be \label{eq:mmptfs}
	S^{}_\text{MM$p$T} = \frac{T}{2} \int d^2 \sigma \, \Bigl( - \p_\sigma X^u \, \p_\sigma X^u + \p_\tau X^{A'} \p_\tau X^{A'} + \lambda_A \, \p_\tau X^A - \lambda_1 \, \p_\sigma X^0 \Bigr)\,,
\ee
which reproduces the action~\eqref{eq:mmpmot} that we derived intrinsically from T-dualizing DLCQ M$p$T. 
Recall that $A = 0\,, \, \cdots, \, p+1$\,, $u = 2\,, \, \cdots, \, p+1$\,, and $A' = p+2\,,\, \cdots, \, 9$\,. 

\subsubsection{Multicritical D-brane} \label{sec:mdbe}

We have learned that MM$p$T and DLCQ M($p$+1)T are T-dual to each other, where the spatial $X^1$ circle in MM$p$T is mapped to the lightlike circle in DLCQ M($p$+1)T. Here, $p \geq 0$\,. We have also learned that M($p$\,+1)T arises from a critical RR ($p$\,+2)-form limit in Section~\ref{eq:rmgtsltd}. 
The critical $B$-field in the MM$p$T prescription~\eqref{eq:bphicppo} comes from T-dualizing the infinite boost limit that leads to the DLCQ of the relevant type II theory. Moreover, the critical RR ($p$\,+1)-form in the MM$p$T prescription~\eqref{eq:bphicppo} comes from T-dualizing the critical RR ($p$\,+2)-form in the decoupling limit that leads to M($p$+1)T. 
A more thorough derivation of such a decoupling limit of type II superstring theory that leads to MM$p$T is given in~\cite{udlstmt, longpaper}. Here, we content ourselves with a simple motivation for why such a multicritical limit is necessary, by focusing on the example of a probe D$p$-brane in MM$p$T. We start with the D$p$-brane action in conventional type IIB superstring theory,
\be \label{eq:hsdz}
	\hat{S}^{}_\text{D$p$} = - T^{}_\text{D$p$} \int d^{p+1} \sigma \, e^{-\hat{\Phi}} \sqrt{- \det \Bigl( \p_\alpha \hat{X}^\mu_{\phantom{\dagger}} \, \p_\beta \hat{X}_\mu + \hat{B}_{\alpha\beta} + F_{\alpha\beta} \Bigr)} + T^{}_\text{D$p$} \int \hat{C}^{(p+1)}\,,
\ee
where we set all the RR fields to zero except the RR ($p$+1)-form. Here, $\hat{B}_{\alpha\beta}$ is the pullback of the $B$-field to the worldsheet of the D$p$-brane. For simplicity, we take the ansatze for the $B$-field and RR ($p$\,+1)-form that are akin to the ones in Eq.~\eqref{eq:bphicppo}, with
\be \label{eq:bcppo}
	\hat{B} = - \CB \, \omega \, dX^0 \wedge dX^1\,,
		\qquad%
	\hat{C}^{(p+1)} = \CC \, \omega^2 \, dX^0 \wedge dX^2 \wedge \cdots \wedge dX^{p+1}.
\ee
Choose the static gauge $X^0 = \tau\,, \, X^2 = \sigma^1\,, \, \cdots, \, X^{p+1} = \sigma^p$\,, and then plug the MM$p$T prescription~\eqref{eq:repxzouap} for the embedding coordinates into the D$p$-brane action, we find the following expression at large $\omega$\,:
\begin{align} \label{eq:mclod}
\begin{split}
	\hat{S}^{}_\text{D$p$} = & \, T^{}_\text{D$p$} \int d^{p+1} \sigma \, \frac{\omega^{\frac{p+2}{2}} }{e^{\hat{\Phi}}} \, \biggl[ - 1 - \frac{\CP}{\omega} + \frac{\CP^2 + \p_\tau X^1 \, \p_\tau X^1 - \p_i X^{A'} \p_i X^{A'}}{2 \, \omega^2} \, + O\bigl(\omega^{-3}\bigr) \biggr] \\[6pt]
	& + T^{}_\text{D$p$} \int d^{p+1} \sigma \, \omega^2 \, \CC\,,
		\qquad\qquad\,%
	\CP \equiv \frac{1}{2} \, \bigl( 1 - \CB^2 \bigr) \, \p_i X^1 \, \p_i X^1\,,
\end{split}
\end{align}
where $\p_i = \p / \p\sigma^i$ with $i = 1\,, \, \cdots, \, p$ denoting the spatial directions on the worldvolume. 
For the $\omega \rightarrow \infty$ to be well defined and nontrivial, it is clear that we have to set
\be \label{eq:hphcz}
	e^{\hat{\Phi}} = \omega^{\frac{p-2}{2}} \, e^\Phi\,,
		\qquad%
	\CB^2 = 1\,,
		\qquad%
	\CC = e^{-\Phi}\,.
\ee
Plugging Eq.~\eqref{eq:hphcz} into the ansatze~\eqref{eq:bcppo}, we recover the prescription~\eqref{eq:bphicppo}. 
In the $\omega \rightarrow \infty$ limit, we find the following D$p$-brane action in MM$p$T:
\be \label{eq:dzb}
	S^{}_\text{D$p$} = \frac{T_\text{D$p$}}{2} \int d^{p+1}\sigma \, e^{-\Phi} \Bigl( \p_\tau X^1 \, \p_\tau X^1 - \p_i X^{A'} \p_i X^{A'} \Bigr)\,,
\ee
which contains a nonrelativistic particle state with its dynamics confined in the $X^1$ direction. It is important that both the $B$-field and RR ($p$+1)-form become critical, such that the divergences in $\omega$ in the D$p$-brane action~\eqref{eq:mclod} are cancelled. This means that MM$p$T is defined on the background of a critical F1-D0 (marginally) bound state\,\footnote{The light excitations in MM$p$T are $\frac{1}{4}$-BPS states~\cite{bpslimits}.}. 

On the other hand, MM$p$T${}^*$ that we have introduced in Section~\ref{sec:sdlcq} arises from the same reparametrizations in Eq.~\eqref{eq:repxzouap} of the embedding coordinates and in Eq.~\eqref{eq:bphicppo} of the other background fields, but with the replacement~\cite{Bidussi:2023rfs}
\be \label{eq:reptau01}
	X^0 \rightarrow i \, X^1\,,
		\qquad%
	X^1 \rightarrow i \, X^0\,,
\ee
which leads to the new set of reparametrizations 
\begin{subequations}
\begin{align}
	\hat{X}^0 & = i \, \omega \, X^1\,, 
		\qquad%
	\hat{X}^1 = i \, X^0\,,
		\qquad%
	\hat{X}^u = \omega^{1/2} \, X^u\,,
		\qquad%
	\hat{X}^{A'} = \omega^{-1/2} \, X^{A'}, \\[4pt]
	\hat{B} & = - \omega \, dX^0 \wedge dX^1\,, 
		\qquad\qquad\qquad\quad\!%
	\hat{\Phi} = \Phi + \tfrac{1}{2} \, \bigl( p - 2 \bigr) \, \ln \omega\,, \\[4pt]
	\hat{C}^{(p+1)} & = i \, \omega^2 \, e^{-\Phi} \, dX^1 \wedge \cdots \wedge dX^{p+1}\,.
\end{align}
\end{subequations}
Here, the critical RR ($p$+1)-form develops an imaginary $\omega$ divergence. The above parametrization can be rewritten using the equivalent prescription below:
\begin{subequations} \label{eq:mmpts}
\begin{align}
	\hat{X}^0 & = X^0\,, 
		\qquad%
	\hat{X}^1 = \omega \, X^0\,,
		\qquad%
	\hat{X}^u = \omega^{1/2} \, X^u\,,
		\qquad%
	\hat{X}^{A'} = \omega^{-1/2} \, X^{A'}, \\[4pt]
	\hat{B} & = - \omega \, dX^0 \wedge dX^1\,, 
		\qquad\qquad\qquad\!%
	\hat{\Phi} = \Phi + \tfrac{1}{2} \, i \, \pi + \tfrac{1}{2} \, \bigl( p - 2 \bigr) \, \ln \omega\,, \\[4pt]
	\hat{C}^{(p+1)} & = \omega^2 \, e^{-\Phi} \, dX^1 \wedge \cdots \wedge dX^{p+1}\,,
\end{align}
\end{subequations}
such that the $\omega \rightarrow \infty$ limit is unchanged. Here, $u = 2\,, \, \cdots, \, p+1$ and $A' = p+2\,,\, \cdots, \, 9$\,. The imaginary term in $\hat{\Phi}$ implies that we are now in the type II${}^*$ version of MM$p$T, and the critical background is a bound state that contains an F1-string and a $(p+1)$-dimensional S-brane. In view of its connection to type II${}^*$ theories, We refer to this longitudinal Carrollian version of MM$p$T as MM$p$T${}^*$.

The prescription in Eq.~\eqref{eq:bphicppo} also makes it possible to study general D-brane configurations in MM$p$T by taking the relevant limits of conventional D-brane effective actions. For example, the MM0T limit of the D2-branes in type IIA superstring theory gives rise to a Galilean version of noncommutative Yang-Mills theory~\cite{smtcft}. Such a D2-brane configuration in MM0T is T-dual to the DLCQ of D3-brane in M1T, which carries the conventional noncommutative Yang-Mills theory. See related discussion in Section~\ref{sec:zblttast} in M0T. 

\subsection{Multicritical String in Nambu-Goto Formulation}

We have discussed the MM$p$T string in the Polyakov formulation and showed that it arises from a multicritical limit of type II superstring theory. To further solidify this multicritical limit, we now examine how it is applied to the Nambu-Goto formulation and how it is consistent with the MM$p$T string action~\eqref{eq:mmpmot}. 

Undo the conformal gauge in the MM$p$T string action~\eqref{eq:mmpmot}, we find the Polyakov formulation
\begin{align} \label{eq:expmmptfc}
\begin{split}
    S^{}_\text{MM$p$T} = \frac{T}{2} \int d^2 \sigma \, e \, \Bigl( - e^\alpha{}^{}_1 \, e^\beta{}^{}_1 \, \p^{}_\alpha X^u \, \p^{}_\beta X^u & + e^\alpha{}^{}_0 \, e^\beta{}^{}_0 \, \p^{}_{\alpha} X^{A'} \p^{}_{\beta} X^{A'} \\
	& + \lambda^{}_A \, e^\alpha{}^{}_0 \, \p_\alpha X^A - \lambda^{}_1 \, e^\alpha{}^{}_1 \, \p^{}_\alpha X^0 \Bigr)\,,
\end{split}
\end{align}
where $A = 0\,, \, \cdots, \, p+1$\,, $u = 2\,, \cdots, \, p+1$\,, and $A' = p+2\,, \cdots, \, 9$\,. 
In order to derive the Nambu-Goto formulation, we integrate out the Lagrange multipliers $\lambda_A$ in this Polyakov action, which imposes the conditions
\be
	e^\alpha{}^{}_0 \, \p^{}_\alpha X^0 = e^\alpha{}^{}_0 \, \p^{}_\alpha X^u = 0\,,
		\qquad%
	e^\alpha{}^{}_0 \, \p^{}_\alpha X^1 = e^\alpha{}^{}_1 \, \p^{}_\alpha X^0\,.
\ee 
These conditions are solved by 
\be
	e^\alpha{}^{}_0 = - \Gamma^1 \, \epsilon^{\alpha\beta} \, \p^{}_\beta X^0\,,
		\quad\,%
	e^\alpha{}^{}_1 = \epsilon^{\alpha\beta} \Bigl( - \Gamma^0 \, \p^{}_\beta X^0 + \Gamma^1 \, \p^{}_\beta X^1 \Bigr)\,,
		\quad\,%
	\p^{}_\alpha X^u = \Gamma^u \, \p^{}_\alpha X^0\,,
\ee
where we have introduced the arbitrary coefficients $\Gamma^A$\,. 
Plugging these solutions back into Eq.~\eqref{eq:expmmptfc} gives rise to the Nambu-Goto-like action,
\begin{align} \label{eq:ngmmpt}
	S^{}_\text{NG} = \frac{T}{2} \int d^2 \sigma \, \biggl[ \, \tau \, \Gamma_u \, \Gamma^u - \tau^{-1} \, \Bigl( \epsilon^{\alpha\beta} \, \p_\alpha X^0 \, \p_\beta X^{A'} \Bigr)^{\!2} \, \biggr]\,.
\end{align}
Here, $\tau = \epsilon^{\alpha\beta} \, \p^{}_\alpha X^0 \, \p^{}_\beta X^1$\,.

Next, we show that the same action~\eqref{eq:ngmmpt} also arises from applying the MM$p$T limit to the Nambu-Goto formulation of the conventional string, which we write as in Eq.~\eqref{eq:mptngv0}, with
\be \label{eq:rewsng}
	\hat{S}^{}_\text{NG} = - \frac{T}{2} \int d^2 \sigma \, \biggl[ v - v^{-1} \, \det \Bigl( \p_{\alpha} \hat{X}^\mu \, \p_\beta \hat{X}_\mu \Bigr) \biggr] - T \int \hat{B}\,. 
\ee
We have turned on the $B$-field, as required by the decoupling limit for MM$p$T.  
Reparametrizing the embedding coordinates as in Eq.~\eqref{eq:repxzouap}, we find
\begin{align}
\begin{split}
    	\det \hat{G}_{\alpha\beta} & = - \omega^3 \, \bigl( \epsilon^{\alpha\beta} \, \p_\alpha X^0 \, \p^{}_\beta X^u \bigr)^2 - \omega^2 \, \Bigl[ \tau^2 - \tfrac{1}{2} \bigl( \epsilon^{\alpha\beta} \, \p_\alpha X^u \, \p_\beta X^v \bigr)^2 \Bigr]
	- \omega \, \CK + O (\omega^0)\,, \\[6pt]
	\CK & = \bigl( \epsilon^{\alpha\beta} \, \p_\alpha X^0 \, \p_\beta X^{A'} \bigr)^2 - \bigl( \epsilon^{\alpha\beta} \, \p_\alpha X^1 \, \p_\beta X^u \bigr)^2\,.
\end{split}
\end{align}
Introduce the Lagrange multiplier $\lambda_u$ and the antisymmetric two-tensor $\lambda_{uv}$ to replace
\begin{subequations}
\begin{align}
	\frac{\omega^3}{v} \, \bigl( \epsilon^{\alpha\beta} \, \p_\alpha X^0 \, \p_\beta X^u \bigr)^2 & \rightarrow \lambda^{}_u \, \epsilon^{\alpha\beta} \, \p_\alpha X^0 \, \p_\beta X^u - \frac{v}{4 \, \omega^3} \, \lambda^{}_u \, \lambda^u\,, \\[4pt]
    \frac{\omega^2}{v} \, \bigl( \epsilon^{\alpha\beta} \, \p^{}_\alpha X^u \, \p^{}_\beta X^v \bigr)^2 & \rightarrow \lambda^{}_{uv} \, \epsilon^{\alpha\beta} \, \p^{}_\alpha X^u \, \p^{}_\beta X^v - \frac{v}{4 \, \omega^2} \, \lambda^{}_{uv} \, \lambda^{uv}\,.
\end{align}
\end{subequations}
Finally, use Eq.~\eqref{eq:bphicppo} to write $\hat{B} = - \omega \, dX^0 \wedge dX^1$ and then integrate $v$ out, we find that Eq.~\eqref{eq:rewsng} becomes
\begin{align}
	\hat{S}^{}_\text{NG} & = - T \int d^2 \sigma \, \sqrt{\omega^2 \, \tau^2 + \omega \, \CK + O \bigl(\omega^0\bigr)} \notag \\[4pt]
	& \quad - T \int \Bigl( - \omega \, dX^0 \wedge dX^1 + \lambda_u \, dX^0 \wedge dX^u + \lambda^{}_{uv} \, dX^u \wedge dX^v \Bigr)\,. 
\end{align}
In the $\omega \rightarrow \infty$ limit, we find the Nambu-Goto action for the MM$p$T string,
\begin{align} \label{eq:sngmmptf}
\begin{split}
	S^{}_\text{MM$p$T} = & \frac{T}{2} \int d^2\sigma \, \frac{1}{\tau} \, \biggl[ \Bigl( \epsilon^{\alpha\beta} \, \p_\alpha X^1 \, \p_\beta X^u \Bigr)^2 - \Bigl( \epsilon^{\alpha\beta} \, \p_\alpha X^0 \, \p_\beta X^{A'} \Bigr)^2 \biggr] \\[4pt]
	& - T \int \Bigl( \lambda_u \, dX^0 \wedge dX^u + \lambda_{uv} \, dX^u \wedge dX^v \Bigr)\,.
\end{split}
\end{align}
Further integrating out $\lambda^{}_u$ and $\lambda^{}_{uv}$ imposes the constraints,
\be
	\epsilon^{\alpha\beta} \, \p_\alpha X^0 \, \p_\beta X^u = 0\,,
		\qquad%
	\epsilon^{\alpha\beta} \, \p_\alpha X^u \, \p_\beta X^v = 0\,,
\ee
which are solved by $\p_\alpha X^u = \Gamma^u \, \p_\alpha X^0$ for an arbitrary function $\Gamma^u$\,. Plugging this solution back into Eq.~\eqref{eq:sngmmptf} reproduces Eq.~\eqref{eq:ngmmpt}. 

The phase-space formulation for the Nambu-Goto action~\eqref{eq:sngmmptf} is
\be \label{eq:psammpt}
	S_\text{p.s.} = \int d^2\sigma \, \biggl[ P^{}_\mu \, \p^{}_\tau X^\mu - \frac{\chi}{2 \, T} \Bigl( P^{}_{A'} \, P^{}_{A'} + 2 \, T \, P^{}_1 \, \p^{}_\sigma X^0 + T^2 \, \p^{}_\sigma X^u \, \p^{}_\sigma X^u \Bigr) - \rho \, P^{}_\mu \, \p^{}_\sigma X^\mu \biggr]\,.
\ee
Integrating out $P_\mu$\,, $\chi$\,, and $\rho$ in this phase-space action leads back to the Nambu-Goto action~\eqref{eq:ngmmpt}. 
Instead, after integrating out $P^{}_{A'}$ in Eq.~\eqref{eq:psammpt}, and under the changes of variables in Eq.~\eqref{eq:wscge0}, \emph{i.e.} $\chi = e / \bigl(e_\sigma{}^1 \bigr){}^2$ and $\rho = e_\tau{}^1 / e_\sigma{}^1$\,, 
together with the redefinitions
\begin{subequations}
\begin{align}
	P^{}_0 & = \frac{T}{2} \, \Bigl( \lambda^{}_0 \, e^{}_\sigma{}^1 + \lambda^{}_1 \, e^{}_\sigma{}^0 \Bigr)\,, 
		\qquad%
	P^{}_1 = \frac{T}{2} \, \lambda^{}_1 \, e^{}_\sigma{}^1\,, \\[4pt]
	P^{}_u & = \frac{T}{2} \, \biggl[ \lambda^{}_u \, e^{}_\sigma{}^1 + e^{}_\sigma{}^0 \biggl( e^\alpha{}^{}_1 \, \tau^{}_\alpha{}^u + \frac{\tau^{}_\sigma{}^u}{e^{}_\sigma{}^1} \biggr) \biggr]\,,
\end{align}
\end{subequations}
we find that the Polyakov action~\eqref{eq:expmmptfc} for the MM$p$T string is recovered. 

The above results can be easily transferred to the longitudinal Carrollian version~\eqref{eq:mmpmotc}, \emph{i.e.} MM$p$T${}^*$, by using the map~\eqref{eq:reptau01}. See detailed discussions in~\cite{Bidussi:2023rfs} for the case of MM$0$T${}^*$. 

\subsection{T-Duality of Strings in Multicritical Matrix \texorpdfstring{$p$}{p}-Brane Theory} \label{sec:tdmmpt}

To further understand the relation between MM$p$T and DLCQ M$q$T, we now study the T-duality transformations of the MM$p$T string action. Here, $p \geq 0$ and $q \neq 0$\,. We will first study the inverse of the duality transformation in Section~\ref{sec:sdlcq}, and show that the longitudinal spatial T-duality transformation in the $X^1$ isometry maps MM$p$T to DLCQ M($p$\,+1)T. We will also show that the timelike T-duality transformation in the $X^0$ isometry maps MM$p$T to DLCQ M(-$p$\,-1)T. This allows us to prove the related statements in~\cite{udlstmt}, intrinsically using the worldsheet formalism. Furthermore, the longitudinal T-duality transformation in an $X^u$ isometry maps MM$p$T to MM($p$\,-1)T, while the transverse T-duality transformation in an $X^{A'}$ isometry maps MM$p$T to MM($p$\,+1)T. This latter set of T-dualities along the $X^u$ and $X^{A'}$ directions are very similar to the ones considered in Section~\ref{sec:fdlcq}, so we will not repeat these derivations here. Instead, we will focus on the T-dualizing $X^1$ and $X^0$ in MM$p$T.

A summary of these T-duality transformations has been given in Fig.~\ref{fig:mmtltd}.  In the decoupling limit that leads to MM$p$T, the target space coordinates scale in $\omega$ as prescribed in Eq.~\eqref{eq:repxzouap}, which we also transcribe in the following table: 
\vspace{3mm}
\begin{center}
\begin{tabular}{c | c | c | c | c }
\hline
	& $\mathbf{X^0}$ & $\mathbf{X^1}$ & $\mathbf{X^u \,\, (u = 2, \cdots, p+1)}$ & $\mathbf{X^{A'} \,\, (A' = 2, \cdots, p+1)}$ \\ 
 		\hline
	& $\omega$ & 1 & $\omega^{1/2}$ & $\omega^{-1/2}$ \\ 
		 \hline
	\textbf{F1-string} & $\times$ & $\times$ & &  \\ 
		\hline
	\textbf{D\emph{p}-brane} & $\times$ & & $\times$ & \\
		\hline
\end{tabular}
\end{center} 
\vspace{3mm}
We indicate in the table that MM$p$T arises from the decoupling limit with a critical background (marginally) bound F1-D$p$ state that consists of an F1-string extending in the $X^1$ directions and a D$p$-brane extending in the $X^u$ directions.

\vspace{3mm}

\noindent $\bullet$~\emph{Spacelike T-duality from MM$p$T to M($p$+1)T.} We start with T-dualizing $X^1$\,. Gauging the isometry in $X^1$, the MM$p$T action~\eqref{eq:expmmptfc} becomes
\begin{align} \label{eq:gaugedmmptf}
\begin{split}
	S^{}_\text{gauged} = - \frac{T}{2} \int d^2 \sigma \, \Bigl( \p_\sigma X^u \, \p_\sigma X^u - \p_\tau X^{A'} \, \p_\tau X^{A'} & - \lambda_u \, \p_\tau X^u - \lambda_0 \, \p_\tau X^0 + \lambda_1 \, \p_\sigma X^0 \\[4pt]
	& - \lambda_1 \, D_\tau X^1 - 2 \, \tilde{X}^+ \epsilon^{\alpha\beta} \p_\alpha v_\beta \Bigr)\,,
\end{split}
\end{align}
where $u = 2\,, \cdots,\, p+1$\,. 
Here, $D_\alpha X^1 = \p_\alpha X^1 + v_\alpha$ and $\tilde{X}^+$ is the Lagrange multiplier imposing that $v_\alpha$ is pure gauge. %
Integrating out $v_\tau$ imposes the condition $\lambda_1 = 2 \, \p_\sigma \tilde{X}^+$\,. Further define $\lambda_- = - \lambda_0$\,, $\lambda_+ = 2 \, (v_\sigma + \p_\sigma X^1 )$\,, and $\tilde{X}^- = X^0$\,, we find the dual action
\begin{align} \label{eq:mmptdf}
\begin{split}
    S_\text{dual} = - \frac{T}{2} \int d^2 \sigma \, \Bigl( - 2 \, \p_\sigma \tilde{X}^- \, \p_\sigma \tilde{X}^+ & + \p_\sigma X^u \, \p_\sigma X^u - \p_\tau X^{A'} \p_\tau X^{A'} \\[4pt]
    & - \lambda_- \, \p_\tau \tilde{X}^- - \lambda_+ \, \p_\tau \tilde{X}^+ - \lambda_u \, \p_\tau X^u \Bigr)\,,
\end{split}
\end{align}
where we have omitted a topological term associated with winding Wilson lines. Define
\begin{align} \label{eq:redxpml}
	\tilde{X}^\pm & = \frac{1}{\sqrt{2}} \, \bigl( \tilde{X}^0 \pm \tilde{X}^1 \bigr)\,,
		&%
	\lambda_\pm & = \frac{1}{\sqrt{2}} \, \bigl( \tilde{\lambda}_0 \pm \tilde{\lambda}_1 \bigr)\,,
\end{align}
we find that the dual action~\eqref{eq:mmptdf} becomes, 
\begin{align} \label{eq:smppotd}
	S^\text{\scalebox{0.8}{DLCQ}}_\text{M($p$\,+1)T} = \frac{T}{2} \int d^2 \sigma \, \Bigl( - \p_\sigma \tilde{X}^A \, \p_\sigma \tilde{X}_A + \p_\tau X^{A'} \, \p_\tau X^{A'} 
		+ \tilde{\lambda}_A \, \p_\tau \tilde{X}^A \Bigr)\,,
\end{align}
Here, $A = 0\,, \cdots, \, p+1$\,, $\tilde{X}^A = (\tilde{X}^0, \, \tilde{X}^1, \, X^u)$\,, and $\tilde{\lambda}_A = (\tilde{\lambda}_0\,, \, \tilde{\lambda}_1\,, \, \lambda_u)$\,. This is in form the M($p$\,+1)T string action~\eqref{eq:m0tst}, with the dual coordinate $\tilde{X}^0$ being the time direction and $\tilde{X}^1$ a spatial direction. The lightlike coordinate $\tilde{X}^-$ is compactified over a circle of radius dual to the original $X^1$ radius in MM$p$T. We thus conclude that this dual theory is DLCQ M($p$\,+1)T.

\vspace{3mm}

\noindent $\bullet$~\emph{Timelike T-duality from MM$p$T to M(-$p$\,-1)T.} We now move on to dualize $X^0$ in MM$p$T. Gauging the isometry time direction $X^0$\,, the MM$p$T string action~\eqref{eq:expmmptfc} becomes
\begin{align} \label{eq:gaugedmmptf2}
\begin{split}
	S^{}_\text{gauged} = - \frac{T}{2} \int d^2 \sigma \, \Bigl( \p_\sigma X^u \, \p_\sigma X^u & - \p_\tau X^{A'} \, \p_\tau X^{A'} - \lambda_u \, \p_\tau X^u - \lambda_1 \, \p_\tau X^1 \\[4pt]
		& - \lambda_0 \, D_\tau X^0 + \lambda_1 \, D_\sigma X^0 - 2 \, \tilde{X}^- \epsilon^{\alpha\beta} \p_\alpha v_\beta \Bigr)\,.
\end{split}
\end{align}
Here, $D_\alpha X^0 = \p_\alpha X^0 + v_\alpha$\,. Integrating out $v_\alpha$ gives $\lambda_0 = 2 \, \p_\sigma \tilde{y}$ and $\lambda_1 = 2 \, \p_\tau \tilde{y}$\,, which leads to the dual action,
\be \label{eq:dlcqmmpt}
	S^{}_\text{dual} = \frac{T}{2} \int d^2 \sigma \, \Bigl( - \p_\sigma X^u \, \p_\sigma X^u - 2 \, \p_\tau \tilde{X}^+ \, \p_\tau \tilde{X}^- + \p_\tau X^{A'} \p_\tau X^{A'} + \lambda_u \, \p_\tau X^u \Bigr)\,,
\ee
where $\tilde{X}^+ = X^1$\,. Note that $u=2\,, \cdots\,, \, p+1$\,, which contains $p$ spatial directions. Further define
\be
	\tilde{X}^\pm = \frac{1}{\sqrt{2}} \bigl( \tilde{X}^0 \pm \tilde{X}^1 \bigr)\,,
		\qquad%
	\tilde{X}^\CA = \bigl( X^0\,, \, X^1\,, \, X^{A'} \bigr)\,,
		\qquad%
	X^{\CA'} = X^u\,,  
\ee
we find that the dual action~\eqref{eq:dlcqmmpt} now becomes
\be
	S^\text{\scalebox{0.8}{DLCQ}}_\text{M(-$p$\,-1)T} = \frac{T}{2} \int d^2 \sigma \, \Bigl( - \p_\sigma X^{\CA'} \, \p_\sigma X^{\CA'} + \p_\tau X^{\CA} \, \p_\tau X^{}_{\CA} + \lambda^{}_{\CA'} \, \p_\tau X^{\CA'} \Bigr)\,.
\ee
This is in form the M(-$p$\,-1)T string action that we have introduced in Eq.~\eqref{eq:mmpt}, with the dual circle being lightlike along $\tilde{X}^-$ direction. Therefore, we find that this dual action describes the DLCQ M(-$p$\,-1)T string. 

\vspace{3mm}

As expected, T-dualizing M($p$\,+1)T and M(-$p$\,-1) with $p\geq0$ along a lighlike isometry gives back MM$p$T. The above T-dual relations allow us to define DLCQ M$p$T using MM$p$T, where the latter theory arises from a multicritical field limit of type II superstring theory and does \emph{not} involve any lightlike circle. As we have discussed in Section~\ref{sec:mdbe}, this multicritical field limit is a BPS limit that zooms in on a background F1-D$p$ configuration in type II superstring theory. 

\subsection{Chiral Worldsheet and Galilean Scattering Equation} \label{sec:gse}

Finally, we consider the phase-space action~\eqref{eq:psammpt} with $p = 0$ for the MM0T (and MM0T${}^*$) string, now in the ambitwistor string gauge (see Section~\ref{sec:ast}). This consideration will allow us to propose a new scattering equation that encodes the kinematics of Galilean particles. 

We start with recording the phase-space action for the MM0T string below:
\be \label{eq:psammpt1}
	S_\text{p.s.} = \int d^2\sigma \, \biggl[ P^{}_\mu \, \p^{}_\tau X^\mu - \frac{\chi}{2 \, T} \Bigl( P^{}_{A'} \, P^{}_{A'} + 2 \, T \, P^{}_1 \, \p^{}_\sigma X^0 \Bigr) - \rho \, P^{}_\mu \, \p^{}_\sigma X^\mu \biggr]\,.
\ee
In the ambitwistor gauge, we have $\chi = 0$ and $\rho = 1$\,, which leads to the following chiral worldsheet theory:
\be \label{eq:mmptca}
	S^{}_\text{chrial} = \int d^2 \sigma \, P_\mu \, \bar{\p} X^\mu\,,
		\qquad%
	P_{A'} \, P_{A'} + T \, P^{}_1 \, \p X^0 = 0\,.
\ee
In the static gauge with $X^0 = \tau + \sigma$\,, the dispersion relation takes the form $P^{}_1 = P_{A'} \, P_{A'} / (2 \, T)$\,, which does not involve the energy. This is reminiscent of what we have discussed at the end of Section~\ref{sec:wsgf} for the zero modes of the M0T string. However, using the map~\eqref{eq:reptau01} with $X^0 \rightarrow i \, X^1$ and $X^1 \rightarrow i \, X^0$\,, together with $P_0 \rightarrow - i \, P_1$ and $P_1 \rightarrow - i \, P_0$\,,
we find that the MM$0$T${}^*$ string action takes the same form as the chiral action~\eqref{eq:mmptca}, but now the Hamiltonian constraint becomes
\be
	P^{}_{A'} P^{}_{A'} + T \, P^{}_0 \, \p X^1 = 0\,. 
\ee
Compactifying the spatial direction $X^1$ is over a circle of radius $R$\,, we find that the zero mode of $X^1$ satisfies $\p_\sigma X^1 = w \, R$\,, with $w$ the winding number encoding how many times the close string wraps around the $X^1$ circle. We then find the Galilean dispersion relation for the zero modes,
\be
	P^{}_0 = \frac{P^{}_{\!A'} \, P^{}_{\!A'}}{2 \, T \, w \, R}\,.
\ee
Following the same derivation that leads to the scattering equation~\eqref{eq:scatteringequation}, we find a Galilean version below:
\be \label{eq:gvse}
    	\sum_{j \neq i} \frac{w^{(i)} \, \varepsilon^{(j)} + w^{(j)} \, \varepsilon^{(i)} - k^{(i)} \cdot k^{(j)}}{\sigma^{}_i - \sigma^{}_j} = 0\,,
\ee
supplemented with the conservation laws for the energy $\varepsilon^{(i)} = 2 \, T \, R \, P^{(i)}_0$\,, the spatial momentum $k^{(i)}_{A'} = P^{(i)}_{A'}$\,, and the winding number $w^{(i)}$\,. Here, the superscript ``$(i)$" refers to the Galilean particle label.  
The appearance of such Galilean scattering equation in MM0T${}^*$ is expected: similarly as we have explained in Section~\ref{sec:tdmmpt}, MM0T${}^*$ is T-dual to DLCQ M(-1)T, where the time direction in MM$p$T${}^*$ is mapped to a lightlike direction in DLCQ M(-1)T. Moreover, we have also shown in Section~\ref{sec:ast} that the scattering equation~\eqref{eq:scatteringequation} arises from the ambitwistor sector of M(-1)T. Therefore, the new scattering equation~\eqref{eq:gvse} is simply the DLCQ version of its relativistic counterpart~\eqref{eq:scatteringequation}. For example, if $|w^{(i)}| = m$ for all the particles, then Eq.~\eqref{eq:gvse} describes the scatterings between Galilean particles with mass $m$\,, and the conservation of string windings translates to the particle number conservation. It is therefore expected that replacing the relativistic scattering equation in the CHY formulae with Eq.~\eqref{eq:gvse} leads to a stringy way of computing field-theoretical amplitudes with Galilean symmetries. 

\section{Generalization to Curved Backgrounds} \label{sec:gcb}

Until now, we have discussed different classes of decoupling limits in type II superstring theory, focusing on the fundamental string in flat target space. These decoupling limits are defined by reparametrizing the embedding coordinates together with various background gauge fields. We have classified these decoupling limits by T-dualizing the non-vibrating string sigma model in flat target space. We have shown that the non-vibrating string resides in Matrix 0-brane theory (M0T), whose light excitations are captured by the D0-branes instead of the fundamental string. The dynamics of these light-excited M0T D0-branes is described by the BFSS Matrix theory. In this section, we generalize the results obtained in this paper to arbitrary curved backgrounds. Due to the foliation structure that commonly exists in the target space under these decoupling limits, the target space geometry becomes non-Riemannian. 

\subsection{Non-Vibrating Strings in Background Fields}

We start with the non-vibrating string, \emph{i.e.}~the M0T string, as the simplest example. We have derived its Polyakov string formulation in Eq.~\eqref{eq:m0tspf0}, which reads
\begin{align} \label{eq:m0tspf00}
    S^{}_\text{M0T} & = \frac{T}{2} \int d^2 \sigma \, e \, \biggl[ \Bigl( e^\alpha{}^{}_1 \, \p^{}_\alpha X^0 \Bigr)^2 + \Bigl( e^\alpha{}^{}_0 \, \p^{}_\alpha X^i \Bigr)^2 + \lambda \, e^\alpha{}^{}_0 \, \p^{}_\alpha X^0 \biggr]\,.
\end{align}
This action is invariant under the global Galilei symmetry, which includes the temporal and spatial translations parametrized by $\Theta^0$ and $\Theta^i$, respectively, spatial rotations parametrized by $\Lambda^{ij}$, and Galilean boost parametrized by $\Lambda^i$. These transformations act on the embedding coordinates (infinitesimally) as
\be \label{eq:gt}
	\delta X^0 = \Theta^0\,,
		\qquad%
	\delta X^i = \Theta^i + \Lambda^{ij} X^j + \Lambda^{i} \, X^0\,,
\ee  
supplemented with an appropriate transformation of the Lagrange multiplier $\lambda$\,. Define the generators associated with the temporal and spatial translations, spatial rotations, and Galilean boost as $H$, $P_i$\,, $J_{ij}$\,, and $G_i$\,, respectively, we find that they constitute the Galilei algebra defined by the following non-vanishing commutators:
\begin{subequations} \label{eq:ggcr}
\begin{align}
	\bigl[ G_i\,, \, H \bigr] & = P_i\,, \\[4pt]
	\bigl[ P_i\,, \, J_{jk} \bigr] & = \delta_{ij} \, P_k - \delta_{ik} \, P_j\,, \\[4pt]
	\bigl[ G_i\,, \, J_{jk} \bigr] & = \delta_{ij} \, G_k - \delta_{ik} \, G_j\,, \\[4pt]
	\bigl[ J_{ij}\,, \, J_{k\ell} \bigr] & = \delta^{}_{jk} \, J^{}_{i\ell} - \delta^{}_{ik} \, J^{}_{j\ell} + \delta^{}_{i\ell} \, J^{}_{jk} - \delta^{}_{j\ell} \, J^{}_{ik}\,.
\end{align}
\end{subequations}
Since the M0T string action is exactly invariant under the Galilean boost, instead of being invariant up to a total derivative, it seems that there is \emph{no} central extension, which implies the vanishing commutator $[G_i\,, \, P_j] = 0$\,. However, the Galilei algebra is extended to be the Bargmann algebra when we also consider the M0T D0-particle. Undo the static gauge in Eq.~\eqref{eq:sdzic}, we find the D0-particle is described by the action 
\be \label{eq:dzfb}
	S^{}_\text{D0} = \frac{1}{2 \, g_\text{s} \, \sqrt{\alpha'}} \int d\tau \, \frac{\dot{X}^i \, \dot{X}^i}{\dot{X}^0}\,.
\ee
Here, $g_s = e^\Phi$ is the string coupling, which we take to be constant for now, and $\dot{X}^\mu = \p_\tau X^\mu$\,. Under the Galilean boost $\delta X^i = \Lambda^i \, X^0$\,, the D0-particle action is invariant up to a boundary term, and the commutator between the spatial momentum and the Galilean boost generators now receives a central extension, with
\be \label{eq:ce}
	\bigl[ G_i\,, \, P_j \bigr] = \delta_{ij} \, N\,,
\ee
where $N$ corresponds to the conservation law of the particle number. The Galilei algebra is now extended to be the Bargmann algebra, which is defined by the non-vanishing generators in Eqs.~\eqref{eq:ggcr} and \eqref{eq:ce}. 

Gauging the Bargmann algebra then gives rise to various curved geometric background fields in the target space. In particular, we denote the target space gauge fields associated with the temporal and spatial translations as $\tau_\mu$ and $E_\mu{}^i$\,, respectively, with $\tau_\mu$ temporal vielbein and $E_\mu{}^i$ the spatial vielbein. We will also call $\tau_\mu$ the longitudinal vielbein and $E_\mu{}^i$ the transverse vielbein, from the perspective of the background D0-particle in the definition of M0T as a decoupling limit of type IIA superstring theory. These vielbein fields transform under the Galilean boost as
\be
	\delta^{}_\text{\scalebox{0.8}{G}} \tau^{}_\mu = 0\,,
		\qquad%
	\delta^{}_\text{\scalebox{0.8}{G}} E^{}_\mu{}^i = \Lambda^i \, \tau^{}_\mu\,.
\ee
The central charge $N$ is associated with a gauge potential $m_\mu$\,, which also transforms non-trivially under the Galilean boost,
\be
	\delta^{}_\text{\scalebox{0.8}{G}} m^{}_\mu = \Lambda^i \, E^{}_\mu{}^i\,. 
\ee
The mass operator $N$ generates an additional symmetry transformation parametrized by $\Sigma$\,, which acts non-trivially on $m^{}_\mu$ as
\be \label{eq:dnmm}
	\delta^{}_\text{\scalebox{0.8}{N}} m^{}_\mu = D_\mu \Sigma\,.
\ee
Here, $D_\mu$ is covariantized with respect to the background dilatation symmetry~\cite{Bergshoeff:2021bmc}. This dilatation symmetry has the origin from the decoupling limit that leads to M0T, which is defined by the $\omega \rightarrow \infty$ limit of type IIA superstring theory parametrized as in Eq.~\eqref{eq:npresxzigc}. This limit remains unchanged when $\omega$ is replaced with $\omega \, \Delta$\,, with $\Delta$ an arbitrary function. Consequently, the dilatation transformations are given by
\be \label{eq:dtt}
	e^\Phi \rightarrow \Delta^{-\frac{3}{2}} \, e^\Phi\,,
		\qquad%
	\tau^{}_\mu \rightarrow \Delta^{\frac{1}{2}} \, \tau^{}_\mu\,,
		\qquad%
	E^{}_\mu{}^i \rightarrow \Delta^{-\frac{1}{2}} \, E^{}_\mu{}^i\,,
		\qquad%
	m^{}_\mu \rightarrow \Delta^{-\frac{3}{2}} \, m^{}_\mu\,,
\ee
while all the other fields transform trivially under the dilatation symmetry. Moreover, following the Carrollian parametrization of the worldsheet defined via Eq.~\eqref{eq:wsgg0}, we also find the induced dilatation weights for the worldsheet zweibein fields, with
\be \label{eq:dttew}
	e^0 \rightarrow \Delta^{-1} \, e^0\,,
		\qquad%
	e^1 \rightarrow e^1\,.
\ee
Since the gauge parameter $\Sigma$ in Eq.~\eqref{eq:dnmm} has the same dilatation weight as $m^{}_\mu$\,, we find $D_\mu \Sigma = \p_\mu \Sigma - \Sigma \, \p_\mu \Phi$\,. The above vielbein fields together with the gauge field $m_\mu$ encode the \emph{Newton-Cartan geometry}~\cite{Hartong:2022lsy}, which is used to covariantize Newtonian gravity. 

In terms of the background fields that we have introduced, we find that the M0T string action in arbitrary background fields is given by
\begin{align} \label{eq:smztcb}
    S^{}_\text{M0T} & = \frac{T}{2} \int d^2 \sigma \, e \Biggl[ \bigl( e^\alpha{}^{}_1 \, \tau^{}_\alpha \bigr)^2 + e^\alpha{}^{}_0 \, e^\beta{}^{}_0 \, H^{}_{\alpha\beta} + \lambda \, e^\alpha{}^{}_0 \, \tau^{}_\alpha \Biggr] - T \int B\,,
\end{align}
Here, $\CT_\alpha = \p_\alpha X^\mu \, \CT_\mu$ defines the pullbacks of any background field to the string worldsheet, and we have defined the manifestly boost invariant quantity
\be
	H^{}_{\mu\nu} = E^{}_{\mu\nu} - \tau^{}_\mu \, m^{}_\nu - \tau^{}_\nu \, m^{}_\mu\,,
\ee
where $E^{}_{\mu\nu} = E^{}_\mu{}^i \, E^{}_\nu{}^i$\,. 
However, $H_{\mu\nu}$ is not invariant under the central charge transformation, which implies that the Lagrange multiplier $\lambda$ must transform as
\be
	\delta^{}_\text{\scalebox{0.8}{N}} \lambda = 2 \, e^\alpha{}^{}_0 \, D^{}_{\!\alpha} \Sigma\,,
\ee
such that $S_\text{M0T}$ is invariant off-shell under the Bargmann symmetry. Now, $\lambda$ is invariant under the Galilean boost. 
The $B$-field is introduced in the action~\eqref{eq:smztcb} in the standard way, and it transforms trivially under the Bargmann symmetry. The M0T string is also invariant under the dilatation transformations in Eqs.~\eqref{eq:dtt} and \eqref{eq:dttew} upon assigning an appropriate dilatation weight to the Lagrange multiplier $\lambda$\,, with
$\lambda \rightarrow \Delta^{-\frac{1}{2}} \, \lambda$\,. 

Similarly, we find that the curved target space generalization of the M0T D0-particle action~\eqref{eq:dzfb} is given by
\be \label{eq:dzfbcb}
	S^{}_\text{D0} = \frac{1}{2 \, \sqrt{\alpha'}} \int d\tau \, e^{-\Phi} \, \frac{\dot{X}^\mu \, \dot{X}^\nu \, H_{\mu\nu}}{\dot{X}^\mu \, \tau_\mu} + \frac{1}{\sqrt{\alpha'}} \int C^{(1)}\,,
\ee 
where the D0-particle is also coupled to the RR one-form background field $C^{(1)}$\,. 
This action is also invariant under the Bargmann symmetry transformations. In particular, under the central charge transformation, this D0-particle action is invariant up to a boundary term. 

In the above representation, the M0T string and D0-particle action are built using the Galilean boost invariant quantity $H_{\mu\nu}$, which is not invariant under the central charge transformation~\eqref{eq:dnmm}. Upon the redefinition of the Lagrange multiplier $\lambda$ in the M0T string action~\eqref{eq:smztcb} as $\lambda \rightarrow \lambda + 2 \, e^\alpha{}^{}_0 \, m^{}_\alpha$\,, we write the M0T string action as
\be \label{eq:smztcb1}
	S^{}_\text{M0T} = \frac{T}{2} \int d^2 \sigma \, e \Biggl[ \bigl( e^\alpha{}^{}_1 \, \tau^{}_\alpha \bigr)^2 + e^\alpha{}^{}_0 \, e^\beta{}^{}_0 \, E^{}_{\alpha\beta} + \lambda \, e^\alpha{}^{}_0 \, \tau^{}_\alpha \Biggr] - T \int B\,.
\ee
Now, $\lambda$ also transforms nontrivially under the Galilean boost as $\delta_\text{\scalebox{0.8}{G}} \lambda = - 2 \, e^\alpha{}^{}_0 \, E^{}_\alpha{}^i \, \Lambda^i$\,. Similarly, replacing $C^{(1)} \rightarrow C^{(1)} - e^{-\Phi} \, m^{}_\mu \, dX^\mu$\,, the D0-particle action~\eqref{eq:dzfbcb} becomes
\be \label{eq:dzfbcb1}
	S^{}_\text{D0} = \frac{1}{2 \, \sqrt{\alpha'}} \int d\tau \, e^{-\Phi} \, \frac{\dot{X}^\mu \, \dot{X}^\nu \, E_{\mu\nu}}{\dot{X}^\mu \, \tau_\mu} + \frac{1}{\sqrt{\alpha'}} \int C^{(1)}\,,
\ee 
where $C^{(1)}$ now transforms nontrivially under the Galilean boost as $\delta_\text{\scalebox{0.8}{G}} C^{(1)} = - dX^\mu \, E^{}_\mu{}^i \, \Lambda^i$\,.

The fact that the $m_\mu$ dependence can be absorbed into a redefinition of the Lagrange multiplier in the M0T string action is the reason why the fundamental string is not sensitive to the Bargmann extension of the symmetry algebra. This is made more manifest in the curved generalization of the Nambu-Goto formulation~\eqref{eq:snvs}, 
\begin{align}
    S^{\text{\scalebox{0.8}{NG}}}_\text{M0T}  = - T \int d^2 \sigma \, \sqrt{\epsilon^{\alpha\beta} \, \epsilon^{\gamma\delta} \, \tau^{}_\alpha \, \tau^{}_\gamma \, H^{}_{\beta\delta}} 
    = - T \int d^2 \sigma \, \sqrt{\epsilon^{\alpha\beta} \, \epsilon^{\gamma\delta} \, \tau^{}_\alpha \, \tau^{}_\gamma \, E^{}_{\beta\delta}}\,,
\end{align}
where the dependence on $m_\mu$\,, which is the gauge field associated with the central charge in the Bargmann algebra, simply drops out.

\subsection{Decoupling Limits Revisited} \label{sec:dlrd}

Before we proceed to the study of the fundamental string actions in general (multicritical) Matrix $p$-brane theories, we first review the associated decoupling limits that are derived from the target space perspective in~\cite{udlstmt}, which will provide us with the appropriate background fields to build the fundamental string actions. 

\vspace{3mm}

\noindent $\bullet$~\textbf{Matrix $0$-brane theory.} Recall the reparametrization~\eqref{eq:npresxzigc} of type IIA superstring theory in flat target space, which we transcribe below:
\be \label{eq:npresxzigc1}
	\hat{X}^0 = \omega^{1/2} \, X^0\,,
		\qquad%
	\hat{X}^i = \omega^{-1/2} \, X^i\,,
		\qquad%
	\hat{\Phi} = \Phi - \tfrac{3}{2} \, \ln \omega\,,
		\qquad%
	\hat{C}^{(1)} = \omega^2 \, e^{-\Phi} \, dX^0\,.
\ee
In the $\omega \rightarrow \infty$ limit, we are led to corner of M0T in flat spacetime. In terms of the target space geometric data, including the temporal vielbein $\tau_\mu$ and spatial vielbein $E_\mu{}^i$\,, dilaton field $\Phi$\,, $B$-field, and RR potentials $C^{(q)}$ in M0T, it is straightforward to generalize Eq.~\eqref{eq:npresxzigc1} to arbitrary background fields as follows~\cite{udlstmt}:
\be \label{eq:mztdlcb}
	\hat{G}^{}_{\mu\nu} = - \omega \, \tau^{}_\mu \, \tau^{}_\nu + \omega^{-1} \, E_{\mu\nu}\,,
		\qquad%
	\hat{\Phi} = \Phi - \tfrac{3}{2} \, \ln \omega\,,
		\qquad%
	\hat{C}^{(1)} = \omega^2 \, e^{-\Phi} \, \tau + C^{(1)}\,,
\ee
with $\hat{G}_{\mu\nu}$ the background metric, $\hat{\Phi}$ the dilaton, and $\hat{C}^{(q)}$ the RR one-form potential in the IIA theory. 
Reparametrizing the bosonic background fields in type IIA superstring theory as in~\eqref{eq:mztdlcb}, its $\omega \rightarrow \infty$ limit gives rise to M0T in curved spacetime. Since the metric $\hat{G}_{\mu\nu}$ in the IIA theory becomes singular in the $\omega \rightarrow \infty$ limit, the target space geometry becomes non-Riemannian, equipped with a codimension-one foliation structure. Formally, the curved spacetime generalization can be achieved by the replacements $dX^0 \rightarrow \tau$ and $dX^i \rightarrow E^i$\,.
\vspace{3mm}

\noindent $\bullet$~\textbf{(Multicritical) Matrix $p$-brane theory ($p\,\geq 0$).} We now consider the other decoupling limits related to the one leading to M0T via T-duality transformations. We have divided these decoupling limits into two different groups, referred to as Matrix $p$-brane Theory (M$p$T) and Multicritical Matrix $p$-brane Theory (MM$p$T) as in~\cite{udlstmt}. The related reparametrizations of type II superstring theory are given below:
\begin{enumerate}[(1)]

\item

\emph{Matrix $p$-brane theory} ($p \geq 0$). For M$p$T, in flat spacetime, the reparametrizations are given in Eqs.~\eqref{eq:mptsps} and \eqref{eq:cphicp}, with
\begin{subequations}
\begin{align}
	\hat{X}^A & = \sqrt{\omega} \, X^A,  		
		&%
	\hat{\Phi} & = \Phi + \frac{p-3}{2} \, \ln \omega\,, \\[4pt]		
	\hat{X}^{A'} & = \frac{X^{A'}}{\sqrt{\omega}}\,,	
		&%
	 \hat{C}^{(p+1)} & = \frac{\omega^2}{e^\Phi} \, dX^0 \wedge \cdots \wedge dX^p\,,
\end{align}
\end{subequations}
where $A = 0\,, \, \cdots\,, \, p$ and $A' = p+1\,, \, \cdots, \, 9$\,. Performing the $\omega \rightarrow \infty$ limit in type II superstring theory leads to the associated M$p$T. 
In curved backgrounds, we make the replacements
\be \label{eq:dxaap}
	dX^A \rightarrow \tau^A\,,
		\qquad%
	dX^{A'} \rightarrow E^{A'}.
\ee
Further define
\be \label{eq:deftaue}
	\tau^{}_{\mu\nu} = \tau^{}_\mu{}^A \, \tau^{}_\nu{}^B \, \eta^{}_{AB}\,,
		\qquad%
	E^{}_{\mu\nu} = E^{}_\mu{}^{A'} E^{}_\nu{}^{A'}.
\ee
Following~\cite{udlstmt, longpaper}, we find the curved background generalizations for M$p$T, 
\begin{subequations} \label{eq:rpmpt2}
\begin{align}
	\hat{G}^{}_{\mu\nu} & = \omega \, \tau^{}_{\mu\nu} + \frac{E^{}_{\mu\nu}}{\omega}\,, 
		&%
	 \hat{C}^{(p+1)} & = \frac{\omega^2}{e^\Phi} \, \tau^0 \wedge \cdots \wedge \tau^p + C^{(p+1)}\,, \\[4pt]
	\hat{\Phi} & = \Phi + \frac{p-3}{2} \, \ln \omega\,.
\end{align}
\end{subequations}
Taking the $\omega \rightarrow \infty$ limit in the type II theory leads to the associated M$p$T in arbitrary (bosonic) background fields.

\item

\emph{Multicritical Matrix $p$-brane theory} ($p \geq 0$). The decoupling limit that leads to MM$p$T is defined by the prescriptions in Eqs.~\eqref{eq:repxzouap} and \eqref{eq:bphicppo}, with
\begin{subequations}
\begin{align}
	  	\hat{X}^0 & = \omega \, X^0\,, 
			\qquad%
		\hat{X}^1 = X^1\,, 
			&%
		\hat{\Phi} & = \Phi + \tfrac{p-2}{2} \ln \omega\,, \\[8pt]
		\hat{X}^u & = \sqrt{\omega} \, X^u\,,
			&%
		\hat{B} & = - \omega \, dX^0 \wedge dX^1\,, \\
	 	\hat{X}^{A'} & = \frac{X^{A'}}{\sqrt{\omega}}\,,
			&%
		\hat{C}^{(p+1)} & = \frac{\omega^2}{e^\Phi} \, dX^0 \wedge dX^2 \wedge \cdots dX^{p+1}\,,
\end{align}
\end{subequations}
where $u = 2\,, \, \cdots, \, p+1$ and $A' = p+2\,, \, \cdots, \, 9$\,. Here, $\hat{B}_{\mu\nu}$ is the Kalb-Ramond field in the relevant type II theory. Following the replacements in Eq.~\eqref{eq:dxaap}, we find that the curved generalization for the above MM$p$T presecription is~\cite{udlstmt, longpaper},
\begin{subequations}
\begin{align}
	\hat{G}^{}_{\mu\nu} & = - \omega^2 \, \tau^{}_\mu{}^0 \, \tau^{}_\nu{}^0 + \tau^{}_\mu{}^1 \, \tau^{}_\nu{}^1 
		+ \omega \, \tau^{}_\mu{}^u \, \tau^{}_\nu{}^u + \omega^{-1} \, E^{}_{\mu\nu}\,, \\[2pt]
	\hat{C}^{(p+1)} & = \frac{\omega^2}{e^\phi} \, \tau^0 \wedge \tau^2 \wedge \cdots \tau^{p+1} + \omega \, \tau^0 \wedge \tau^1 \wedge C^{(p-1)} + C^{(p+1)}\,,  \\[2pt]
	\hat{C}^{(q)} & = \omega \, \tau^0 \wedge \tau^1 \wedge C^{(q-2)} + C^{(q)}\,, \quad\, q \neq p+1\,, \\[6pt]
	\hat{B} & = - \omega \, \tau^0 \wedge \tau^1 + B\,, 
		\qquad%
	\hat{\Phi} = \Phi + \tfrac{p-2}{2} \ln \omega\,.
\end{align}
\end{subequations}
Taking the $\omega \rightarrow \infty$ limit in the relevant type II theory leads to MM$p$T in arbitrary (bosonic) background fields.

\end{enumerate}

\vspace{3mm}

\noindent $\bullet$~\textbf{Matrix $p$-brane theory with $p < 0$.} The above discussion can also be further extended to M$p$T with $p < 0$\,, which in the flat target space is defined as the $\omega \rightarrow \infty$ of type II superstring theory parametrized as in Eq.~\eqref{eq:lpmptnp}, with
\begin{subequations}
\begin{align}
	\hat{X}^A & = \frac{X^A}{\sqrt{\omega}}\,,
		&%
	\hat{\Phi} & = \Phi + \frac{i \pi}{2} + \frac{q-4}{2} \, \ln \omega\,, \\[4pt]
	\hat{X}^{A'} & = \sqrt{\omega} \, X^{A'}\!,
		&%
	\hat{C}^{(q)} & = \frac{\omega^2}{e^\Phi} \, dX^1 \wedge \cdots \wedge X^q\,,
\end{align}
\end{subequations}
where $A = 0\,, \, q+1\,, \, \cdots, \, 9$\,, $A' = 1\,, \, \cdots, \, q$\,, and $q = - p - 1 \geq 0$\,. In the special case where $q = 0$\,, we have $\hat{C}^{(0)} = \omega^2 \, e^{-\Phi}$ being critical. The curved background generalization for M$p$T with $p < 0$ is then given by the following prescriptions~\cite{udlstmt, longpaper}:
\begin{subequations}
\begin{align}
	\hat{G}_{\mu\nu} & = \frac{\tau_{\mu\nu}}{\omega} + \omega \, E_{\mu\nu}\,, 
		\qquad%
	\hat{\Phi} = \Phi + \frac{i \pi}{2} + \frac{p-4}{2} \, \ln \omega\,, \\[4pt]
	\hat{C}^{(q)} & = \frac{\omega^2}{e^\Phi} \, \tau^1 \wedge \cdots \wedge \tau^q + C^{(q)}\,, 
		\quad%
	q = - p - 1\,.
\end{align}
\end{subequations}
Here, $\tau_{\mu\nu}$ and $E_{\mu\nu}$ are defined in Eq.~\eqref{eq:deftaue}, but now $A = 0\,, \, q+1\,, \, \cdots, \, 9$ and $A' = 1\,, \, \cdots, \, q$\,. The $\omega \rightarrow \infty$ limit of the relevant type II superstring theory leads to M$p$T with $p < 0$\,, which is of type II${}^*$. In the case where $p < -1$\,, the target space is Carroll-like. In the special case where $p = -1$\,, \emph{i.e.}~$q = 0$\,, we have $\hat{C}^{(0)} = \omega^2 \, e^{-\Phi}$\,, which leads to M(-1)T, or tensionless string theory, in arbitrary (bosonic) background fields. 

\vspace{3mm}

\noindent $\bullet$~\textbf{Carrollian version of multicritical Matrix $p$-brane theory ($p \geq 0$).} We have learned that MM$p$T${}^*$ arises from performing the replacement~\eqref{eq:reptau01} in M$p$T, with $X^0 \rightarrow i \, X^1$ and $X^1 \rightarrow i \, X^0$, such that the $X^0$-$X^1$ sector becomes Carrollian. In flat spacetime, the MM$p$T${}^*$ prescriptions are given by Eq.~\eqref{eq:mmpts}, with
\begin{subequations}
\begin{align}
	  	\hat{X}^0 & = X^0\,, 
			\qquad%
		\hat{X}^1 = \omega \, X^1\,, 
			&%
		\hat{\Phi} & = \Phi + \tfrac{i \pi}{2} + \tfrac{p-2}{2} \ln \omega\,, \\[8pt]
		\hat{X}^u & = \sqrt{\omega} \, X^u\,,
			&%
		\hat{B} & = - \omega \, dX^0 \wedge dX^1\,, \\[2pt]
	 	\hat{X}^{A'} & = \frac{X^{A'}}{\sqrt{\omega}}\,,
			&%
		\hat{C}^{(p+1)} & = \frac{\omega^2}{e^\Phi} \, dX^1 \wedge dX^2 \wedge \cdots dX^{p+1}\,,
\end{align}
\end{subequations}
where $u = 2\,, \, \cdots, \, p+1$ and $A' = p+2\,, \, \cdots, \, 9$\,. The curved version is given by
\begin{subequations} \label{eq:mmptcbp}
\begin{align}
	\hat{G}^{}_{\mu\nu} & = - \tau^{}_\mu{}^0 \, \tau^{}_\nu{}^0 + \omega^2 \,  \tau^{}_\mu{}^1 \, \tau^{}_\nu{}^1 
		+ \omega \, \tau^{}_\mu{}^u \, \tau^{}_\nu{}^u + \omega^{-1} \, E^{}_{\mu\nu}\,, 
		&%
	\hat{\Phi} & = \Phi + \tfrac{i\pi}{2} + \tfrac{p-3}{2} \ln \omega\,, \\[4pt]
	\hat{C}^{(p)} & = \frac{\omega^2}{e^\phi} \, \tau^1 \wedge \tau^2 \wedge \cdots \tau^{p} - \omega \, \tau^0 \wedge \tau^1 \wedge C^{(p-2)} + C^{(p)}\,, 
		&%
	\hat{B} & = - \omega \, \tau^0 \wedge \tau^1 + B\,, \\[6pt]
	\hat{C}^{(q)} & = - \omega \, \tau^0 \wedge \tau^1 \wedge C^{(q-2)} + C^{(q)}\,, \quad
	q \neq p\,.  
\end{align}
\end{subequations}
The $\omega \rightarrow \infty$ limit of the reparametrized type II superstring theory defines MM$p$T${}^*$\,.  

\subsection{Fundamental Strings in Background Fields}

We are now ready to generalize the fundamental string actions that arise from various decoupling limits that we have discussed through the paper to be in arbitrary curved (bosonic) background fields, using the geometric data introduced in Section~\ref{sec:dlrd}. Here, we only discuss the coupling to the geometric background fields encoded by the vielbein fields $\tau_\mu{}^A$ and $E_\mu{}^{A'}$, together with the $B$-field. We collect the relevant string actions below.

\vspace{3mm}

\noindent $\bullet$~\textbf{Matrix $p$-brane theory.} We start with the cases where $p \geq 0$\,. The curved-background generalization for the Polyakov formulation~\eqref{eq:hspmpt20} of the M$p$T string is given by
\begin{align} \label{eq:mptspfvt}
    	S^\text{\scalebox{0.8}{(P)}}_\text{M$p$T} = \frac{T}{2} \int d^2 \sigma \, e \, \Bigl( - e^\alpha{}^{}_1 \, e^\beta{}^{}_1 \, \tau^{}_{\alpha\beta} + e^\alpha{}^{}_0 \, e^\beta{}^{}_0 \, E_{\alpha\beta} + \lambda^{}_A \, e^\alpha{}^{}_0 \, \tau^{}_\alpha{}^A \Bigr)
     	- T \int B\,,
     		\quad%
	p \geq 0\,.
\end{align}
We have defined the following pullbacks from the target space to the string worldsheet:
\be
	\tau^{}_{\alpha\beta} = \p^{}_\alpha X^\mu \, \p^{}_\beta X^\nu \, \tau^{}_{\mu\nu}\,,
		\qquad%
	E^{}_{\alpha\beta} = \p^{}_\alpha X^\mu \, \p^{}_\beta X^\nu \, E^{}_{\mu\nu}\,,
		\qquad%
	B = dX^\mu \wedge dX^\nu \, B^{}_{\mu\nu}\,.
\ee
Here, $\tau_{\mu\nu}$ and $E_{\mu\nu}$ are defined in Eq.~\eqref{eq:deftaue}, with $A = 0 \,, \, \cdots, \, p$ and $A' = p+1\,, \, \cdots, \, 9$\,. 
The relevant Nambu-Goto action~\eqref{eq:ngf} is now generalized to be
\be \label{eq:ngfcsg}
	S^\text{\scalebox{0.8}{(NG)}}_\text{M$p$T} = - T \int d^2 \sigma \, \sqrt{- \epsilon^{\alpha\beta} \, \epsilon^{\gamma\delta} \, \tau^{}_{\alpha\gamma} \, E^{}_{\beta\delta}} - T \int \Bigl( B + \lambda^{}_{AB} \, \tau^A \wedge \tau^B \Bigr)\,,
		\quad%
	p \geq 0\,.
\ee
where $\lambda^{}_{AB}$ is an anti-symmetric two-tensor. The above results hold when $p \geq 0$\,. In the special case where $p = 0$\,, we have the Polyakov action,
\begin{align} \label{eq:m0tscb}
	S^\text{\scalebox{0.8}{(P)}}_\text{M0T} = \frac{T}{2} \int d^2 \sigma \, e \, \Bigl( e^\alpha{}^{}_1 \, e^\beta{}^{}_1 \, \tau^{}_\alpha \, \tau^{}_\beta + e^\alpha{}^{}_0 \, e^\beta{}^{}_0 \, E^{}_{\alpha\beta} + \lambda^{}_A \, e^\alpha{}^{}_0 \, \tau^{}_\alpha{}^A \Bigr)
     - T \int B\,,
\end{align}
and the Nambu-Goto action
\be
	S^\text{\scalebox{0.8}{(NG)}}_\text{M0T} = - T \int d^2 \sigma \, \sqrt{\epsilon^{\alpha\beta} \, \epsilon^{\gamma\delta} \, \tau^{}_{\alpha} \tau^{}_{\gamma} \, E^{}_{\beta\delta}} - T \int B\,.
\ee
which respectively generalize Eqs.~\eqref{eq:m0tspf0} and \eqref{eq:snvs} to curved backgrounds. 

When $p < 0$\,, the relevant string actions can be obtained by using the shortcut of formally swapping the roles played by $\tau^A$ and $E^{A'}$ in the above string actions, which leads to the Polyakov action
\begin{align} \label{eq:mptspfvts}
    	S^\text{\scalebox{0.8}{(P)}}_\text{M$p$T} = \frac{T}{2} \int d^2 \sigma \, e \, \Bigl( - e^\alpha{}^{}_1 \, e^\beta{}^{}_1 \, E^{}_{\alpha\beta} + e^\alpha{}^{}_0 \, e^\beta{}^{}_0 \, \tau^{}_{\alpha\beta} + \lambda^{}_{A'} \, e^\alpha{}^{}_0 \, E^{}_\alpha{}^{A'} \Bigr)
     	- T \int B\,,
     		\quad%
	p < 0\,,
\end{align}
that generalizes Eq.~\eqref{eq:mmpt}. Here, $\tau_{\mu\nu}$ and $E_{\mu\nu}$ are still the same as in Eq.~\eqref{eq:deftaue}, but now with $A = 0\,, \, q+1\,, \, \cdots, \, 9$\,, $A' = 1\,, \, \cdots, \, q$\,, and $q = - p - 1 \geq 0$\,. The Nambu-Goto action is
\be \label{eq:ngfcsgplz}
	S^\text{\scalebox{0.8}{(NG)}}_\text{M$p$T} = - T \int d^2 \sigma \, \sqrt{- \epsilon^{\alpha\beta} \, \epsilon^{\gamma\delta} \, E^{}_{\alpha\gamma} \, \tau^{}_{\beta\delta}} - T \int \Bigl( B + \lambda^{}_{A'B'} \, E^{A'} \wedge E^{B'} \Bigr)\,,
		\quad%
	p < 0\,,
\ee
which generalizes Eq.~\eqref{eq:ngfcs}. When $p < - 1$\,, we obtain Carrollian string theory. Note that, formally, the Carroll-like invariant M$p$T strings with $p < -1$ can be obtained from the Galilei-like invariant M$p$T strings with $p \geq 0$ by swapping the roles played by $\tau^A$ and $E^{A'}$~\cite{Barducci:2018wuj, Bergshoeff:2022qkx}. In the special case where $p = -1$\,, \emph{i.e.}~$q = 0$\,, we find the ILST tensionless string action in the Polyakov formulation,
\begin{align} \label{eq:mptspfvts2}
    S^{}_\text{M(-1)T} = \frac{T}{2} \int d^2 \sigma \, e \, e^\alpha{}^{}_0 \, e^\beta{}^{}_0 \, \tau^{}_{\alpha\beta}
     - T \int B\,. 
\end{align}
In M(-1)T, we have $A = 0\,, \, \cdots, \, 9$ in the definition of $\tau^{}_{\mu\nu} = \tau^{}_\mu{}^A \, \tau^{}_\nu{}^B \, \eta^{}_{AB}$\,, and the target space is Lorentzian. 

\vspace{3mm}

\noindent $\bullet$~\textbf{Multicritical Matrix $p$-brane theory.} The curved-background generalization for the Polyakov formulation~\eqref{eq:expmmptfc} of the MM$p$T string is given by
\begin{align} \label{eq:expmmptfccs}
\begin{split}
    S^\text{\scalebox{0.8}{(P)}}_\text{MM$p$T} = - \frac{T}{2} \int d^2 \sigma \, e \, \Bigl( e^\alpha{}^{}_1 \, e^\beta{}^{}_1 \, \tau^{}_\alpha{}^u \, \tau^{}_\beta{}^u & - e^\alpha{}^{}_0 \, e^\beta{}^{}_0 \, E^{}_{\alpha\beta} \\
	& - \lambda^{}_A \, e^\alpha{}^{}_0 \, \tau^{}_\alpha{}^A + \lambda^{}_1 \, e^\alpha{}^{}_1 \, \tau^{}_\alpha{}^0 \Bigr) - T \int B\,,
\end{split}
\end{align}
where $u = 2\,, \cdots, \, p+1$ and $E_{\mu\nu} = E_\mu{}^{A'} E_\nu{}^{A'}$, with $A' = p+2\,, \cdots, \, 9$\,. 
The relevant Nambu-Goto action~\eqref{eq:sngmmptf} is now generalized to be
\be \label{eq:sngmmptfcb}
	S^\text{\scalebox{0.8}{(NG)}}_\text{MM$p$T} = \frac{T}{2} \int d^2\sigma \, \tau \, \Bigl[\bigl( \tau^\alpha{}^{}_0 \, \tau^{}_\alpha{}^u \bigr)^2 - \tau^\alpha{}^{}_1 \, \tau^\beta{}^{}_1 \, E^{}_{\alpha\beta} \Bigr] - T \int \Bigl( B + \lambda_u \, \tau^0 \wedge \tau^u + \lambda_{uv} \, \tau^u \wedge \tau^v \Bigr)\,.
\ee
Here, $\tau^\alpha{}^{}_a = \tau^{-1} \, \epsilon^{\alpha\beta} \, \epsilon_{ab} \, \tau^{}_\beta{}^b$\,, where $\tau = \epsilon^{\alpha\beta} \, \tau^{}_\alpha{}^0 \, \tau^{}_\beta{}^1$\,. Here, $a = 0\,, \, 1$\,, and the Levi-Civita $\epsilon^{}_{ab}$ is defined via $\epsilon^{}_{01} = 1$\,. 

The MM$p$T${}^*$ string can be obtained from the MM$p$T string by replacing $\tau^a \rightarrow - i \, \epsilon^a{}^{}_b \, \tau^b$ as in~\cite{Bidussi:2023rfs}, such that the Polyakov formulation is
\begin{align} \label{eq:expmmptfccss}
\begin{split}
    S^\text{\scalebox{0.8}{(P)}}_\text{MM$p$T${}^*$} = - \frac{T}{2} \int d^2 \sigma \, e \, \Bigl( e^\alpha{}^{}_1 \, e^\beta{}^{}_1 \, \tau^{}_\alpha{}^u \, \tau^{}_\beta{}^u & - e^\alpha{}^{}_0 \, e^\beta{}^{}_0 \, E^{}_{\alpha\beta} \\
	& - \lambda^{}_A \, e^\alpha{}^{}_0 \, \tau^{}_\alpha{}^A + \lambda^{}_0 \, e^\alpha{}^{}_1 \, \tau^{}_\alpha{}^1 \Bigr) - T \int B\,,
\end{split}
\end{align}
where we have also performed the replacement $\lambda^{}_a \rightarrow - i \, \epsilon^{}_a{}^b \, \lambda^{}_b$\,. The Nambu-Goto formulation is given by
\be
	S^\text{\scalebox{0.8}{(NG)}}_\text{MM$p$T} = \frac{T}{2} \int d^2\sigma \, \tau \, \Bigl[ \tau^\alpha{}^{}_0 \, \tau^\beta{}^{}_0 \, E^{}_{\alpha\beta}  - \bigl( \tau^\alpha{}^{}_1 \, \tau^{}_\alpha{}^u \bigr)^2 \Bigr] - T \int \Bigl( B + \lambda_u \, \tau^1 \wedge \tau^u + \lambda_{uv} \, \tau^u \wedge \tau^v \Bigr)\,,
\ee
 where we have also performed the replacement $\lambda_u \rightarrow - i \, \lambda_u$\,. 

\subsection{Buscher Rules}

We have discussed various T-duality transformations that map between various Polyakov string actions in flat target space through Sections~\ref{sec:fdlcq}$-$\ref{sec:tnddlcq}. We now generalize these T-dual relations to curved backgrounds and derive the Buscher rules for the background geometric data and Kalb-Ramond fields. 

\subsubsection{T-duality in Matrix \texorpdfstring{$p$}{p}-brane theories}

Performing T-duality transformations along spatial isometries in the M0T string action~\eqref{eq:m0tscb} in curved spacetime gives rise to the M$p$T string sigma model in general background fields\,\footnote{This is under the condition that there is \emph{no} internal $B$-field in the compactification.},
\begin{align} \label{eq:mptspf}
    	S^{}_\text{M$p$T} = \frac{T}{2} \int d^2 \sigma \, e \, \Bigl( - e^\alpha{}^{}_1 \, e^\beta{}^{}_1 \, \tau^{}_{\alpha\beta} + e^\alpha{}^{}_0 \, e^\beta{}^{}_0 \, E_{\alpha\beta} + \lambda^{}_A \, e^\alpha{}^{}_0 \, \tau^{}_\alpha{}^A \Bigr)
     	- T \int B\,,
     		\qquad%
	p \geq 0\,.
\end{align}
Here, $\tau_{\mu\nu} = \tau^{}_\mu{}^A \, \tau^{}_\nu{}^B \, \eta^{}_{AB}$ with $A = 0\,, \cdots, \, p$ and $E^{}_{\mu\nu} = E^{}_\mu{}{}^{A'} E^{}_\mu{}{}^{A'}$, $A' = p+1\,, \cdots, 9$\,. The target space geometry has a codimension-($p$+1) foliation stucture and is referred to as the \emph{$p$-brane Newton-Cartan geometry}, where the usual local Lorentz boost is now broken to the local $p$-brane Galilean boost, \emph{i.e.} $\delta_\text{\scalebox{0.8}{G}} \tau^A = 0$ and $\delta_\text{\scalebox{0.8}{G}} E^{A'} = \Lambda_A{}^{A'} \tau^A$. 
We have noted that the action~\eqref{eq:mptspf} generalizes Eq.~\eqref{eq:hspmpt20} to curved background fields. In the special case with the indices $A = 0$ and $A' = 1\,, \cdots, 9$\,, the M$p$T string action~\eqref{eq:mptspf}  reproduces the M$0$T string action~\eqref{eq:m0tscb}. We now derive the Buscher rules for T-dualities relating the M$p$T strings. 

\vspace{3mm}

\noindent $\bullet$~{\emph{Longitudinal spatial T-duality from M$p$T to M\emph{($p$\,-1)}T.}} We start with the T-duality transformation along a longitudinal spatial isometry in the M$p$T string action~\eqref{eq:mptspf}. Consider the Killing vector $k^\mu$ satisfying
\be \label{eq:defk}
	k^\mu \, \tau_\mu{}^{\CA} = 0\,, 
		\qquad%
	k^\mu \, \tau_\mu{}^p \neq 0\,,
		\qquad%
	k^\mu \, E_\mu{}^{A'} = 0\,,
\ee
where $\CA = 0\,, \cdots\,, \, p-1$\,. 
In the coordinates $X^\mu = ( y\,, X^i )$ adapted to $k^\mu$, where $y$ is defined via $\p_y = k^\mu \, \p_\mu$\,, we have
\be 
	\tau_y{}^{\CA} = 0\,, 
		\qquad%
	\tau_y{}^p \neq 0\,,
		\qquad%
	E_y{}^{A'} = 0\,.
\ee
We perform the T-duality transformation by gauging the isometry $y$ as in the M$p$T string action~\eqref{eq:mptspf}, followed by integrating out the associated U(1) gauge potential. The T-dual action is
\begin{align} \label{eq:mptspfd}
    \tilde{S}^{}_\text{M($p$\,-1)T} & = \frac{T}{2} \int \!\! d^2 \sigma \, e \, \Bigl( - e^\alpha{}^{}_1 \, e^\beta{}^{}_1 \, \tilde{\tau}^{}_{\alpha\beta} + e^\alpha{}^{}_0 \, e^\beta{}^{}_0 \, \tilde{E}_{\alpha\beta} + \lambda^{}_u \, e^\alpha{}^{}_0 \, \tau^{}_\alpha{}^u \Bigr) - T \int B\,,
\end{align}
where $\tilde{\tau}_{\mu\nu} = \tau^{}_\mu{}^\CA \, \tau^{}_\nu{}^\CB \, \eta^{}_{\CA\CB}$ and $\tilde{E}^{}_{\mu\nu} = \tilde{E}^{}_\mu{}^{\CA'} \tilde{E}^{}_\nu{}^{\CA'}$, $\CA' = p\,, \cdots, 9$\,. This dual action describes the fundamental string in M($p$\,-1)T. 
The Buscher rules for the vielbein fields and $B$-field associated with this T-duality map from M$p$T to M($p$\,-1)T are given by
\begin{subequations} \label{eq:lstdb}
\begin{align}
	\tilde{E}_{yy} = \frac{1}{\tau_{yy}}\,, 
		\qquad%
	\tilde{E}_{yi} & = \frac{B_{yi}}{\tau_{yy}}\,,
		\qquad%
	\tilde{E}_{ij} = E_{ij} + \frac{B_{yi} \, B_{yj}}{\tau_{yy}}\,, \\[4pt]
	\tilde{B}_{yi} & = \frac{\tau_{yi}}{\tau_{yy}}\,,
		\qquad%
	\tilde{B}_{ij} = B_{ij} + \frac{B_{yi} \, \tau_{yj} - B_{yj} \, \tau_{yi}}{\tau_{yy}}\,.
\end{align}
\end{subequations}
Note that the dual isometry $\tilde{y}$ satisfies
\be
	\tilde{\tau}^{}_y{}^\CA = 0\,,
		\qquad%
	\tilde{E}^{}_y{}^p = \frac{1}{\tau^{}_y{}^p}\,,
		\qquad%
	\tilde{E}^{}_y{}^{A'} = 0\,.
\ee
This implies that $\tilde{y}$ is transverse now. 

\vspace{3mm}

\noindent $\bullet$~{\emph{Transverse T-duality from M$p$T to M\emph{($p$+1)}T.}} Next, we present the T-duality transformation of the M$p$T string sigma model \eqref{eq:mptspf} along a transverse isometry $y$\,, with
\be
	\tau_y{}^{A} = 0\,,
		\qquad%
	E_y{}^{p+1} \neq 0\,,
		\qquad%
	E_y{}^{\CA'} = 0\,,
\ee
where $\CA' = p+2\,, \cdots\,, 9$\,. 
We have gone to the adapted coordinates $X^\mu = (y\,, X^i)$\,. The T-dual action is
\begin{align} \label{eq:mptspftt}
    \tilde{S}^{}_\text{M($p$\,+1)T} & = \frac{T}{2} \int \!\! d^2 \sigma \, e \, \Bigl( - e^\alpha{}^{}_1 \, e^\beta{}^{}_1 \, \tilde{\tau}^{}_{\alpha\beta} + e^\alpha{}^{}_0 \, e^\beta{}^{}_0 \, \tilde{E}_{\alpha\beta} + \tilde{\lambda}^{}_u \, e^\alpha{}^{}_0 \, \tilde{\tau}^{}_\alpha{}^u \Bigr) - T \int \tilde{B}\,,
\end{align}
where 
\begin{align}
	\tilde{\tau}^{}_{\mu\nu} & = \tilde{\tau}^{}_\mu{}^\CA \, \tilde{\tau}^{}_\nu{}^\CB \, \eta^{}_{\CA\CB}\,,
        		\qquad%
    	\tilde{\tau}_\mu{}^A = \tau_\mu{}^A\,.
\end{align}
Here, $\CA = (A\,, \, p+1)$\,, \emph{i.e.}~$\CA = 0\,, \cdots\,, \, p+1$\,. Moreover, $\tilde{E}_{\mu\nu} = \tilde{E}_\mu{}^{\CA'} \, \tilde{E}_\nu{}^{\CA'}$, $\CA' = p+2 \,, \cdots\,, 9$\,. The action~\eqref{eq:mptspftt} describes the fundamental string in M($p$\,+1)T. The Buscher rules for the vielbein fields and $B$-fields associated with this T-duality map from M$p$T to M($p$\,+1)T are given below:
\begin{subequations} \label{eq:ttdr}
\begin{align}
	\tilde{\tau}^{}_y{}^{p+1} & = \frac{1}{E^{}_y{}^{p+1}}\,, 
		&%
	\tilde{\tau}^{}_i{}^{p+1} & = \frac{B^{}_{yi}}{E^{}_{y}{}^{p+1}}\,,
		&%
	\tilde{\tau}^{}_y{}^A & = \tilde{E}^{}_{y\mu} = 0\,, \\[4pt]
	\tilde{E}^{}_{ij} & = E^{}_i{}^{u'} \, E^{}_j{}^{u'}\,,
		&%
	\tilde{B}^{}_{yi} & = \frac{E^{}_{yi}}{E^{}_{yy}}\,,
		&%
	\tilde{B}^{}_{ij} & = B^{}_{ij} + \frac{B^{}_{yi} \, E^{}_{yj} - B^{}_{yj} \, E^{}_{yi}}{E^{}_{yy}}\,.	
\end{align}
\end{subequations}
The dual isometry $\tilde{y}$ satisfies
\be
	\tilde{\tau}^{}_y{}^{A} = 0\,,
		\qquad%
	\tilde{\tau}^{}_y{}^{p+1} \neq 0\,,
		\qquad%
	\tilde{E}^{}_y{}^{\CA'} = 0\,,
\ee
\emph{i.e.} $\tilde{y}$ is now longitudinal. The longitudinal and transverse T-dual relations between different M$p$Ts are illustrated in Fig.~\ref{fig:tdmpt}. 

\vspace{3mm}

\noindent $\bullet$~{\emph{Timelike T-duality from M$p$T to M\emph{(-$p$\,-1)}T.}} We have shown that, in flat spacetime, the M$p$T string is mapped to the M(-$p$\,-1)T string via a timelike T-duality transformation in Section~\ref{sec:otdr}. See Figure~\ref{fig:tltd} for a road map. Now, we consider the generalization of these results to curved backgrounds. We T-dualize the M$p$T string action~\eqref{eq:mptspf} with $p>0$ along the timelike isometry $y$ that satisfies
\be
    \tau_y{}^0 \neq 0\,, 
        \qquad%
    \tau_y{}^{\CA'} = 0\,,
        \qquad%
    E_y{}^{A'} = 0\,, 
\ee
where $\CA' = 1\,, \, \cdots, \, p$ and $A' = p+1\,, \, \cdots, \, 9$\,. The dual string action is
\begin{align} \label{eq:dualmmp}
    \tilde{S}^{}_\text{M(-$p$\,-1)T} = \frac{T}{2} \int d^2\sigma \, e \, \Bigl( - e^\alpha{}^{}_1 \, e^\beta{}^{}_1 \, \tilde{E}^{}_{\alpha\beta} + e^\alpha{}^{}_0 \, e^\beta{}^{}_0 \, \tilde{\tau}^{}_{\alpha\beta} + \lambda^{}_{\CA'} \, e^\alpha{}^{}_0 \, \tilde{E}^{}_\alpha{}^{\CA'} \Bigr) - T \int \tilde{B}\,,
\end{align}
where $\tilde{\tau}^{}_{\mu\nu} = \tilde{\tau}^{}_\mu{}^\CA \, \tilde{\tau}^{}_\nu{}^\CB \, \eta^{}_{\CA\CB}$\,, $\CA = 0\,, \, p+1\,, \, \cdots, \, 9$\,, and $\tilde{E}^{}_{\mu\nu} = \tau^{}_{\mu}{}^{\CA'} \, \tau^{}_\nu{}^{\CA'}$\,. The Buscher rules are akin to the ones in Eq.~\eqref{eq:lstdb}, with
\begin{subequations} 
\begin{align}
	\tilde{\tau}^{}_{yy} = \frac{1}{\tau^{}_{yy}}\,, 
		\qquad%
	\tilde{\tau}^{}_{yi} & = \frac{B^{}_{yi}}{\tau^{}_{yy}}\,,
		\qquad%
	\tilde{\tau}^{}_{ij} = E^{}_{ij} + \frac{B^{}_{yi} \, B^{}_{yj}}{\tau^{}_{yy}}\,. \\[4pt]
	\tilde{B}^{}_{yi} & = \frac{\tau^{}_{yi}}{\tau^{}_{yy}}\,,
		\qquad%
	\tilde{B}^{}_{ij} = B^{}_{ij} + \frac{B^{}_{yi} \, \tau^{}_{yj} - B^{}_{yj} \, \tau^{}_{yi}}{\tau^{}_{yy}}\,.
\end{align}
\end{subequations}
The T-dual action~\eqref{eq:dualmmp} describes the M(-$p$\,-1)T string in Eq.~\eqref{eq:mmpt}. In the case where $p = 0$\,, $\{ \CA' \}$ is an empty set and the above T-duality transformation maps the M0T string  to the M(-1)T string.

On the other hand, we start with the M$p$T string action~\eqref{eq:mmpt} with $p < -1$ with a target space timelike isometry $y$ that satisfies
\be
    \tau^{}_y{}^0 \neq 0\,,
        \qquad%
    \tau^{}_y{}^{\CA'} \! = 0\,,
        \qquad%
    E^{}_y{}^{A'} = 0\,,
\ee
where $\CA' = q+1\,, \cdots, \, 9$\,, $A' = 1 \,, \cdots, \, q$\,, and $p + q = -1$\,.
We have defined the adapted coordinates $X^\mu = (y\,, X^i)$\,.
T-dualizing along $y$ gives rise to the dual string action,
\begin{align} \label{eq:tsmpot}
    \tilde{S}^{}_\text{M$q$T} = \frac{T}{2} \int d^2\sigma \, e \, \Bigl( - e^\alpha{}^{}_1 \, e^\beta{}^{}_1 \, \tilde{\tau}^{}_{\alpha\beta} + e^\alpha{}^{}_0 \, e^\beta{}^{}_0 \, \tilde{E}^{}_{\alpha\beta} + \tilde{\lambda}^{}_\CA \, e^\alpha{}^{}_0 \, \tilde{\tau}^{}_\alpha{}^\CA \Bigr) - T \int \tilde{B}\,.
\end{align}
Here,
$\tilde{\tau}^{}_{\mu\nu} = \tilde{\tau}^{}_\mu{}^\CA \, \tilde{\tau}^{}_\mu{}^\CB \, \eta^{}_{\CA\CB}$\,, $\CA = 0\,, \cdots, \, q$\,, $\tilde{E}^{}_{\mu\nu} = \tau^{}_\mu{}^{\CA'} \, \tau^{}_\nu{}^{\CA'}$, and $\tilde{\lambda}^{}_\CA = \bigl( \tilde{\lambda}^{}_0\,, \, \lambda^{}_{A'} \bigr)$\,.
The Buscher rules are
\begin{subequations} \label{eq:ttdrmp}
\begin{align}
	\tilde{\tau}^{}_y{}^0 & = \frac{1}{\tau^{}_y{}^0}\,, 
		&%
	\tilde{\tau}^{}_i{}^0 & = \frac{B^{}_{yi}}{\tau^{}_{y}{}^0}\,,
		&%
	\tilde{\tau}^{}_y{}^{A'} & = \tilde{E}^{}_{y\mu} = 0\,, 
        \qquad%
    \tilde{\tau}^{}_i{}^{A'} = E^{}_i{}^{A'},
    \\[4pt]
	\tilde{E}^{}_{ij} & = \tau^{}_i{}^{\CA'} \, \tau^{}_j{}^{\CA'}\,,
		&%
	\tilde{B}^{}_{yi} & = \frac{\tau^{}_{yi}}{\tau^{}_{yy}}\,,
		&%
	\tilde{B}^{}_{ij} & = B^{}_{ij} + \frac{B^{}_{yi} \, \tau^{}_{yj} - B^{}_{yj} \, \tau^{}_{yi}}{\tau^{}_{yy}}\,,	
\end{align}
\end{subequations}
which are akin to the transverse T-duality Buscher rules in Eq.~\eqref{eq:ttdr}. The T-dual action~\eqref{eq:tsmpot} describes the M(-$p$\,-1)T string (see Eq.~\eqref{eq:mptspf}). In the case where $p = -1$, $\{A'\}$ is an empty set and the above T-duality transformation maps the M(-1)T string, \emph{i.e.}~the tensionless string, to the M0T string. 

\vspace{3mm}

The above discussion proves the T-dual relation between M$p$T and M(-$p$\,-1)T for any integer $p$\,. The spacelike T-duality transformations in M$p$T with $p < 0$ are in form very similar to the analogous ones in M$p$T with $p \geq 0$\,, which we will not repeat here.

\subsubsection{T-dualities in multicritical Matrix \texorpdfstring{$p$}{p}-brane theories}

In Section~\ref{sec:tnddlcq}, we considered the lightlike T-duality transformation in DLCQ M($p$+1)T, which is then mapped to MM$p$T. In retrospective, since MM$p$T arises from a well-defined BPS limit of type II superstring theory, it may be used to DLCQ M($p$+1)T via a T-duality transformation along a spacelike or timelike circle. The exotic lightlike compactification only arises on the T-dual side in DLCQ M($p$+1)T. In the following, we generalize the discussions in Section~\ref{sec:tdmmpt} to curved backgrounds. See Fig.~\ref{fig:mmtltd} for a summary of the results. 

\vspace{3mm}

\noindent $\bullet$~\emph{Spacelike T-duality from MM$p$T to DLCQ M($p$+1)T.} Consider MM$p$T string action with a spatial isometry $y$ that is longitudinal to the background critical F1-string, such that
\be
	\tau^{}_y{}^0 = 0\,,
		\qquad%
	\tau^{}_y{}^1 \neq 0\,,
		\qquad%
	\tau^{}_y{}^u = 0\,,
		\qquad%
	E^{}_y{}^{A'} = 0\,,
\ee
where $u=2\,, \cdots, \, p+1$ and $A' = p+2\,, \cdots, \, 9$\,. 
T-dualizing $y$ in the MM$p$T string action~\eqref{eq:expmmptfccs} gives the dual DLCQ M($p$\,+1)T string action (see Eq.~\eqref{eq:mptspf}),
\begin{align} \label{eq:mmptspf}
\begin{split}
    S^\text{\scalebox{0.8}{DLCQ}}_\text{M($p$\,+1)T} = \frac{T}{2} \int d^2 \sigma \, e \, \biggl[ - e^\alpha{}^{}_1 \, e^\beta{}^{}_1 \, \tilde{\tau}^{}_{\alpha\beta} & + e^\alpha{}^{}_0 \, e^\beta{}^{}_0 \, E_{\alpha\beta} \\[4pt]
	& + e^\alpha{}^{}_0 \, \Bigl( \tilde{\lambda}^{}_- \, \tilde{\tau}^{}_\alpha{}^- + \tilde{\lambda}^{}_+ \, \tilde{\tau}^{}_\alpha{}^+ + \lambda^{}_u \, \tau^{}_\alpha{}^u \Bigr) \biggr]
     - T \int \tilde{B}\,,
\end{split}
\end{align}
where $\tilde{\tau}^{}_{\mu\nu} = - \tilde{\tau}^{}_\mu{}^- \, \tilde{\tau}^{}_\nu{}^+ - \tilde{\tau}^{}_\mu{}^+ \, \tilde{\tau}^{}_\nu{}^- + \tau^{}_{\mu}{}^u \, \tau^{}_\nu{}^u$\,. The Buscher rules are
\begin{subequations}
\begin{align}
	\tilde{\tau}^{}_y{}^- & = 0\,, 
		&
	\tilde{\tau}^{}_i{}^- & = \tau^{}_i{}^0\,, 
		&
	\tilde{B}^{}_{iy} & = \frac{\tau^{}_i{}^1}{\tau^{}_y{}^1}\,, \\[2pt]
	\tilde{\tau}^{}_y{}^+ & = \frac{1}{\tau^{}_y{}^1}\,,
		&
	\tilde{\tau}^{}_i{}^+ & = \frac{B^{}_{iy}}{\tau^{}_y{}^1}\,,
		&
	\tilde{B}^{}_{ij} & = B^{}_{ij} + \frac{B^{}_{yi} \, \tau^{}_j{}^1 - B^{}_{yj} \, \tau^{}_i{}^1}{\tau^{}_y{}^1}\,,
\end{align}
\end{subequations}
The dual isometry $\tilde{y}$ is lightlike. 

\vspace{3mm}

\noindent $\bullet$~\emph{Timelike T-duality from MM$p$T to DLCQ M\emph{(-$p$\,-1)}T.} In curved backgrounds, we consider MM$p$T with a target space timelike isometry $y$\,,
\be
	\tau^{}_y{}^0 \neq 0\,,
		\qquad%
	\tau^{}_y{}^1 = 0\,,
		\qquad%
	\tau^{}_y{}^u = 0\,,
		\qquad%
	E^{}_y{}^{A'} = 0\,,
\ee
T-dualizing the timelike isometry $y$ in the MM$p$T string action~\eqref{eq:expmmptfc} gives
\begin{align} \label{eq:mmptspm1f}
\begin{split}
    S^\text{DLCQ}_\text{M(-$p$\,-1)T} = \frac{T}{2} \int d^2 \sigma \, e \, \Bigl( - e^\alpha{}^{}_1 \, e^\beta{}^{}_1 \, \tilde{E}^{}_{\alpha\beta} & + e^\alpha{}^{}_0 \, e^\beta{}^{}_0 \, \tilde{\tau}_{\alpha\beta} + \lambda^{}_u \, e^\alpha{}_0 \, \tilde{E}^{}_\alpha{}^u \Bigr)
     - T \int \tilde{B}\,,
\end{split}
\end{align}
with $\tilde{E}_{\mu\nu} = \tilde{E}_{\mu}{}^u \, \tilde{E}_\nu{}^u$\,, $\tilde{E}_\mu{}^u = \tau^{}_\mu{}^u$\,, $\tilde{\tau}^{}_{\mu\nu} = - \tilde{\tau}^{}_\mu{}^- \, \tilde{\tau}^{}_\nu{}^+ - \tilde{\tau}^{}_\mu{}^+ \, \tilde{\tau}^{}_\nu{}^- + E_{\mu\nu}$\,, and the Buscher rules
\begin{subequations}
\begin{align}
	\tilde{\tau}^{}_y{}^- & = 0\,, 
		&
	\tilde{\tau}^{}_i{}^- & = \tau^{}_i{}^1\,, 
		&
	\tilde{B}^{}_{iy} & = \frac{\tau^{}_i{}^0}{\tau^{}_y{}^1}\,, \\[4pt]
	\tilde{\tau}^{}_y{}^+ & = \frac{1}{\tau^{}_y{}^0}\,,
		&
	\tilde{\tau}^{}_i{}^+ & = \frac{B^{}_{iy}}{\tau^{}_y{}^0}\,,
		&
	\tilde{B}^{}_{ij} & = B^{}_{ij} + \frac{B^{}_{yi} \, \tau^{}_j{}^0 - B^{}_{yj} \, \tau^{}_i{}^0}{\tau^{}_y{}^0}\,,
\end{align}
\end{subequations}
The dual isometry $\tilde{y}$ is again lightlike. 

\vspace{3mm}

As expected, T-dualizing DLCQ M($p$\,+1)T and DLCQ M(-$p$\,-1) with $p\geq0$ along a lighlike isometry also gives back MM$p$T. 

\subsection{Spin Matrix Theory and Nonrelativistic Holography} \label{sec:smtnh}

In this final subsection, we discuss an intriguing connection to a near BPS limit of the AdS/CFT correspondence.
We focus on the special case where $p=0$\,, the $\tau^u$ terms disappear in the MM$p$T action~\eqref{eq:expmmptfccs}. Perform the replacement~\eqref{eq:ctg} with $e^0 \rightarrow i \, e^1$ and $e^1 \rightarrow i \, e^0$ mapping from the Carrollian parametrization to the Galilean parametrization of the string worldsheet, together with
$\lambda_A \rightarrow i \, \lambda_A$\,,
we find that MM$p$T string action~\eqref{eq:expmmptfccs} with $p=0$ becomes
\begin{align} \label{eq:mm0t}
	S^{}_\text{MM0T} = - \frac{T}{2} \int d^2 \sigma \, e \, \Bigl[ e^\alpha{}^{}_1 \, e^\beta{}^{}_1 \, E^{}_{\alpha\beta} + \lambda^{}_0 \, e^\alpha{}^{}_1 \, \tau^{}_\alpha{}^0 + \lambda^{}_1 \, \bigl( e^\alpha{}^{}_1 \, \tau^{}_\alpha{}^1 - e^\alpha{}^{}_0 \, \tau^{}_\alpha{}^0 \bigr) \Bigr] - T \int B\,.
\end{align}
Furthermore, choosing $B^{}_{\mu\nu} = ( \tau^{}_{\mu}{}^A \, m^{}_{\nu}{}^B - \tau^{}_{\nu}{}^A \, m^{}_{\mu}{}^B ) \, \epsilon^{}_{AB}$\,,
where $m^A$ is an arbitrary one-form, and replacing $(\tau^0, \, m^1) \rightarrow - (\tau^0, \, m^1)$\,, we find that the action~\eqref{eq:mm0t} becomes
\begin{align} \label{eq:mm0tfc}
\begin{split}
	S^{}_\text{MM0T} = - \frac{T}{2} \int d^2 \sigma \, \Bigl[ & \, e \, e^\alpha{}^{}_1 \, e^\beta{}^{}_1 \, E^{}_{\alpha\beta} + 2 \, \epsilon^{\alpha\beta} \bigl( \tau_\alpha{}^0 \, m_\beta{}^1 - \tau_\alpha{}^1 \, m_\beta{}^0 \bigr) \\[4pt]
	& \qquad + \bigl( - \lambda^{}_0 \bigr) \, \epsilon^{\alpha\beta} \, e^{}_\alpha{}^0 \, \tau^{}_\beta{}^0 + \lambda^{}_1 \, \epsilon^{\alpha\beta} \bigl( e^{}_\alpha{}^0 \, \tau^{}_\beta{}^1 + e^{}_\alpha{}^1 \, \tau^{}_\beta{}^0 \bigr) \Bigr]\,.
\end{split}
\end{align}
In this Galilean worldsheet associated with the prescription~\eqref{eq:wsgg0}, the MM0T action~\eqref{eq:mm0tfc} is identical to the string sigma model for \emph{Spin Matrix Theory} (SMT) (see Eq.~(A.15) in \cite{Harmark:2018cdl}). See~\cite{Oling:2022fft, Baiguera:2023fus} for reviews of SMT and \cite{Bidussi:2023rfs, Baiguera:2022pll} for the latest developments. 

SMT refers to a class of integrable systems that arise from near BPS limits of $\CN = 4$ SYM theory with SU($N$) gauge symmetry on $\mathbb{R} \times S^3$\,, which we define below~\cite{Harmark:2014mpa, Harmark:2008gm}. For a given state in $\CN = 4$ SYM with energy $E$ and a linear sum $Q$ over the Cartan charges of the PSU($2, 2|4$) algebra, we consider the SMT limits,
\be \label{eq:ltemqa}
	\lambda^{}_\text{t} \rightarrow 0\,,
		\qquad%
	E - Q \rightarrow 0\,,
		\qquad%
	\frac{E - Q}{\lambda^{}_\text{t}} \rightarrow \text{fixed}\,,
\ee
where $\lambda_\text{t}$ is the 't Hooft coupling of SYM. 
Note that $N$ is fixed in these limits. 
Here, the Cartan charges contain the angular momenta $S^{}_1$ and $S^{}_2$ on the $S^3$ as well as the R-charges $J_1$, $J_2$ and $J_3$\,. In the SMT limits, $\CN=4$ SYM simplifies to SMTs, which are quantum mechanical systems with a Hilbert space of harmonic oscillators with both spin group indices and extra matrix indices. In the large $N$ limit, SMT becomes a nearest-neighbor spin chain, where the low-energy excitations are magnons. 

The bulk gravity dual to SMT was first proposed in~\cite{Harmark:2017rpg} and then further studied in \emph{e.g.}~\cite{Harmark:2018cdl,  Harmark:2019upf, Harmark:2020vll, Kluson:2021sym, Roychowdhury:2021wte, Bidussi:2023rfs}. We briefly review the essential ingredients for this holographic duality below, 
with slight adaptions towards the notation in the current paper. On the bulk side, we start with the AdS${}_5 \times S^5$ metric in the global coordinates,
\be \label{eq:gp}
	ds^2 = R^2 \, \Bigl( - \cosh^2 \rho \, dt^2 + d\rho^2 + \sinh^2 \! \rho \, d\Omega_3^2 + d\Omega_5^2 \Bigr)\,.
\ee
Here, $R$ is the radius of $S^5$ as well as the AdS scale. The AdS/CFT dictionary states that 
\be \label{eq:ltrlsf}
	\lambda^{}_\text{t} = R^4 / \ell_\text{s}^4\,,
\ee
where $\ell_s$ is the string length. 
The $\CN=4$ SYM angular momenta $\{ S_1\,, S_2 \}$ correspond to the angular momenta on the $S^3$ within AdS${}_5$\,, and the R-charges $\{ J_1\,, J_2\,, J_3 \}$ correspond to the angular momenta on the $S^5$ within AdS${}_5 \times S^5$. We focus on the SMT limit with $S = S_1 + S_2$ and $J = J_1 + J_2 + J_3$\,, in which case the spin group is PSU($1,2|3$). The Hopf fibrations of the $S^3$ and $S^5$ in AdS${}_5 \times S^5$ give, respectively,
\be
	d\Omega_3^2 = \bigl( d\psi + A_1 \bigr)^2 + d\Sigma_1^2\,,
		\qquad%
	d\Omega_5^2 = \bigl( d\chi +A_2 \bigr)^2 + d\Sigma_2^2\,,
\ee
where $\psi$ and $\chi$ are the U(1) fibre coordinates, $A_1$ and $A_2$ are some one-forms whose detailed expressions are not important here, and $d\Sigma_k^2$ are Fubini-Study metrics on $\mathbb{C}$P${}^k$ with $k = 1, 2$\,.
The conserved charges $E$, $S$, and $J$ on the field theory side correspond to the Killing vectors in AdS${}_5 \times S^5$\,, with
\be
	E = i \, \p_t\,,
		\qquad%
	S = - i \, \p_\psi\,,
		\qquad%
	J = - i \, \p_\chi\,.
\ee 
Under the change of variables,\,\footnote{In the up-to-date reference \cite{Harmark:2020vll}, a different coordinate choice is taken, which is more natural from the boundary spin chain perspective. Here, we use a different set of variables such that Eq.~\eqref{eq:emq} is simple. The discussions in this section are insensitive to the details of these coordinate choices.}
\be \label{eq:cov}
	\begin{cases}
        y^- = \tfrac{1}{2} \, \bigl( t + \chi \bigr)\,, \\[4pt]
        x^+ = \tfrac{1}{4} \, \bigl( 2 \, t - \psi - \chi \bigr)\,, \\[4pt]
		x^2 = \tfrac{1}{4} \, \bigl( \psi - \chi \bigr)\,, 
	\end{cases}
		\implies\,\,%
	\begin{cases}
		\p_t = \tfrac{1}{2} \, \bigl( \p_{y^-} + \p_{x^+} \bigr)\,, \\[4pt]
		\p_\psi = \tfrac{1}{4} \, \bigl( \p_{x^2} - \p_{x^+} \bigr)\,, \\[4pt]
		\p_\chi = \tfrac{1}{4} \, \bigl( 2 \, \p_{y^-} - \p_{x^+} - \p_{x^2} \bigr)\,,
	\end{cases}
\ee
it follows that ($Q = S + J$)
\be \label{eq:emq}
	E - Q = i \, \p_{y^-}\,,
		\qquad%
	E + Q = i \, \p_{x^+}\,. 
\ee
In terms of the new variables $\{ y^-, \, x^+, \, x^2 \}$\,, the metric~\eqref{eq:gp} becomes
\be \label{eq:rpm}
	ds^2 = R^2 \, \Bigl[ - 2 \, K \bigl( dy^- + K^-_i \, dx^i \bigr) \, \bigl( dx^+ + K^+_i \, dx^i \bigr) + K_{ij} \, dx^i \, dx^j \Bigr]\,,
\ee
where $i = 2\,, \cdots, 9\,$ denotes the rest of spacetime directions other than the lightlike directions $y^-$ and $x^+$\,. The detailed expressions of $K$\,, $K^\pm_i$, and $K^{}_{ij} = K^{}_i{}^{A'} K^{}_j {}^{A'}$ with $A' = 2\,, \cdots,\,9$ are not important here. From Eqs.~\eqref{eq:ltemqa}, \eqref{eq:ltrlsf}, and \eqref{eq:emq}, we find $R \rightarrow 0$\,, $y^- \rightarrow \infty$ and $y^- \, R^4 \rightarrow \text{fixed}$\,.
This limit can be achieved by first reparametrizing
\be \label{eq:rpryp}
	R^4 = \frac{L^4}{\omega^2}\,,
		\qquad%
	y^- =  \omega^2 \, x^-\,,
\ee
and then taking the $\omega \rightarrow \infty$ limit. Under the reparametrizaion~\eqref{eq:rpryp}, we find that the metric~\eqref{eq:rpm} takes the following form:
\be \label{eq:smtrm}
	ds^2 \rightarrow \Bigl( \omega \, \tau_{\mu\nu} + \omega^{-1} \, E_{\mu\nu} \Bigr) \, dx^\mu \, dx^\nu\,,
		\qquad%
	x^\mu = (x^-\!, \, x^+\!, \, x^i)\,,
\ee
where $\tau_{\mu\nu} = - \tau_\mu{}^- \, \tau_\nu{}^+ - \tau_\mu{}^+ \, \tau_\nu{}^-$ and $E_{\mu\nu} = E_\mu{}^{A'} E_\nu{}^{A'}$. Moreover,
\begin{subequations}
\begin{align}
	\tau^{}_-{}^- & = L \, K\,,
		&%
	\tau^{}_+{}^- & = 0\,,
		&%
	\tau^{}_i{}^- & = \omega^{-2} \, L \, K \, K^-_i\,, \label{eq:tipsl} \\[4pt]
	\tau^{}_-{}^+ & = 0\,,
		&%
	\tau^{}_+{}^+ & = L\,,
		&%
	\tau^{}_i{}^+ & = L \, K^+_i\,, \\[4pt]
	E^{}_-{}^{A'} & = 0\,, 
		&%
	E^{}_+{}^{A'} & = 0\,, 
		&%
	E^{}_i{}^{A'} & = L \, K_i{}^{A'}\,.
\end{align}
\end{subequations}
Furthermore, note that $\psi$ and $\chi$ have the periodic boundary conditions $\psi \sim \psi + 4\pi$ and $\chi \sim \chi + 4\pi$~\cite{Duff:1998us}. 
The change of variables in Eq.~\eqref{eq:cov} and Eq.~\eqref{eq:rpryp} implies
\be
	x^- \sim x^- + \frac{2 \, \pi}{\omega^2}\,,
		\qquad%
	x^+ \sim x^+ \! - 2 \pi\,,
		\qquad%
	y \sim y\,.
\ee
In the $\omega \rightarrow \infty$ limit, we find that
$x^- \sim x^-$ is noncompact, while the other lightlike direction $x^+$ is compactified over a circle. 

We observe that the metric~\eqref{eq:smtrm} matches the M1T reparametrization in Eq.~\eqref{eq:rpmpt2} with $p=1$\,, where there is a two-dimensional longitudinal sector being scaled up by $\omega$ and an eight-dimensional transverse sector being scaled down by $\omega^{-1}$\,. The subleading $\omega^{-2}$ term in $\tau_i{}^-$ from Eq.~\eqref{eq:tipsl} drops out in the $\omega \rightarrow \infty$ limit of the F1-string action. In the SMT limit where $\omega \rightarrow \infty$\,, this bulk geometry matches the one in DLCQ M1T, with the lightlike circle $x^+$. As we have learned earlier in this section (see Fig.~\ref{fig:mmtltd}), after performing a T-duality transformation along this lightlike circle, the bulk geometry maps to the one in MM0T, which arises from the $\omega \rightarrow \infty$ of the metric~\eqref{eq:mmptcbp} when $p = 0$\,, with
\be
	\hat{G}^{}_{\mu\nu} = - \omega^2 \, \tau^{}_\mu{}^0 \, \tau^{}_\nu{}^0 + \tau^{}_\mu{}^1 \, \tau^{}_\nu{}^1 
	+ \omega^{-1} \, E^{}_{\mu\nu}\,, 
\ee
where $\hat{G}_{\mu\nu}$ is the metric field in type IIA superstring theory. 
In this T-dual description, $x^-$ maps to the noncompact time direction and $x^+$ maps to a spatial circle in the MM0T geometry. In addition, a critical $B$-field is induced in the limiting prescription in type IIA superstring theory that leads to MM0T, with
\be
	\hat{B} = - \omega \, \tau^0 \wedge \tau^1 + B\,,
\ee
where $\hat{B}$ is the Kalb-Ramond field in type IIA superstring theory and $B$ is in MM0T. 
Therefore, the string sigma model associated with SMT is described by the MM0T string action~\eqref{eq:mm0t}~\cite{Harmark:2018cdl, Bidussi:2023rfs}. 

It is important to note that, even though the reparametrized metric matches between DLCQ M1T and the SMT limit of AdS${}_5 \times S^5$\,, the reparametrizations of the string coupling and RR two-form in Eq.~\eqref{eq:rpmpt2} do \emph{not} match. We focus on the reparametrization of the string coupling here. In the latter case where the AdS${}_5 \times S^5$ geometry is concerned, before the SMT limit is performed, the AdS/CFT correspondence implies that $4 \pi \, g^{}_\text{s} = \lambda^{}_\text{t} / N$\,,  
where $g^{}_\text{s}$ is the string coupling. From Eqs.~\eqref{eq:ltrlsf} and \eqref{eq:rpryp}, we find 
$g^{}_\text{s} \rightarrow \omega^{-2} \, g^{}_\text{s}$\,.
However, in M1T, the scaling~\eqref{eq:rpmpt2} of the dilaton implies $\hat{g}^{}_\text{s} = \omega^{-1} \, g^{}_\text{s}$\,. Hence, the SMT limit of AdS${}_5 \times S^5$ does \emph{not} quite lead to the corner described by MM0T in the T-dual frame. This is because MM0T arises from T-dualizing a lightlike isometry in DLCQ M1T. In general, (M)M$p$Ts arise from certain BPS limits that zoom in on a background brane configuration. When the AdS/CFT correspondence is concerned, such BPS limits are performed on the field theory side, which correspond to certain near-horizon limits on the gravity side. The relation between the SMT and MM0T string implies that the bulk AdS geometry acquires a DLCQ${}^2$ structure, in addition to the original near-horizon limit from which the AdS geometry arises. Generically, one does not expect that the limiting prescription for the dilaton and RR potentials on the field theory side to match the ones on the gravity side. Moreover, the above observation implies that SMT as a field theory might reside in the DLCQ${}^3$ orbit. It would be interesting to further understand how this expectation could work out in detail.  

\section{S-Duality and Nonrelativistic String Theory} \label{sec:sdnrst}

\begin{figure}[t!]
\centering
\begin{adjustbox}{width=.65\textwidth}
\hspace{-8mm}
\begin{tikzpicture}
    \begin{scope}[scale=1]

        \path[every node/.style=draw, rounded corners=1, line width=1.5pt, minimum width=120pt, minimum height=20pt, font = \small]   
                (-3,0) node {IIB nonrel. string theory}
                (0,3) node {nonrelativistic M-theory}
                (3,0) node {Matrix 1-brane Theory}
                ;

        \path   (0,1.5) node {\begin{minipage}{2cm}
                \includegraphics[width=1\textwidth]{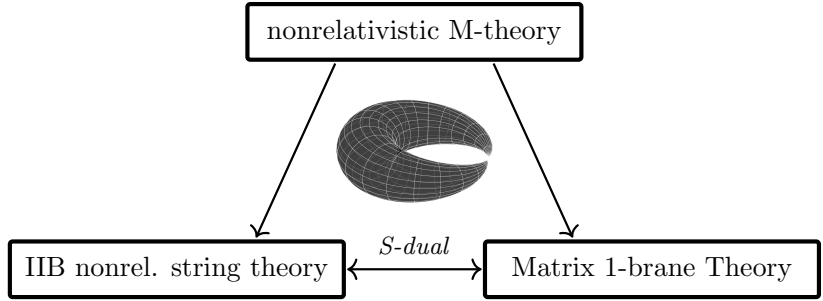}
                \end{minipage}}
                (0,0.3) node [font=\footnotesize] {\emph{S-dual}}
                ;

        \draw[-{>[length=1.3mm]}, thick] (-1,2.6) -- (-2,0.4);
        \draw[-{>[length=1.3mm]}, thick] (1,2.6) -- (2,0.4);
        \draw[{<[length=1.3mm]}-{>[length=1.3mm]}, thick] (-0.85,0) -- (0.85,0);

\end{scope}
\end{tikzpicture}
\end{adjustbox}
\caption{Compactifying nonrelativistic M-theory over a vanishing, pinched two-torus leads to two distinct decoupling limits of type IIB sueprstring theory in ten dimensions: (1) type IIB nonrelativistic superstring theory, where the $B$-field is fined tuned to its critical value to cancel the background fundamental string tension; (2) Matrix 1-brane theory, where the RR two-form is fined tuned to its critical value to cancel the background D1-string tension. These two theories are S-dual to each other.}
\label{fig:mtopt}
\end{figure}

So far, we have showed how different corners of type II superstring theories are related to each other by performing T-duality transformations of various string worldsheet theories. As shown in~\cite{udlstmt}, this duality web of decoupling limits also connects to other important corners via S-duality transformations. Notably, an S-duality transformation of Matrix 1-brane theory leads to (type IIB) \emph{nonrelativistic string theory}~\cite{Dijkgraaf:1997vv, Klebanov:2000pp, Gomis:2000bd, Danielsson:2000gi, Ebert:2023hba, udlstmt}. Nonrelativistic string theory was initially proposed in~\cite{Klebanov:2000pp, Gomis:2000bd, Danielsson:2000gi, Danielsson:2000mu} and has been studied extensively during the recent years~(see \cite{Oling:2022fft} for a review). In the M1T limit, an RR two-form is taken to be critical such that it cancels the infinite background D1-string tension. Under S-duality, this M1T limit is mapped to a different limit where the $B$-field, instead of the RR two-form, becomes critical and cancels the infinite background F1-string tension. This critical $B$-field limit leads to nonrelativistic string theory, where the light closed string excitations are wrapped fundamental strings. In M1T, the light excitations are captured by Matrix string theory, which is $\CN = 8$ SYM in two dimensions, which is S-dualized to the second quantized strings in nonrelativistic string theory~\cite{Dijkgraaf:1997vv, Motl:1997th}. Such nonrelativistic string theory is a unitary and ultra-violet complete string theory that has a Galilean-invariant string spectrum. Moreover, this string theory has a conventional Riemannian worldsheet, which allows for first-principles worldsheet computations using conventional techniques from conformal field theory~\cite{Gomis:2000bd}. Along these lines, nonrelativistic string amplitudes are studied in~\cite{Klebanov:2000pp, Gomis:2000bd, Danielsson:2000gi, Kristiansson:2000xv, Yan:2021hte}, Weyl anomalies are analyzed in~\cite{Gomis:2019zyu, Gallegos:2019icg, Yan:2019xsf, Gomis:2020fui, Yan:2021lbe, Kim:2007hb}, and the corresponding target space supergravity and worldvolume effective theories for the D$p$-branes are constructed in \emph{e.g.}~\cite{Bergshoeff:2018vfn, Gomis:2020fui, Gallegos:2020egk, Bergshoeff:2021bmc, Blair:2021waq, Bergshoeff:2021tfn, Bergshoeff:2022iss, Ebert:2021mfu, Bergshoeff:2023ogz, Ebert:2023hba}. These new advances were made possible due to the improved understanding of the non-Lorentzian geometry\,\footnote{See~\cite{Hartong:2022lsy} for a recent review on non-Lorentzian gravity. See also~\cite{Horava:2009uw, Hartong:2015zia} for a different but related framework.} in the target space~(see \emph{e.g.}~\cite{Gomis:2005pg, Brugues:2006yd, Andringa:2012uz, Harmark:2017rpg, Bergshoeff:2018yvt, Harmark:2018cdl, Harmark:2019upf, Bergshoeff:2019pij, Bidussi:2021ujm, Hartong:2022dsx}; see also \emph{e.g.}~\cite{Ko:2015rha, Morand:2017fnv, Gallegos:2020egk, Blair:2020gng} using double field theory). 

The fact that the light excitations in nonrelativistic string theory are captured by the fundamental strings promotes this corner to have an anchoring role in the duality web of decoupling limits. 
The physical contents, such as scattering amplitudes and effective actions, computed in nonrelativistic string (field) theory can be mapped to the corresponding ones in M$p$T and MM$p$T via a series of duality transformations. 
Intriguingly, studies of worldsheet quantum consistency~\cite{Yan:2021lbe} and target space supersymmetry~\cite{Bergshoeff:2021tfn} in nonrelativistic string theory imply that extra intrinsic constraints are required in the target space non-Lorentzian geometry, which generate various related geometric constraints in M$p$T and MM$p$T via the duality web. It would be highly interesting to revisit the studies of the correspondence between Matrix theory and supergravity~\cite{Taylor:2001vb} as well as Matrix theory beyond flat spacetime, in view of the close interplay between M$p$T and Matrix theory.   

In contrast to the Riemannian worldsheet in nonrelativistic string theory, we have learned from Sections~\ref{sec:wstnrs} that the string worldsheet in M$p$T and MM$p$T is nonrelativistic, and the worldsheet topology is described by a nodal Riemann sphere. In the one-loop case, the nodal Riemann sphere corresponds to the pinched torus. Intriguingly, such pinched torus also naturally arises in the target space, when we uplift nonrelativistic string theory and M1T to M-theory. Such an eleven-dimensional uplift is referred to as \emph{nonrelativistic M-theory}~\cite{Danielsson:2000gi, Gomis:2000bd, Gopakumar:2000ep, Harmark:2000ff, Bergshoeff:2000jn, Garcia:2002fa, Kamimura:2005rz, Blair:2021waq, Ebert:2021mfu, Ebert:2023hba}, where the three-form gauge potential is fine-tuned to its critical value such that it cancels the infinite tension of a background M2-brane. Nonrelativistic M-theory has a target space geometry with a codimension-three foliation structure, and it gives rise to the S-dual type IIB nonrelativistic superstring theory and M1T via a compactification over an \emph{anisotropic} torus with one of its cycles in the sector that is longitudinal to the background M2-brane and the other in the transverse sector~\cite{Ebert:2023hba}. 

From the perspective of IIB superstring theory, the modulus of the two-torus over which relativistic M-theory is compactified is $\hat{\tau} = \hat{C}^{(0)} + i \, e^{-\hat{\Phi}}$\,, where $\hat{C}^{(0)}$ is the RR 0-form and $\hat{\Phi}$ the dilaton in the ten-dimensional IIB theory. In order to zoom in the corner of nonrelativistic string theory, we require the prescription $\hat{C}^{(0)} = C^{(0)}$ and $\hat{\Phi} = \Phi + \ln \omega$~\cite{Bergshoeff:2019pij, Ebert:2021mfu}. In the $\omega \rightarrow \infty$ limit, we find $\hat{\tau} \rightarrow \text{real}$\,. This is of course S-dual to the M1T prescription~\eqref{eq:rpmpt2} with $\hat{C}^{(0)} = C^{(0)}$ and $\hat{\Phi} = \Phi - \ln \omega$~\cite{Ebert:2023hba, udlstmt}. In the $\omega \rightarrow \infty$ limit, we find $\hat{\tau} \rightarrow i \, \infty$\,. We have learned in Section~\ref{sec:wstnrs} that the two-torus becomes pinched under this limit. Therefore, we conclude that IIB nonrelativistic string theory and M1T arise from compactifying nonrelativistic M-theory over a pinched torus. See Figure~\ref{fig:mtopt}. This unconventional compactification is responsible for the salient features of the SL($2\,,\mathbb{Z}$) transformations in IIB nonrelativistic string theory and M1T, such as branched structures~\cite{Bergshoeff:2022iss} and polynomial realizations of SL($2\,,\mathbb{Z}$)~\cite{Bergshoeff:2023ogz, longpaper}.

The detailed connection between the part of the duality web that is accessible via T-dualizing the M0T string in this paper to nonrelativistic string and M-theory will appear in~\cite{longpaper}, which will provide the target space perspective. 

\section{Summary of Main Results and Outlook} \label{sec:conclusion}

Through this paper, we have developed the worldsheet perspective for the duality web of decoupling limits in type II superstring theories. This work provides a complementary proof of the duality web that has been recently studied in~\cite{udlstmt} from the target space perspective. In the following list, we highlight the T-dual relations to different decoupling limits and the related fundamental string sigma models that we have discussed through this paper.
\begin{enumerate}

\item

\emph{Matrix theories}~(Section~\ref{sec:fdlcq} and Fig.~\ref{fig:tdmpt}): In Section~\ref{sec:mtnlsw}, we discussed that the BFSS Matrix theory lives on the bound D0-brane states in Matrix 0-brane theory (M0T), and revealed that the fundamental string in M0T is the non-vibrating string~\cite{Batlle:2016iel}. The Polyakov formulation~\eqref{eq:m0tspf0} of the M0T string has been put forward in Section~\ref{sec:nvst}, where the worldsheet topology is also studied. We have uncovered the fundamental string sigma model~\eqref{eq:mptspf} in Matrix $p$-brane theory (M$p$T) by T-dualizing the M0T string along spatial isometries. The light excitations in M$p$T are the bound D$p$-branes~\cite{udlstmt}, which are described by different Matrix gauge theories~\cite{Gopakumar:2000ep, Harmark:2000ff, Gomis:2000bd, Danielsson:2000gi}. In particular, Matrix string theory~\cite{Dijkgraaf:1997vv, Motl:1997th} is associated with the bound D1-branes in M1T and $\CN = 4$ SYM is associated with the bound D3-branes in M3T. Upon decompactifying the relevant torus, the target space geometry in M$p$T is non-Lorentzian and is described by generalized Newton-Cartan geometry, which admits Galilei-type instead of Lorentz boosts.

\item

\emph{Tensionless string theory}~(Section~\ref{eq:ikktts} and Fig.~\ref{fig:tltd}): We showed that T-dualizing the M0T string along a timelike isometry\,\footnote{A timelike T-duality maps type II superstring theories to the type II${}^*$ theories~\cite{Hull:1998vg}.} leads to the fundamental string sigma model~\eqref{eq:smot0} in M(-1)T. We found that the \text{M(-1)T} string sigma model coincides with the \emph{Isberg-Lindstr\"{o}m-Sundborg-Theodoris} (ILST) tensionless limit of the fundamental string~\cite{Lindstrom:1990qb, Isberg:1993av}. Note that the BFSS Matrix theory in M0T is T-dualized to \emph{Ishibashi-Kawai-Kitazawa-Tsuchiya} (IKKT) Matrix theory in M(-1)T, which lives on a stack of D(-1)-instantons~\cite{Ishibashi:1996xs}. In M(-1)T, the target space geometry is Lorentzian.

\item

\emph{Carrollian string theory}~(Section~\ref{sec:cst} and Fig.~\ref{fig:tltd}): T-dualizing the M(-1)T string along spatial isometries leads to the fundamental string action~\eqref{eq:mmpt} that realizes a Carroll-type boost symmetry in the target space, which is described by a generalized Carrollian geometry as in~\cite{Bergshoeff:2023rkk}. These string actions generalize the Carrollian strings in~\cite{Cardona:2016ytk} and connect this previous work to the duality web of decoupling limits. In contrast to Galilean geometry, where the time is absolute but the space transforms into time under the Galilean boost, the space in Carrollian geometry is absolute but the time transforms into space under the Carrollian boost. While Galilean geometry arises from the infinite speed-of-light limit of Lorentzian geometry, Carrollian geometry arises from the opposite zero speed-of-light limit~\cite{levy1965nouvelle, sen1966analogue, Duval:2014uoa, Bergshoeff:2017btm}. Such Carrollian strings reside in M$p$T with $p < -1$\,, whose dynamics is supposedly encoded by Matrix theories on S(pacelike)-branes~\cite{Hull:1998vg, Gutperle:2002ai} that are T-dual to IKKT Matrix theory. This relation to S-branes, which are localized in time, implies that a Carrollian field theory on certain D-branes in M$p$T with $p < -1$ might only be defined nonperturbatively. Further studies along these lines may shed light on the pathology in the perturbative quantization of Carrollian field theories~\cite{Figueroa-OFarrill:2023qty, deBoer:2023fnj}, which may eventually help us understand celestial holography (see \emph{e.g.}~\cite{Pasterski:2021raf} for a review), in view of its close relation to Carrollian holography~\cite{Donnay:2022aba}. 

\item

\emph{Ambitwistor string theory}~(Section~\ref{sec:ast} and Fig.~\ref{fig:tltd}; Section~\ref{sec:gse} and Fig.~\ref{fig:tltd}): Classically, ambitwistor string theory~\cite{Mason:2013sva} arises from a singular gauge choice in tensionless string theory~\cite{Casali:2016atr, Siegel:2015axg}. Quantum mechanically, this suggests that one zoom in an unconventional twisted vacuum where creation and annihilation operators are flipped~\cite{Casali:2016atr, Bagchi:2020fpr}. The quantum amplitudes in ambitwistor string theory reproduce the Cachazo-He-Yuan (CHY) formulae~\cite{Cachazo:2013hca, Cachazo:2013iea}, which compute field-theoretical amplitudes in the form of string loops. The particle kinematics is encoded by the scattering equations that localize the moduli space of the associated string amplitudes to a set of discrete points~(see \emph{e.g.}~\cite{Geyer:2015bja} for such localizations at loop orders). In the duality map of decoupling limits, the ambitwistor string theory is connected to the M(-1)T, whose fundamental dynamics should be captured by IKKT Matrix theory~\cite{udlstmt}. See more in Section~\ref{sec:ast}. We showed in Section~\ref{sec:gse} that the MM0T string in the ambitwistor string gauge naturally leads to the DLCQ version of the scattering equation, which is potentially useful for constructing CHY formulae for Galilei-invariant field theories.

\item

\emph{Second DLCQ and Spin Matrix Theory}~(Sections~\ref{sec:tnddlcq} and~\ref{sec:smtnh}; see also Fig.~\ref{fig:mmtltd}): In M$p$T with $p \neq 0$\,, there exists a relativistic sector of the target space where at least one spatial direction is on the same footing as the time. It is possible to form a second DLCQ in M$p$T with $p \neq 0$\,. We showed that T-dualizing DLCQ M($p$\,+1)T (or M(-$p$\,-1)) string action with $p \geq 0$ along the lightlike isometry gives rise to the fundamental string~\eqref{eq:expmmptfc} in multicritical Matrix $p$-brane theory (MM$p$T), where the lightlike isometry in M$p$T is mapped to a spacelike (timelike) isometry in MM$p$T. It is shown in~\cite{udlstmt} that MM$p$T arises from a multicritical field limit, where both the background $B$-field and RR $(p+1)$-form are fine-tuned to their critical values, such that they cancel the fundamental string and D$p$-brane tensions in a background bound F1-D$p$ configuration (see also~\cite{longpaper}). See Section~\ref{sec:tnddlcq} for further details. We showed explicitly in Section~\ref{sec:smtnh} that the worldsheet action in MM0T matches the string action associated with a Spin Matrix limit of the AdS/CFT correspondence~\cite{Harmark:2017rpg, Harmark:2018cdl, Harmark:2020vll, Bidussi:2023rfs}. In the boundary $\CN=4$ SYM, such a limit corresponds to a near-BPS limit that leads to Spin Matrix theories~\cite{Harmark:2014mpa}, which generalize spin chains and are also integrable at large $N$\,.

\end{enumerate}
See~\cite{udlstmt} for a road map that summarizes how the above corners are connected in the duality web. 
A study from the target space perspective that expands~\cite{udlstmt} will appear in~\cite{longpaper}, which is complementary to the worldsheet treatment in this paper. In~\cite{udlstmt, longpaper}, the dilaton and RR fields are easier to be accessed. Moreover, derivations of Matrix theories in M$p$T, U-dualities between different decoupling limits of M-theory, the DLCQ of nonrelativistic M-theory, and multicritical M-theory that uplifts MM$p$T have been discussed in~\cite{udlstmt} and will be further detailed in~\cite{longpaper}. It is also interesting to note that the same duality web of decoupling limits can be accessed by mapping out the U-dual orbit by using invariant BPS mass formulae~\cite{bpslimits}, where it is indicated that a third layer of the duality web from performing three consecutive DLCQs\,\footnote{Note that intermediate U-dualities that map the lightlike circle to a spatial circle are required. See~\cite{udlstmt, longpaper}.} in M-theory is possible. Moreover, besides the fundamental strings and the light-excited brane configurations considered in this paper, we are also equipped with the necessary tools for systematically studying other D-brane objects in various decoupling limits. 
Also note that we have only focused on the bosonic sector through the paper. It would be natural to explore the supersymmetrization of the string sigma models considered here, for which useful ingredients can be borrowed from~\cite{Gomis:2004pw}. 

Last but not the least, as we have emphasized through this paper, it is will be interesting to further explore the concrete connections to strings that are nonrelativistic, tensionless, ambitwistor, Carrollian, tropological, \emph{etc}, as well as potential implications of this work for Matrix theory, the AdS/CFT correspondence and flat space holography.

\acknowledgments

We would like to thank Stefano Baiguera, Eric Bergshoeff, Chris Blair, Ritankar Chatterjee, Stephen Ebert, Kevin Grosvenor, Troels Harmark, Henrik Johansson, Johannes Lahnsteiner, Yang Lei, Niels Obers, Gerben Oling, Oliver Schlotterer, Bo Sundborg, Matthew Yu, and Konstantin Zarembo for useful discussions. JG would like to thank Perimeter Institute for their hospitality and suport during this work and 
Galileo Galilei Institute for Theoretical Physics for the hospitality and the INFN for partial support during the completion of this work. 
The research of JG was supported in part by
PID2019-105614GB-C21 and by the State Agency for Research of the Spanish Ministry of Science and Innovation 
through the Unit of Excellence Maria de Maeztu 2020-2023 award to the Institute of Cosmos Sciences (CEX2019-000918-M).
ZY would like to thank Groningen University, Scuola Normale Superiore di Pisa, Soochow University, Universitat de Barcelona, and Uppsala University for their hospitality and stimulating discussions. ZY is supported by the European Union’s Horizon 2020 research and innovation programme under the Marie Sk{\l}odowska-Curie grant agreement No 31003710. Nordita is supported in part by NordForsk.

\newpage

\appendix

\section{Elliptic Curve and Pinched Torus} \label{app:nodalcurve}

\begin{figure}[b!]
	\centering
	\includegraphics[width=0.25\textwidth]{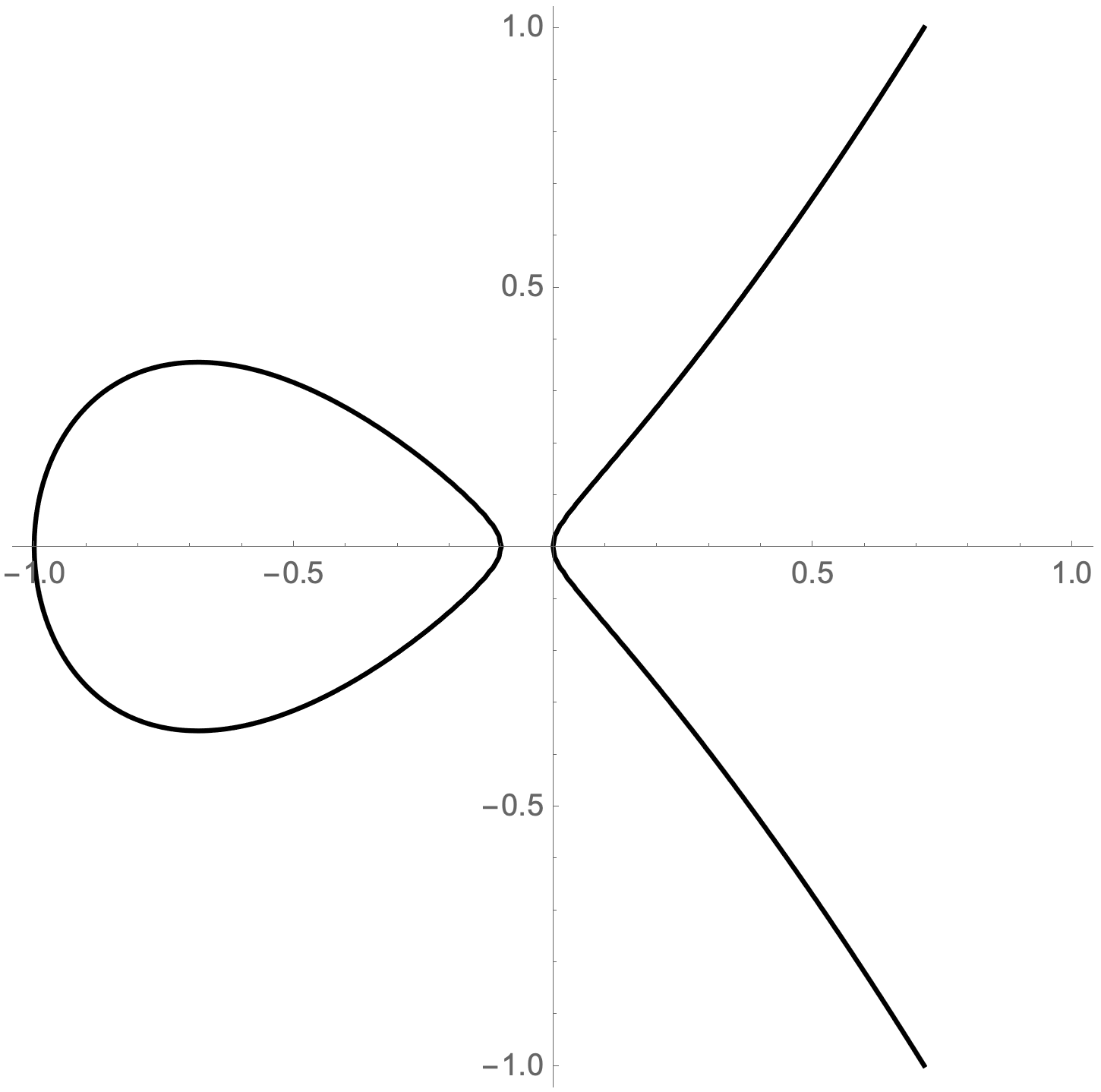}
		\qquad
	\includegraphics[width=0.25\textwidth]{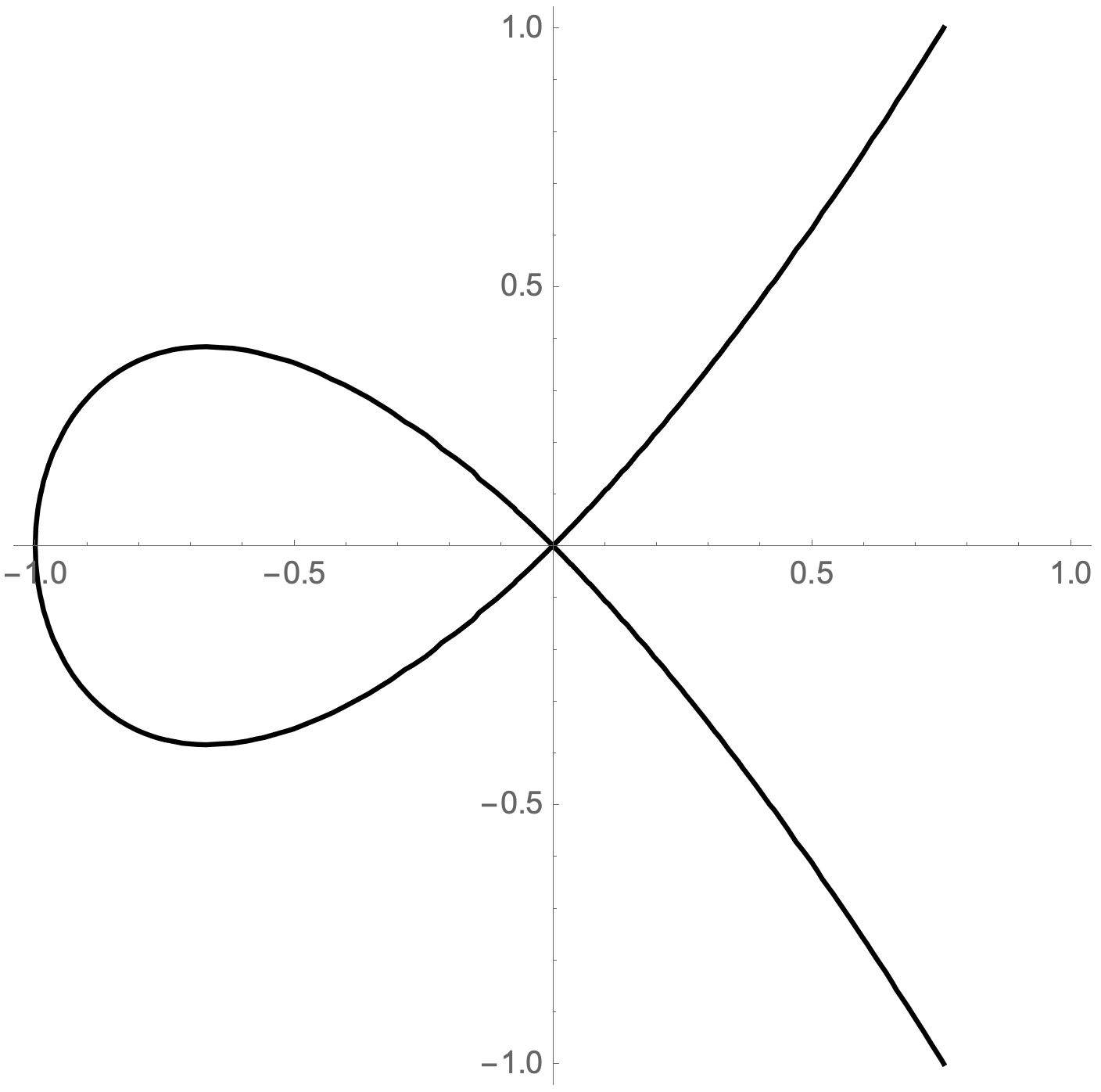}
		\qquad
	\includegraphics[width=0.35\textwidth]{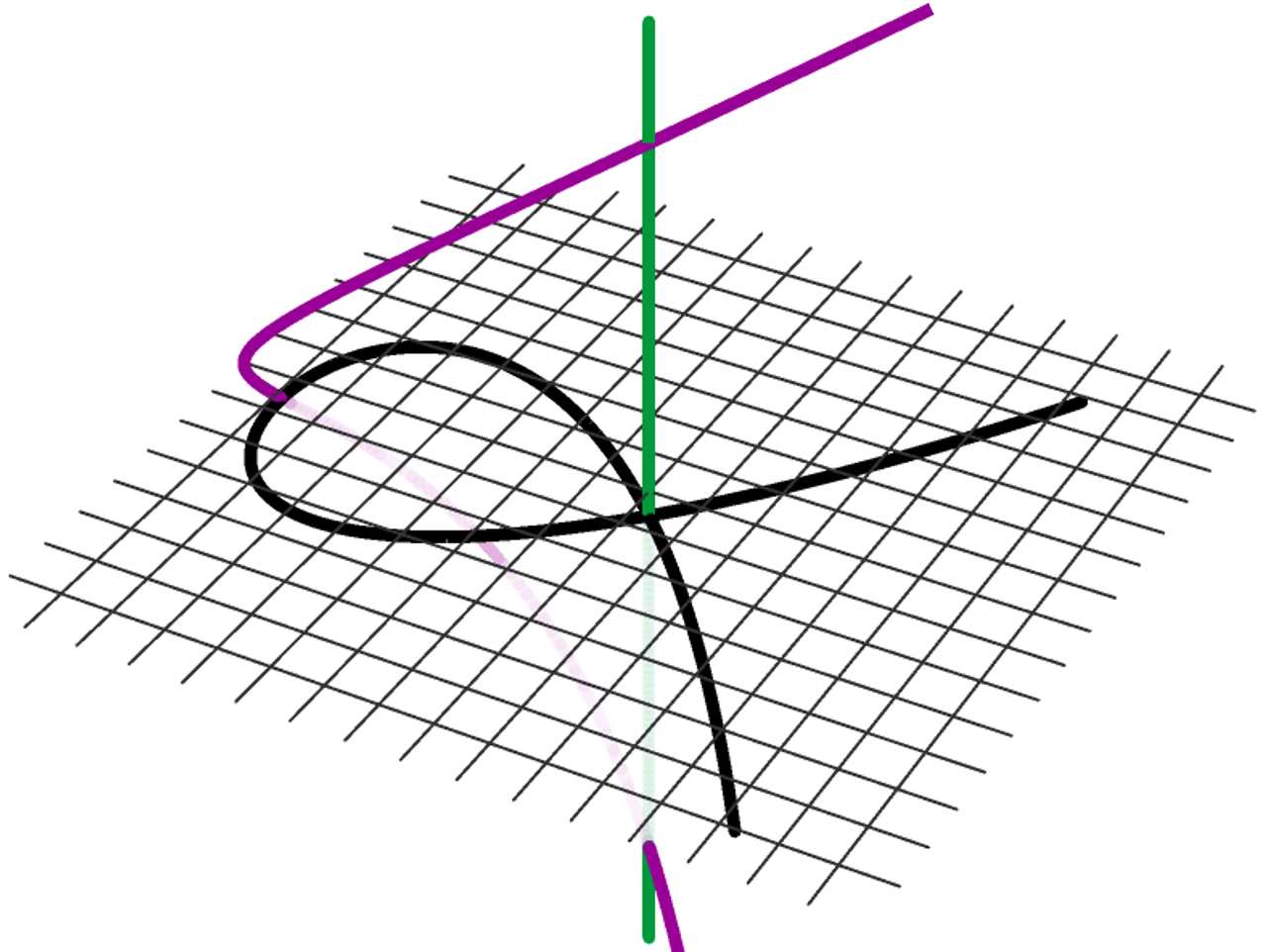}
	\caption{\textbf{Left}: the elliptic curve $y^2 \! = \! x (x+0.1) (x+1)$ that is topologically a torus. \textbf{Middle}: the nodal curve $y^2 \! = \! x^2 \, (x+1)$\,. \textbf{Right}: Blowing up the singularity. The black curve is the original nodal curve. The green line is (I) and the purple curve is (II) in Eq.~\eqref{eq:greencurve}.}
	\label{fig:nodalcurve}
\end{figure}

The relation between a pinched torus and Eq.~\eqref{eq:inftau}, where the modulus $\tau$ of a two-torus is sent to $i \, \infty$\,, becomes more manifest in the language of elliptic curves, which we discuss in this appendix\,\footnote{The discussions here mostly follow the online notes from
\href{https://www.math.purdue.edu/~arapura/graph/nodal.html}{\textcolor{black}{[Purdue]}} and \href{https://web2.ph.utexas.edu/~vadim/Classes/2013f/SW2.pdf}{\color{black}[UTexas]}. See \cite{silverman2009arithmetic} for a more comprehensive exposition of elliptic curves.}. Without loss of generality, we consider the curve
\be \label{eq:ec}
	y^2 = p(x)\,, 
		\qquad
	p (x) = x \, (x + \epsilon) \, (x + 1)\,.
\ee
The variables $x$ and $y$ are complex while $\epsilon$ is real. 
When $\epsilon \neq 0$\,, \emph{i.e.} $p(x)$ has three distinct roots, this elliptic curve is topologically a torus. However, when $\epsilon = 0$\,, \emph{i.e.}~$p(x)$ has a double root, this is a nodal curve that has a singular point at the origin with distinct tangent directions. See Fig.~\ref{fig:nodalcurve} for the plots of the real curves with $\epsilon = 0$ (the nodal case) and $\epsilon = 0.1$\,. 

Starting with the curve \eqref{eq:ec} with a small $\epsilon$\,. Take the $a$ cycle to be a circle of radius $|\epsilon|$ surrounding the poles at $x = 0$ and $x = - \epsilon$\,, and the $b$ cycle surrounding the poles at $x = -\epsilon$ and $x = - 1$\,. Using the single-valued differential $dz = dx / y$\,, the periods are
\be
	\omega_a = \int_a \frac{dx}{y} = - 2\pi\,,
		\qquad%
	\omega_b = \int_b \frac{dx}{y} = - 2 \, i \, \log \bigl( 16 \, \epsilon^{-1} \bigr)\,. 
\ee
The torus modulus is given by
\be
	\tau = \frac{\omega_b}{\omega_a} = \frac{i}{\pi} \log \bigl( 16 \, \epsilon^{-1} \bigr)\,.
\ee
In the $\epsilon \rightarrow \infty$ limit, we find $\tau \rightarrow i \, \infty$ as in Eq.~\eqref{eq:inftau}. Therefore, the nonrelativistic worldsheet in M0T due to the $\omega \rightarrow \infty$ limit is the same as the $\epsilon \rightarrow 0$ limit that leads to the nodal curve.   

\begin{figure}[t!]
	\centering
	\begin{minipage}{2.2cm}
	\includegraphics[width=1\textwidth]{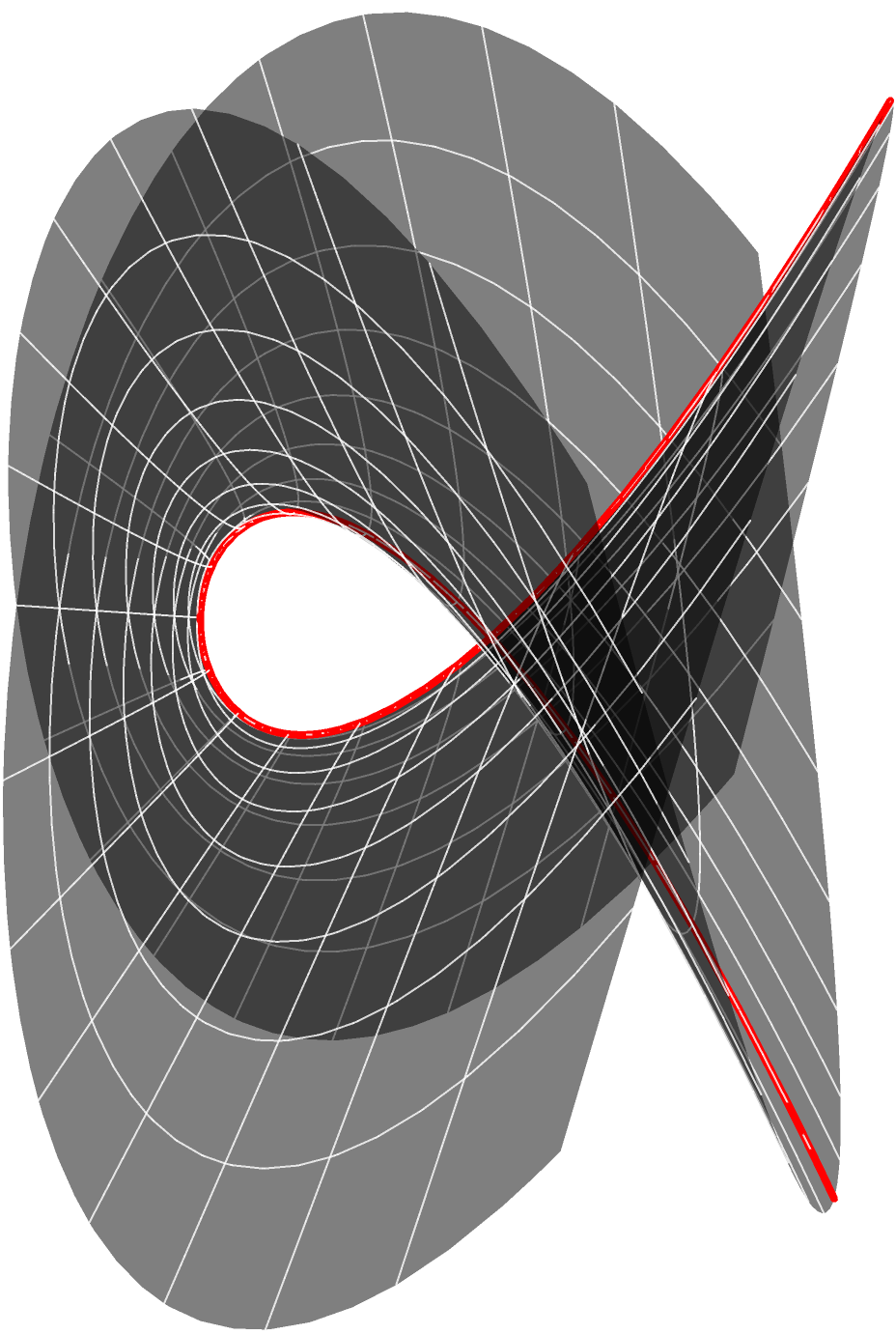}
	\end{minipage}
	\hspace{0cm}
        $\implies$
    \hspace{0cm}
	\begin{minipage}{3cm}
	\includegraphics[width=1\textwidth]{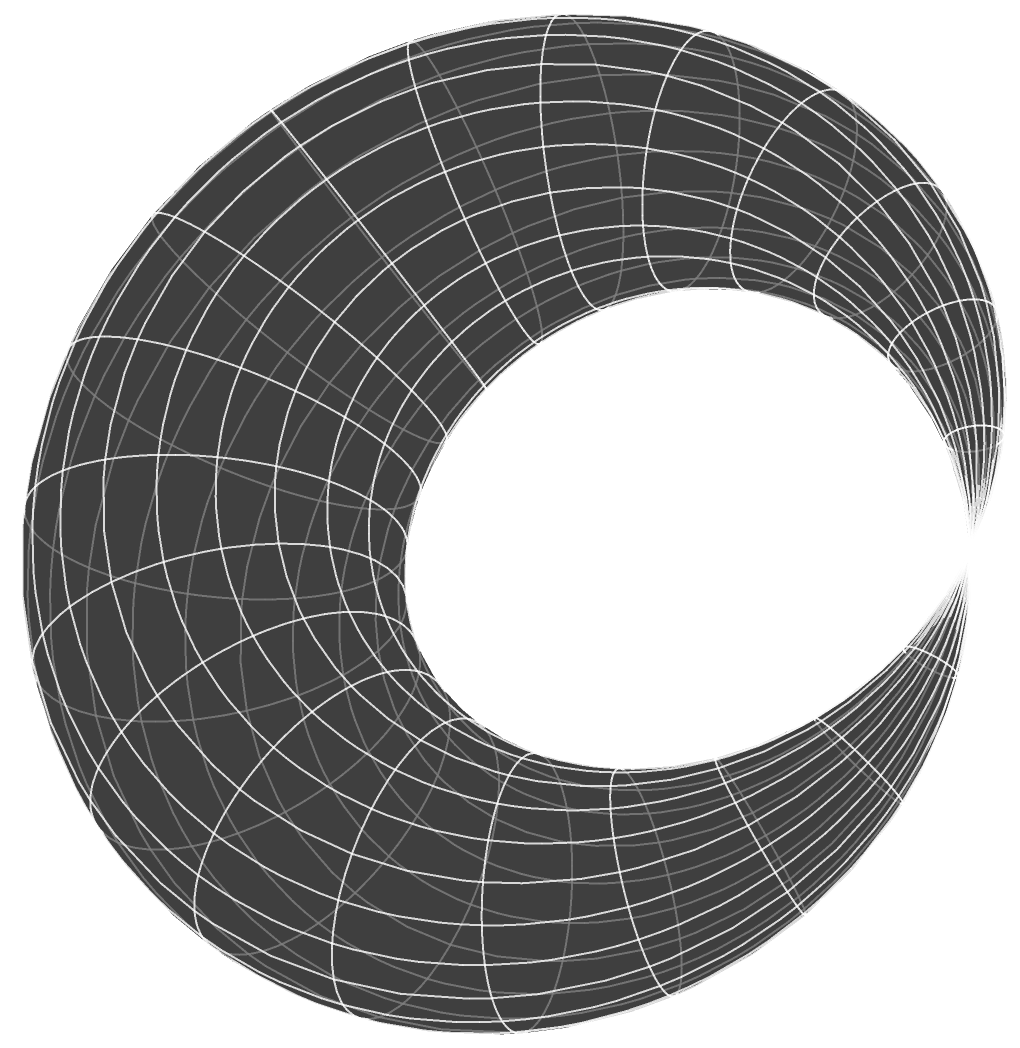}
	\end{minipage}
	\caption{\textbf{Left}: The graph of the blown-up complex nodal curve that is parametrized in terms of $u$ and $v$ in Eq.~\eqref{eq;rexy}. The Cartesian coordinates are $(\text{Re}(x)\,, \text{Im}(x)\,, \text{Re}(y))$\,. The red curve is the real nodal curve. \textbf{Right}: The nodal graph on the left can be deformed into a nodal Riemann sphere after identifying the infinity as a point.}
	\label{fig:ng}
\end{figure}

In order to understand the complex graph defined by the nodal cubic 
\be \label{eq:nodalcurve}
	y^2 = x^2 \, (x+1)\,,
\ee 
we \emph{blow up} the singularity by perform the change of variable $y = t \, x$\,, where the parts of the nodal curve that intersect at the origin are separated, with the variable $t$ representing the slopes. Plugging $y = t \, x$ into Eq.~\eqref{eq:nodalcurve}, we find that the cubic factorizes into two components:  
\be \label{eq:greencurve}
	\text{(I)} \quad x = 0\,, \quad y = 0\,, \quad z = t\,;
		\qquad%
	\text{(II)} \quad x = t^2 - 1\,, \quad y = t^3 - t\,, \quad z = t\,.
\ee
Here, (I) is the green curve and (II) is the purple curve in the plot on the right of Fig.~\ref{fig:nodalcurve}. Note that (II) This is a rational parametrization of the nodal curve.
In the complex case, we write $t = u + i \, v$\,, which implies
\begin{subequations} \label{eq;rexy}
\begin{align}
	\text{Re}(x) &= u^2 - v^2 - 1\,,
		&%
	\text{Re}(y) &= u^3 - 3 \, u \, v^2 - u\,, \\[4pt]
	\text{Im}(x) &= 2 \, u \, v\,,
		&%
	\text{Im}(y) &= 3 \, u^2 \, v - v^3 - v\,.
\end{align}
\end{subequations}
In Fig.~\ref{fig:ng}, we draw the graph of the blown-up complex nodal curve as parametrized in Eq.~\eqref{eq;rexy}, where it is shown that this is topologically a nodal Riemann sphere. This demonstrates that the torus becomes a nodal Riemann sphere in the $\tau \rightarrow i \, \infty$ limit. 

We only focused on the nodal singularity of the elliptic curve here. In the case where $p(x)$ in Eq.~\eqref{eq:ec} takes a different form that has a triple root, the elliptic curve develops a singular point called \emph{cusp} where there is only one tangent direction. In this case, the elliptic curve is topologically a knot.

\newpage

\bibliographystyle{JHEP}
\bibliography{wfdlst_v5}

\end{document}